\documentclass[twocolumn]{aa}
\pdfoutput=1
\usepackage[multidot]{grffile}
\usepackage{graphicx}
\usepackage{txfonts}
\usepackage{natbib}
\bibpunct{(}{)}{;}{a}{}{,}

\begin{document}

\title
{
Background derivation and image flattening: \textsl{getimages}
}


\author
{
A.~Men'shchikov 
}


\institute
{
Laboratoire AIM Paris--Saclay, CEA/DSM--CNRS--Universit{\'e} Paris Diderot, IRFU, Service d'Astrophysique,\\
Centre d'Etudes de Saclay, Orme des Merisiers, 91191 Gif-sur-Yvette, France\\
\email{alexander.menshchikov@cea.fr}
}

\date{Received 3 April 2017 / Accepted 26 July 2017}

\offprints{Alexander Men'shchikov}
\titlerunning{Background derivation and image flattening: \textsl{getimages}}
\authorrunning{Men'shchikov}


\abstract
{ 
Modern high-resolution images obtained with space observatories display extremely strong intensity variations across images on
all spatial scales. Source extraction in such images with methods based on global thresholding may bring unacceptably large numbers
of spurious sources in bright areas while failing to detect sources in low-background or low-noise areas. It would be highly
beneficial to subtract background and equalize the levels of small-scale fluctuations in the images before extracting sources or
filaments. This paper describes \textsl{getimages}, a new method of background derivation and image flattening. It is based on
median filtering with sliding windows that correspond to a range of spatial scales from the observational beam size up to a maximum
structure width $X_{\lambda}$. The latter is a single free parameter of \textsl{getimages} that can be evaluated manually from the
observed image $\mathcal{I}_{\!\lambda}$. The median filtering algorithm provides a background image
$\tilde{\mathcal{B}}_{\lambda}$ for structures of all widths below $X_{\lambda}$. The same median filtering procedure applied to an
image of standard deviations $\mathcal{D}_{\lambda}$ derived from a background-subtracted image $\tilde{\mathcal{S}}_{\lambda}$
results in a flattening image $\tilde{\mathcal{F}}_{\!\lambda}$. Finally, a flattened detection image
$\mathcal{I}_{{\!\lambda}\mathrm{D}}{\,=\,}\tilde{\mathcal{S}}_{\lambda}{/}\tilde{\mathcal{F}}_{\!\lambda}$ is computed, whose
standard deviations are uniform outside sources and filaments. Detecting sources in such greatly simplified images results in much
cleaner extractions that are more complete and reliable. As a bonus, \textsl{getimages} reduces various observational and
map-making artifacts and equalizes noise levels between independent tiles of mosaicked images.
} 
\maketitle


\section{Introduction}
\label{introduction}

Recent high-resolution and high-sensitivity observations of large areas of Galactic star-forming regions with the \emph{Herschel}
Space Observatory \citep{Pilbratt_etal2010} revealed dense interstellar clouds as extremely complex. Multiwavelength images of the
numerous regions observed within the \emph{Herschel} Gould Belt \citep{Andre_etal2010} and HOBYS \citep{Motte_etal2010} surveys
display bright and strongly variable backgrounds (on all spatial scales), blended with the omnipresent filamentary structures and
many hundreds of (crowded) sources of different nature \citep[e.g.,][]{Men'shchikov_etal2010,Andre_etal2014}. The observed extreme
complexity reflects that of the astrophysical reality, which is further exacerbated by its projection onto the sky plane. To
analyze and understand the reality, it is necessary to separate the blended components: sources, filaments, and background (and/or
noise).

The enormous complexity of observed images presents serious problems for the fully automated source extraction methods that apply
global thresholds \citep[e.g., \textsl{sextractor}, \textsl{cutex}, \textsl{getsources},
\textsl{getfilaments};][]{BertinArnouts1996, Molinari_etal2011,Men'shchikov_etal2012,Men'shchikov2013} in order to separate sources
from background and noise. The use of thresholds in terms of the background brightness or fluctuation levels that are computed for
an entire image is based on the assumption that these thresholds are approximately uniform across the image. The observations of
complex backgrounds demonstrate, however, that intensities and their fluctuations both often vary by several orders of magnitude in
different parts of large fields. It is easy to see that any single threshold value would be inappropriate for such images, either
producing many spurious sources or leaving many sources undetected. For a complete and reliable source detection, it would be
highly desirable to simplify the complex observed images by removing signals from irrelevant spatial scales and equalizing the
background fluctuation levels (outside sources) over the images. A solution of this non-trivial problem might require iterations,
as the locations of sources are unknown before source extraction.

An attempt to solve this problem was made in two previous papers \citep[][hereafter referred to as Paper I and Paper
II]{Men'shchikov_etal2012,Men'shchikov2013}, where the multiscale, multiwavelength source and filament extraction methods
\textsl{getsources} and \textsl{getfilaments} were presented. The algorithms employ two iterations for a complete source
extraction. An initial (somewhat deeper) extraction aims to detect all sources (possibly including spurious sources) and estimate
their sizes. After extracting and removing the sources and filaments, the method attempts to equalize fluctuation levels of the
remaining background. A second (and final) extraction uses the \emph{flattened} detection images to identify sources more reliably,
reducing the chances of detecting spurious sources. Although this approach works well for some types of images, it is not universal
and therefore not fully satisfactory. In the most complex images, the initial extraction tends to produce large areas of many
overlapping sources. Extremely complex observed backgrounds, in combination with large crowded areas of overlapping sources, could
make the derived background inexact and hence render that approach inaccurate in some applications.

This paper presents a much better solution of the problem of background derivation and image flattening that does not require any
prior source extraction or knowledge of the locations and sizes of sources. The new method, called \textsl{getimages}, is based on
a simple and straightforward transformation of the observed images using median filtering.

A well-known technique \citep[][]{Tukey1977}, median filtering has been widely used during the past four decades in image
processing, especially to reduce noise or small-scale artifacts. The idea of this image transformation is to replace the value of
each pixel with a median value computed over all pixels in a window centered on the pixel that is being modified. The size of the
sliding window is a free parameter of this technique. For the purpose of reducing sharp pixel-to-pixel intensity jumps, even a
small $3{\,\times\,}3$ pixel window may be sufficient, whereas larger window sizes are necessary to remove wider features.

Two-dimensional median filtering has the very useful property of efficiently truncating any peaks -- those produced by instrumental
noise, intensity fluctuations of a background molecular cloud, filamentary structures, or astrophysical emission sources of any
type. Finding a universal way of removing sources or other structures of all sizes of interest would essentially mean estimating
their background, a major step toward designing a reliable method of flattening background-subtracted images to ensure uniform
fluctuation levels in all their parts. These are the ideas that initiated the development of \textsl{getimages}.

As in Papers I and II, images are represented by capital calligraphic characters (e.g., $\mathcal{A}, \mathcal{B}, \mathcal{C}$) to
distinguish them from other parameters. The median filtering operator using a window of radius $R$ is denoted as
$\mathrm{mf}_{R\,}(\mathcal{I})$, and the standard deviations operator using a circular $n$-pixels window is denoted as
$\mathrm{sd}_{n\,}(\mathcal{I})$. The new background derivation and image flattening method is described in Sect.~\ref{bgsflat} and
discussed in Sect.~\ref{discussion}, the conclusions are outlined in Sect.~\ref{conclusions}, and further details are presented in
Appendices \ref{AppendixA}--\ref{AppendixC}.

\section{Background estimation and flattening}
\label{bgsflat}

\subsection{Median filtering as a structure removal operator}
\label{sourceremoval}

The point-spread functions (PSFs, beams) of modern telescopes are usually represented by two-dimensional Gaussians down to percent
levels, below which the beams become more complicated\footnote{The \emph{Herschel} beams are described in the PACS Observer's
Manual and the SPIRE Handbook at \url{http://herschel.esac.esa.int/Docs}}. Observations show mostly unresolved or slightly resolved
structures with Gaussian profiles. However, some bright and well-resolved structures are better approximated by a Gaussian core
with power-law wings. In addition to the simple Gaussian shapes, this paper therefore also considers profiles with the functional
form used in Papers I and II:
\begin{equation} 
I(\theta) = I_{\mathrm{P}} \left(1+f(\zeta)\,(\theta/\Theta)^2\right)^{-\zeta},
\label{moffatfun}
\end{equation} 
where $I_{\mathrm{P}}$ is the structure peak intensity, $\theta$ the angular distance, $\Theta$ the structure half-width at
half-maximum, $\zeta$ a power-law exponent, and ${f(\zeta){\,=\,}(2^{1/\zeta}-1)}$ adjusts the profile to have the same $\Theta$
for all values of $\zeta$. This function has an almost Gaussian shape in its core, transforming into a power-law profile
${I(\theta){\,\propto\,}\theta^{\;\!-2\,\zeta}}$ for ${\theta{\,\gg\,}\Theta}$. For simplicity, $\zeta{\,=\,}1$ is fixed in this
section, and the term ``power law'' refers to the shapes with ${I(\theta){\,\propto\,}\theta^{\;\!-2}}$ at large $\theta$. The
Gaussian and power-law shapes can represent both starless cores and protostellar envelopes (cf. Appendix~\ref{AppendixA}),
depending on their signal-to-noise (S/N) ratio and on the degree to which they are resolved.

\begin{figure*}
\centering
\centerline{\resizebox{0.3640\hsize}{!}{\includegraphics{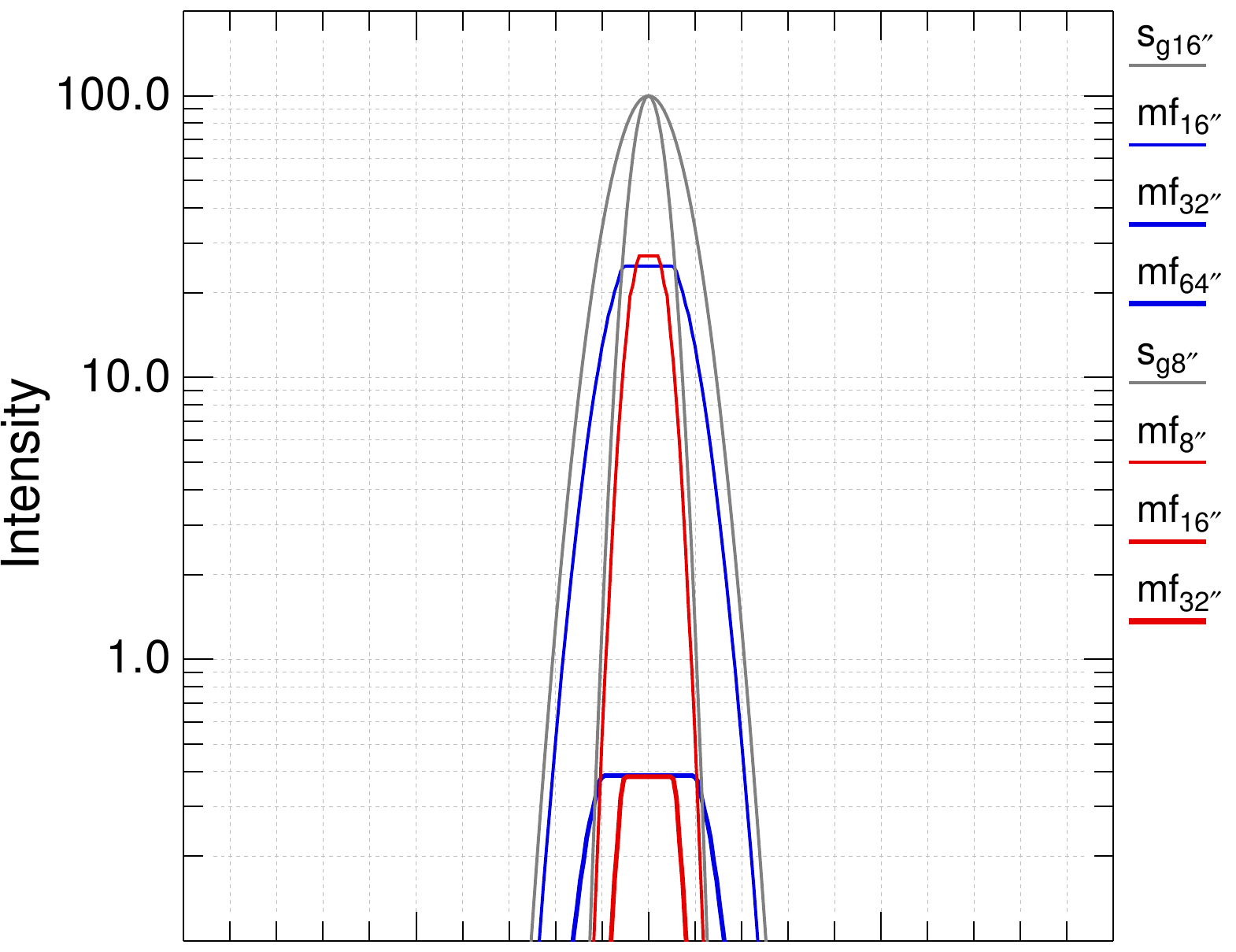}}
            \resizebox{0.3180\hsize}{!}{\includegraphics{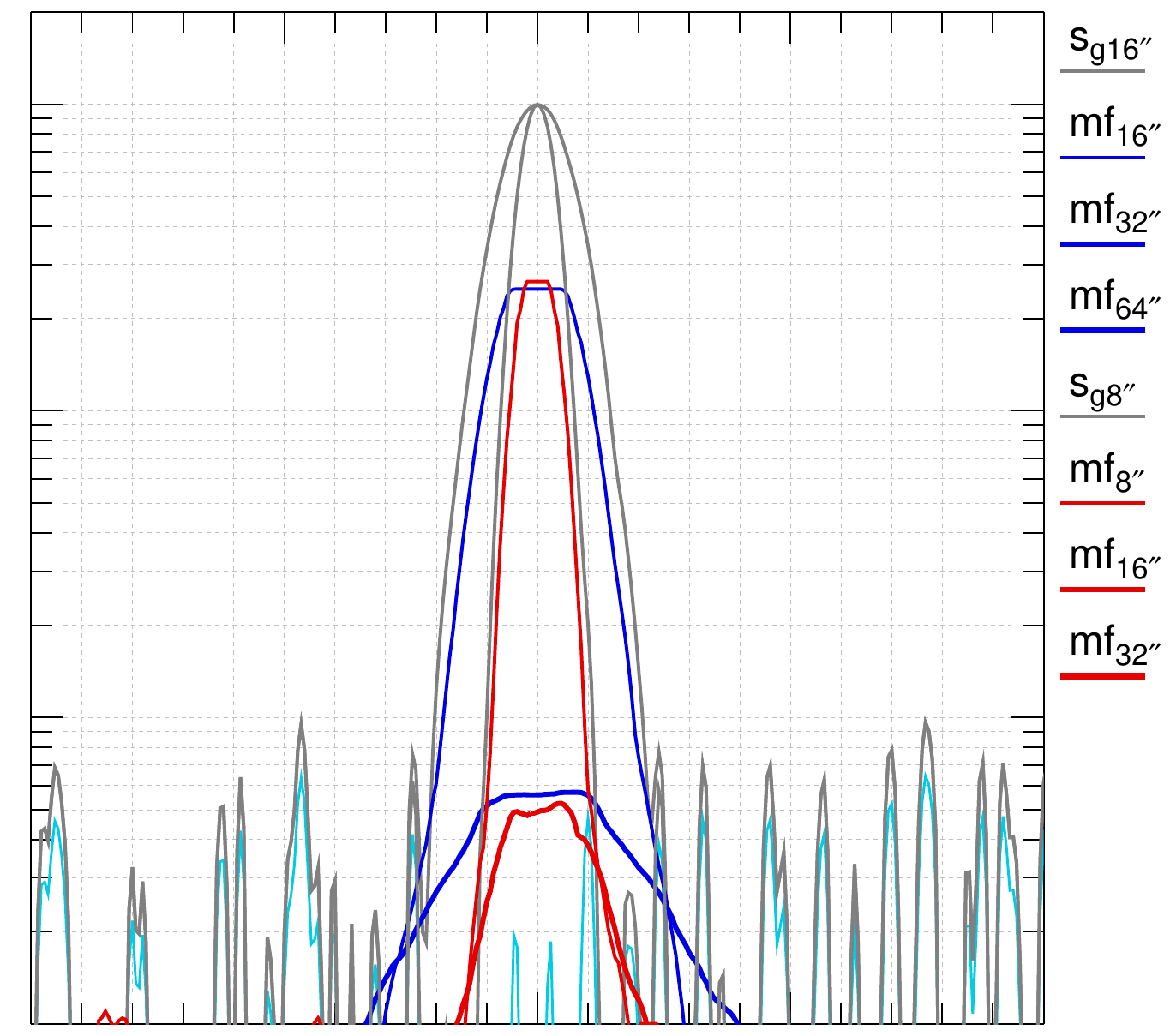}}
            \resizebox{0.3180\hsize}{!}{\includegraphics{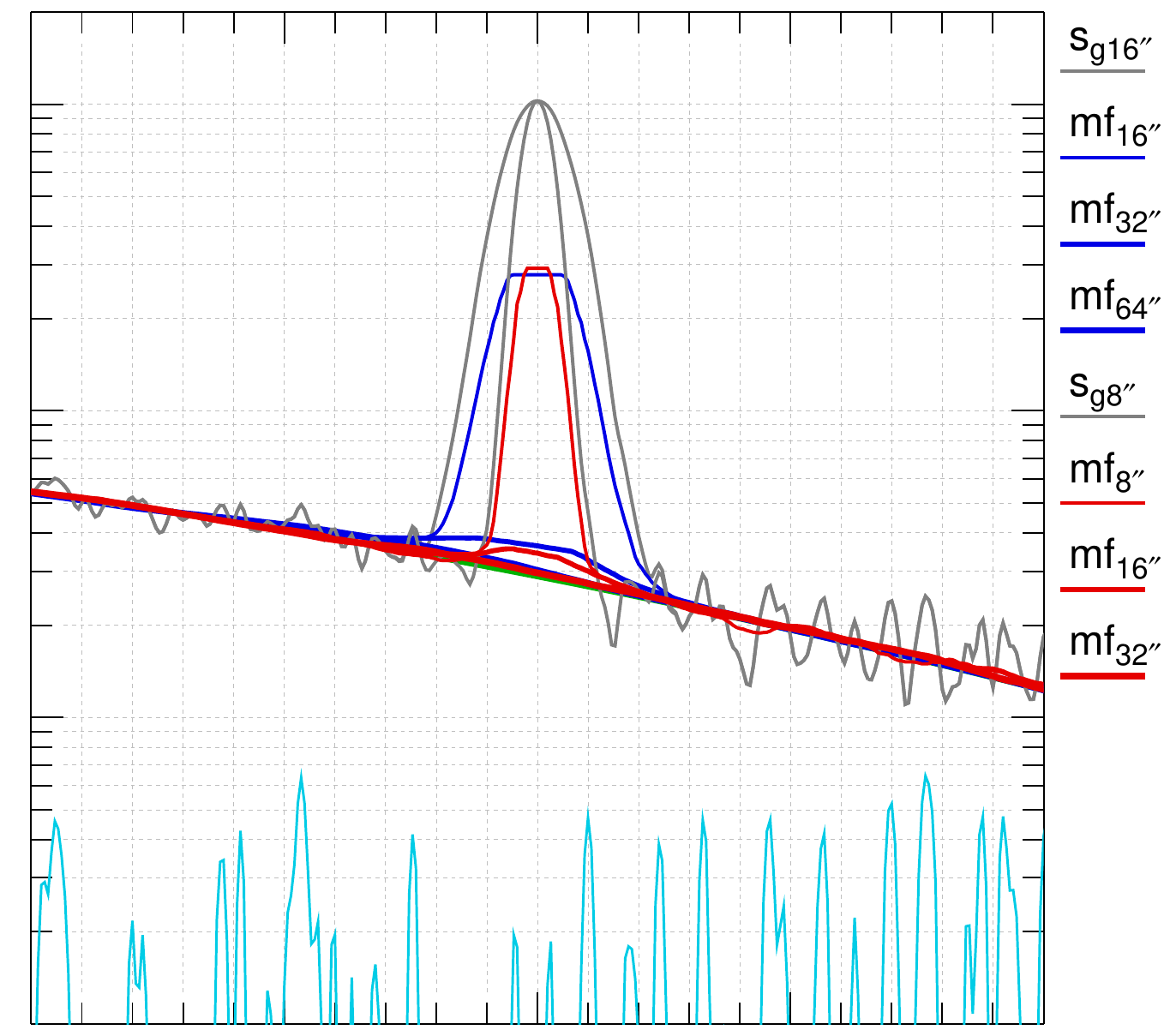}}}
\centerline{\resizebox{0.3640\hsize}{!}{\includegraphics{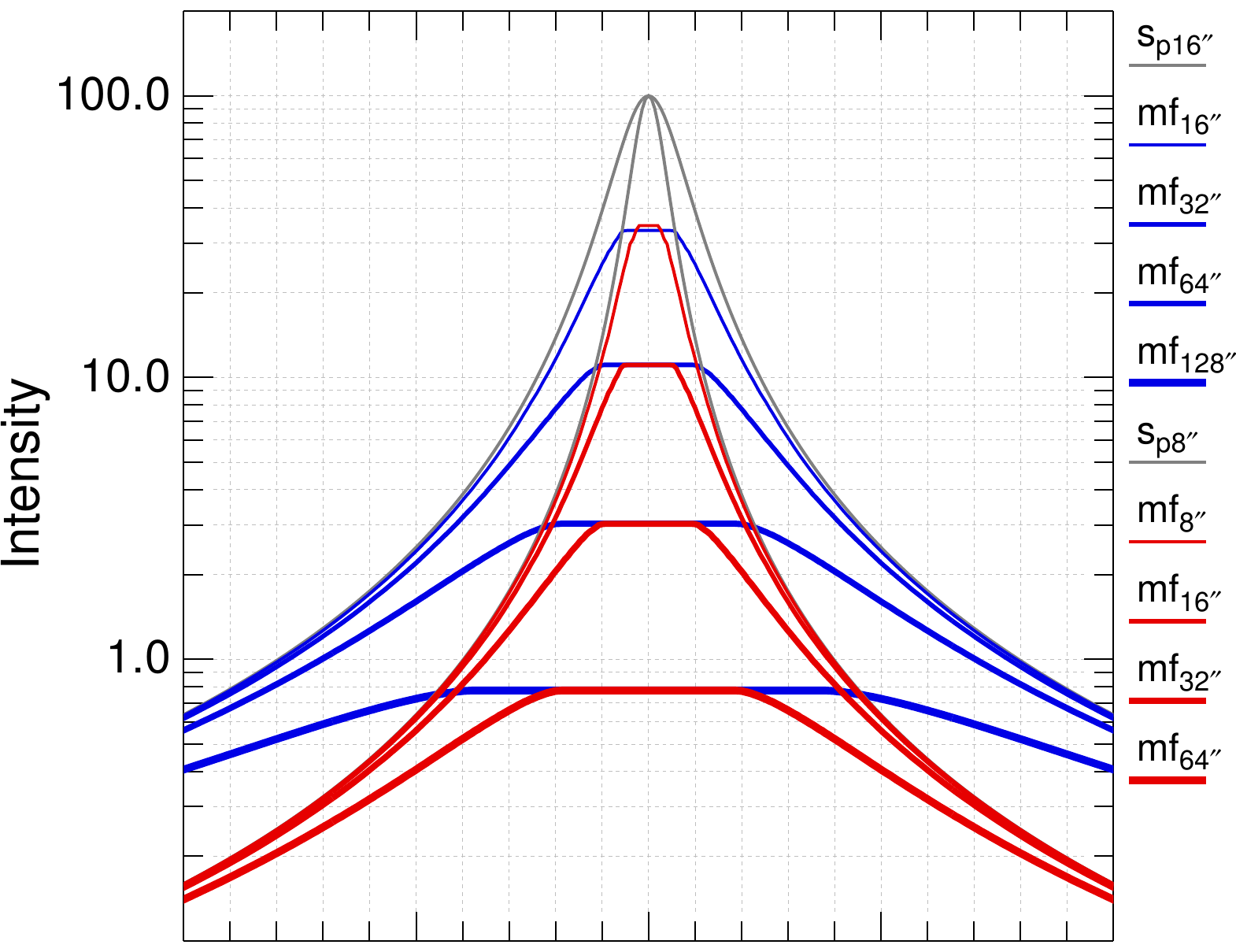}}
            \resizebox{0.3180\hsize}{!}{\includegraphics{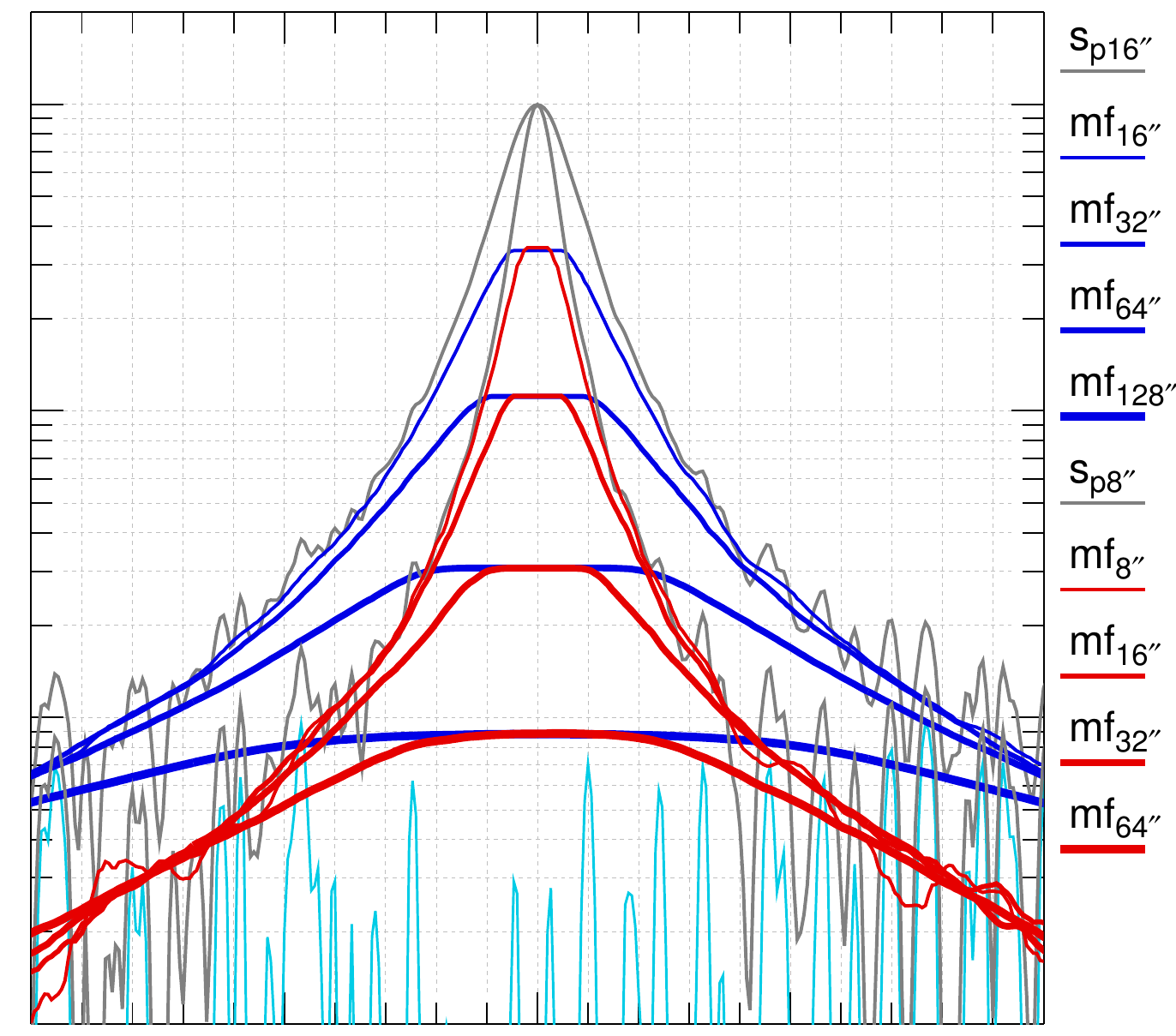}}
            \resizebox{0.3180\hsize}{!}{\includegraphics{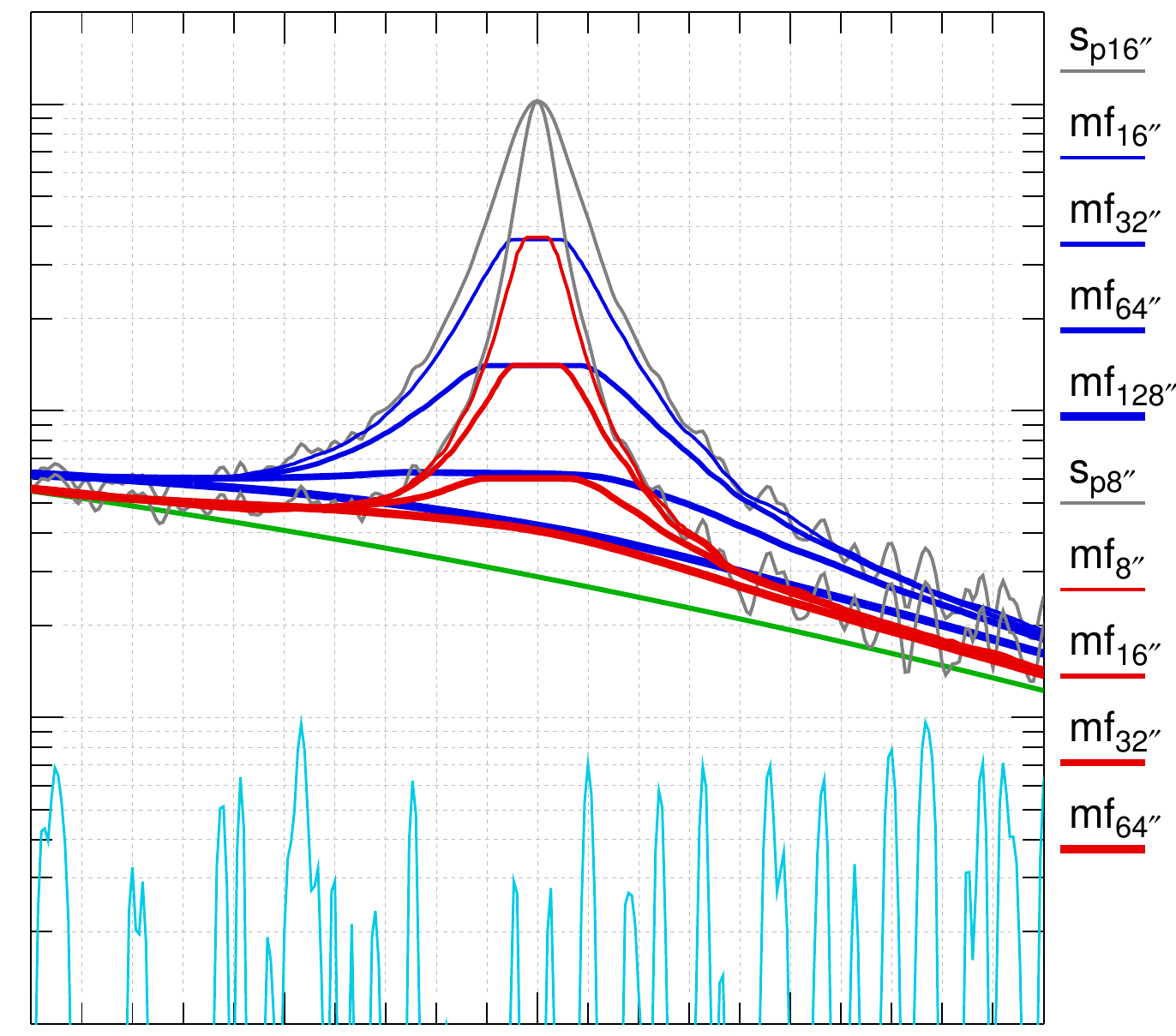}}}
\centerline{\resizebox{0.3640\hsize}{!}{\includegraphics{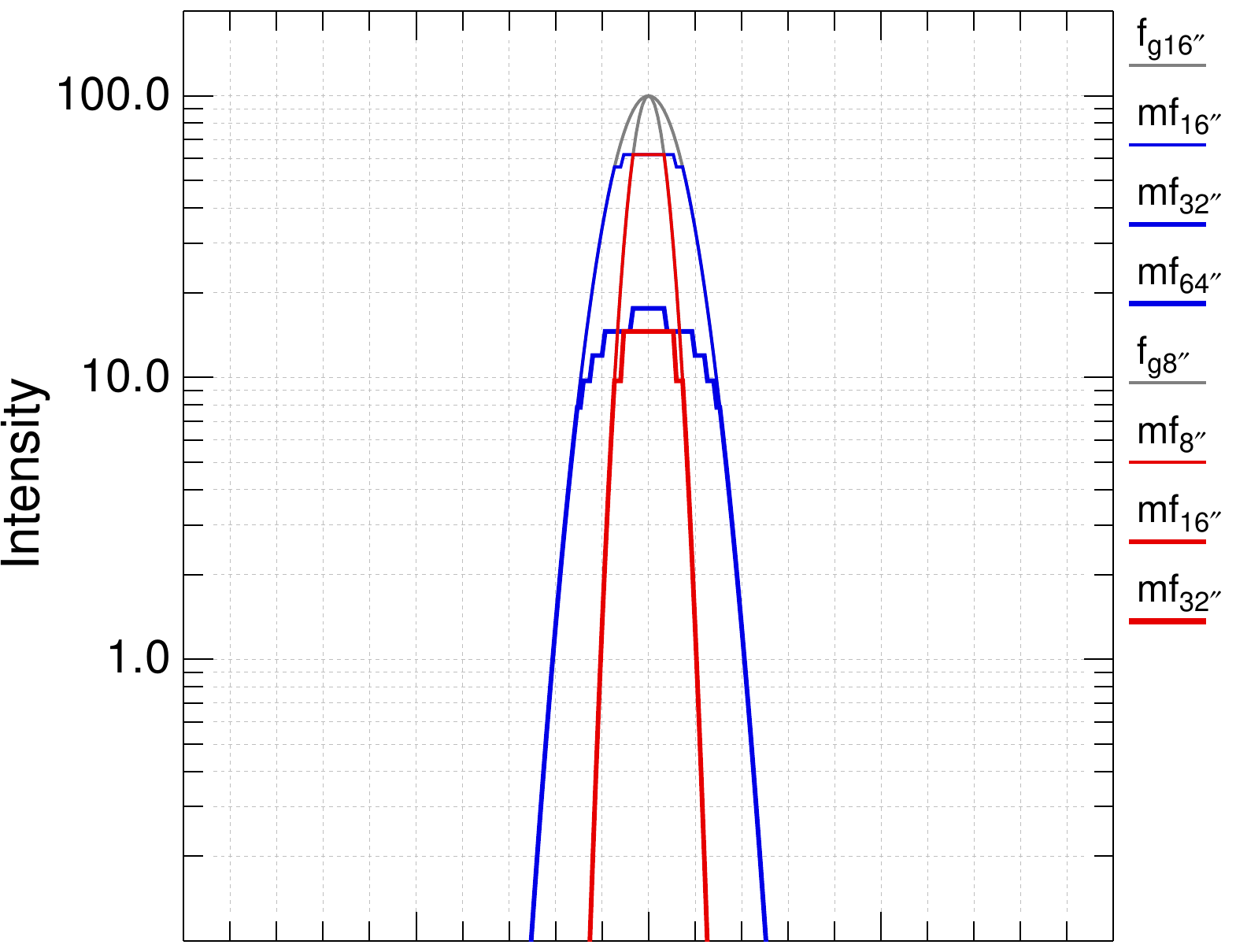}}
            \resizebox{0.3180\hsize}{!}{\includegraphics{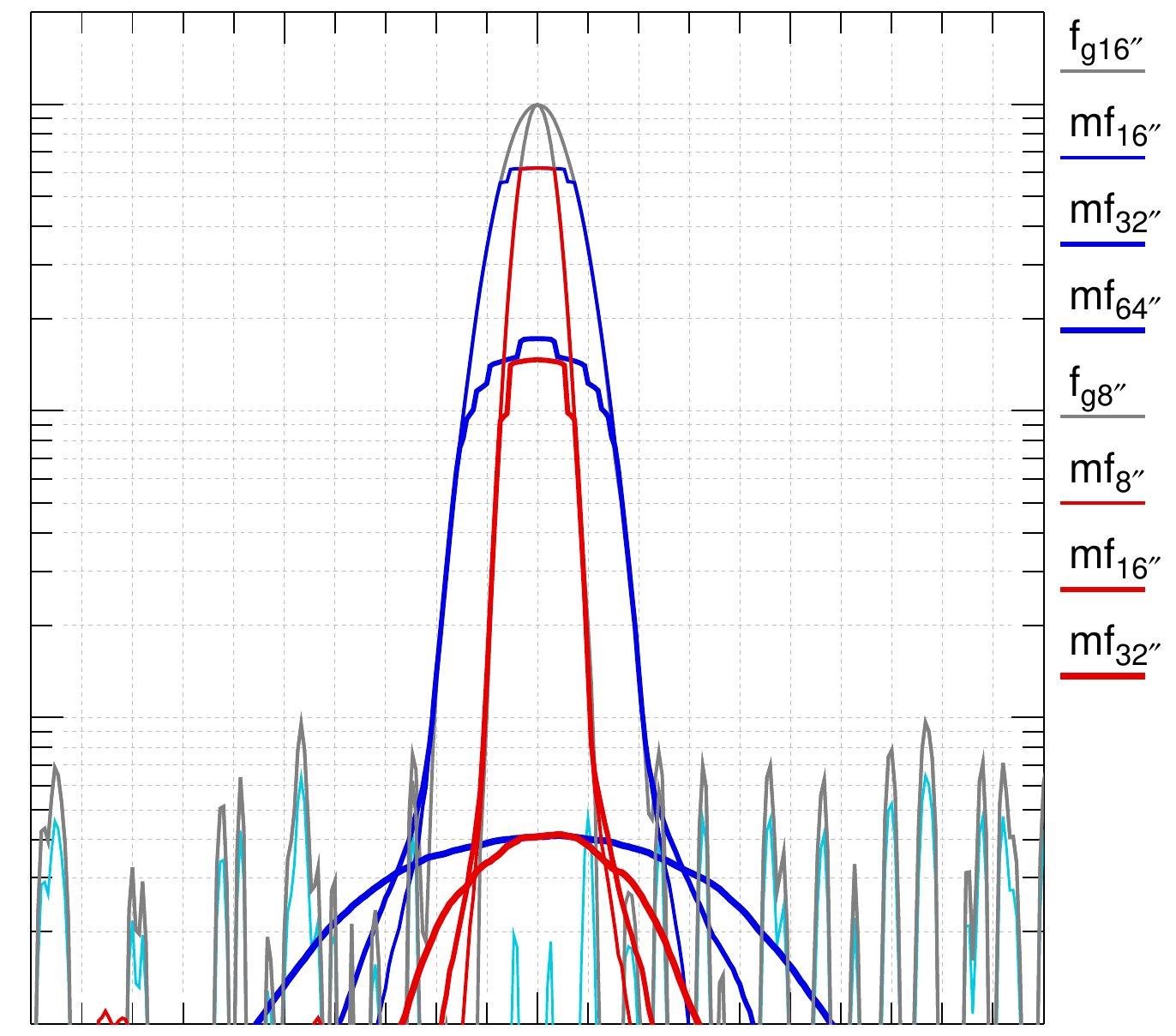}}
            \resizebox{0.3180\hsize}{!}{\includegraphics{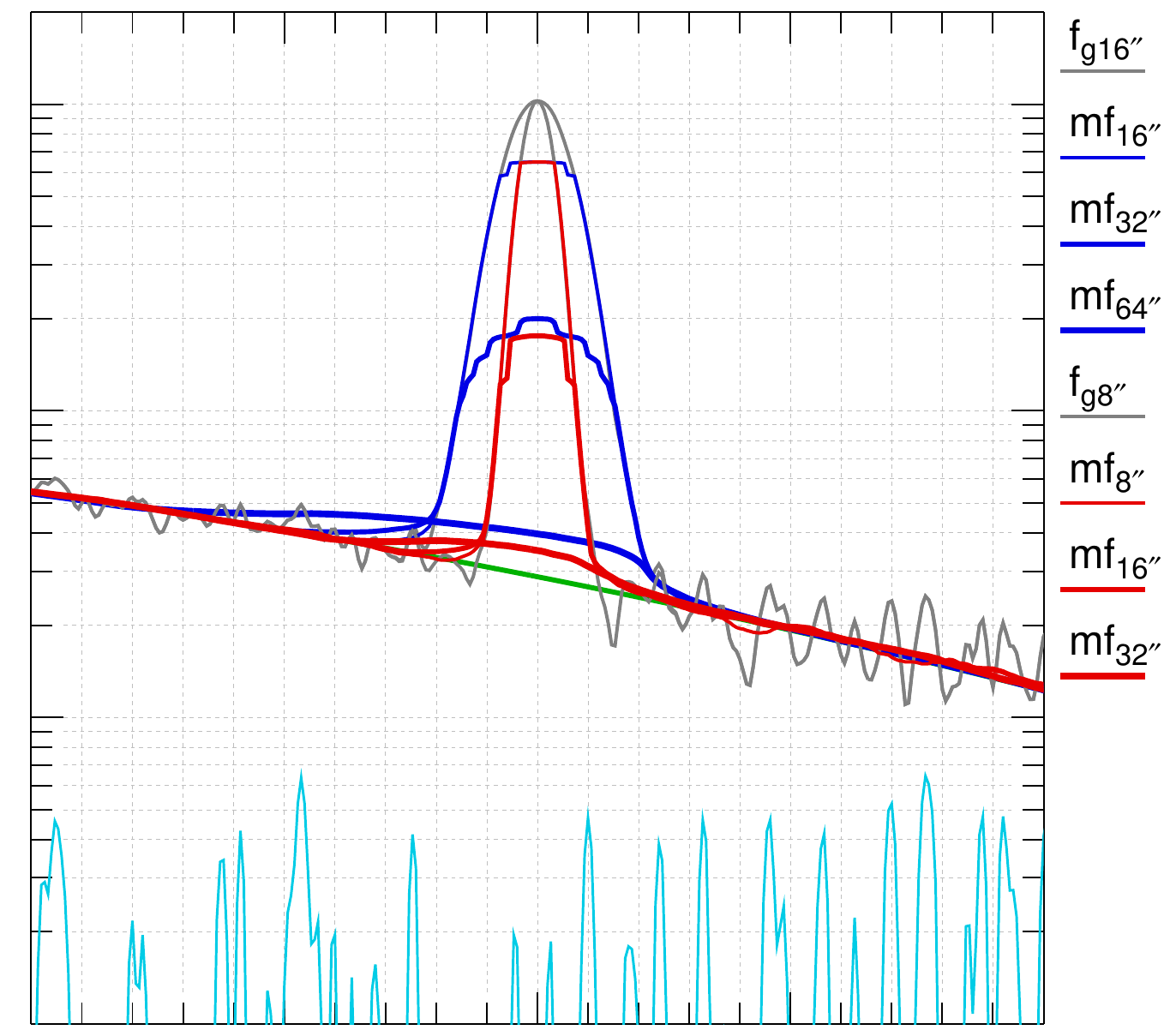}}}
\centerline{\resizebox{0.3640\hsize}{!}{\includegraphics{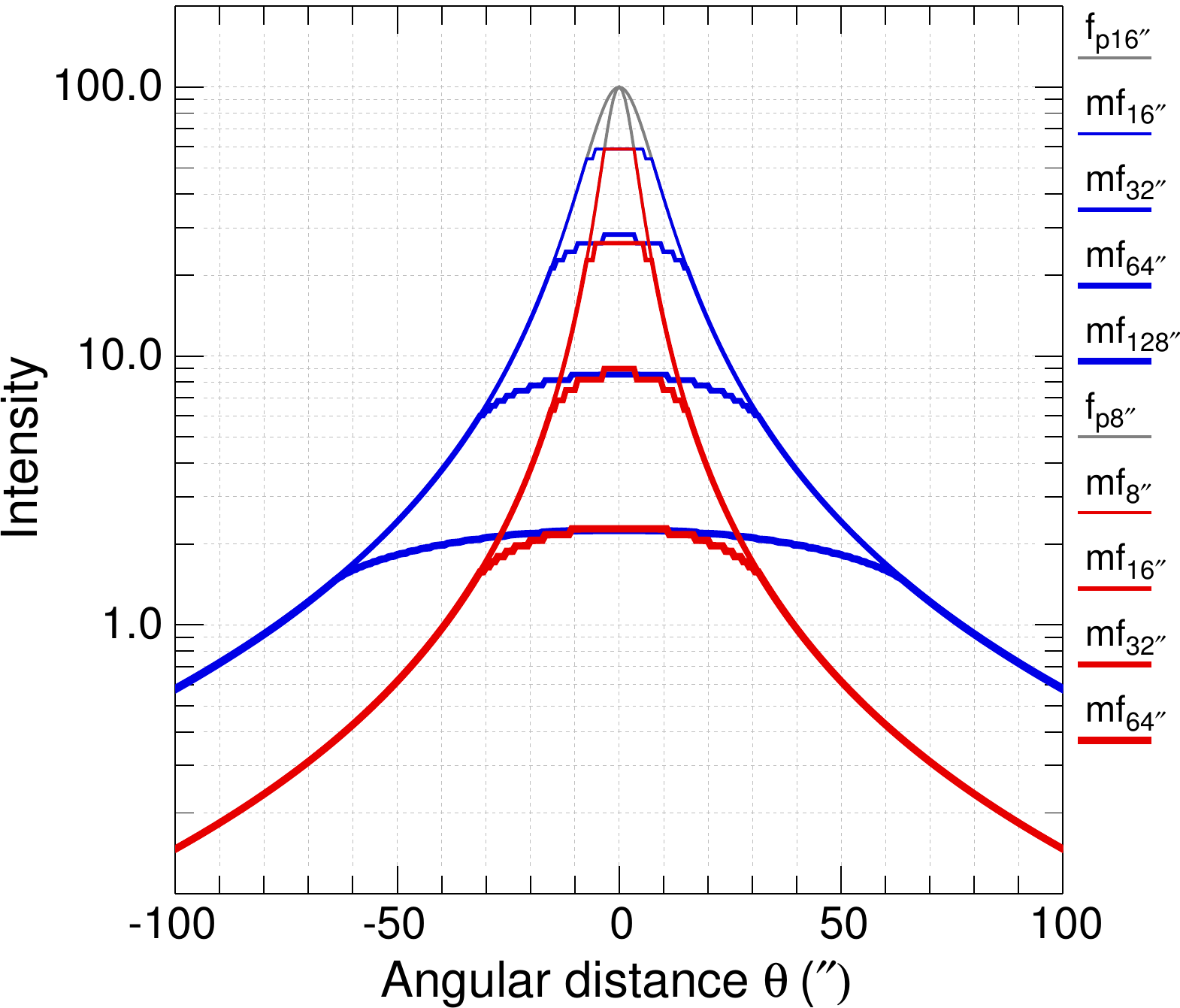}}
            \resizebox{0.3180\hsize}{!}{\includegraphics{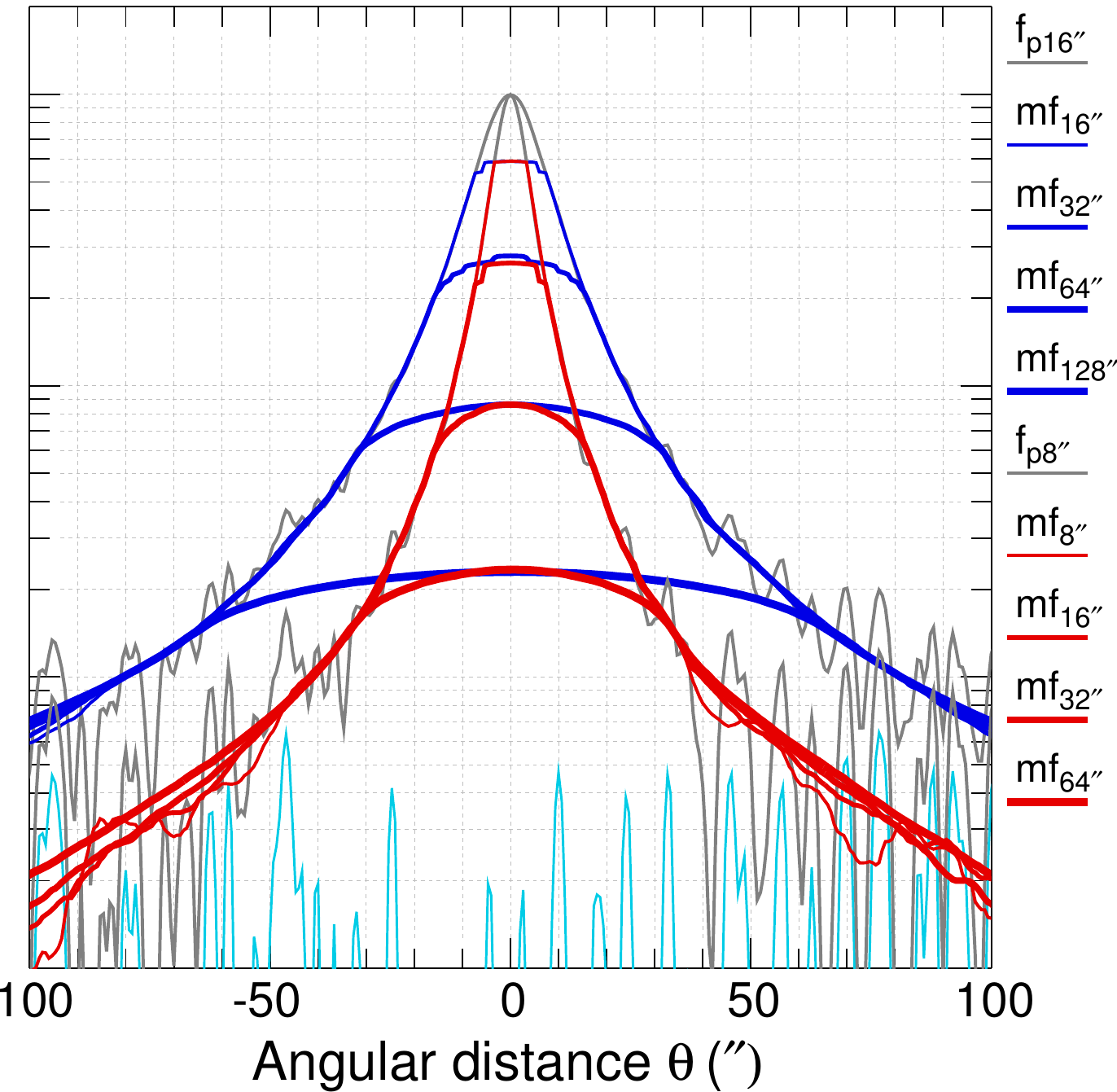}}
            \resizebox{0.3180\hsize}{!}{\includegraphics{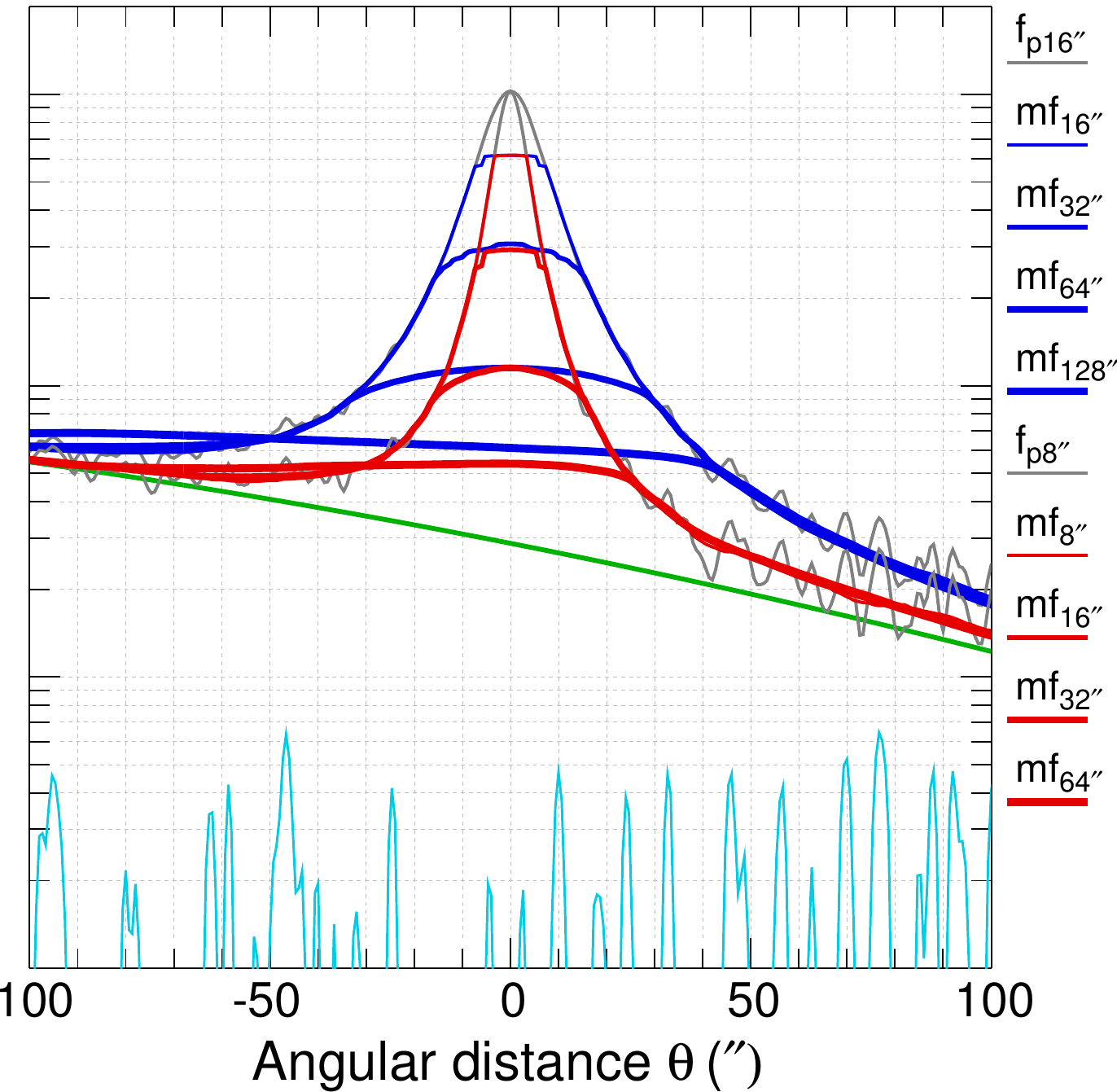}}}
\caption
{ 
Median filtering applied to Gaussian sources $\mathcal{S}_{\mathrm{G}}$ (\emph{upper row}), power-law sources
$\mathcal{S}_{\mathrm{P}}$ (\emph{second row}), Gaussian filaments $\mathcal{F}_{\mathrm{G}}$ (\emph{third row}), and power-law
filaments $\mathcal{F}_{\mathrm{P}}$ (\emph{bottom row}) with sizes $H$ (FWHM) of $8${\arcsec} and $16${\arcsec} (\emph{left}) and
to the same images after addition of random Gaussian noise $\mathcal{N}$ with $\sigma{\,=\,}0.33$ ({\emph{middle}}) and background
$\mathcal{B}$ modeled as a large Gaussian with FWHM of $512${\arcsec} (\emph{right}). The radii $R$ of circular sliding windows
given by the indices on the curve labels correspond to the relative radii $W{\,=\,}R/H{\,=\,}\{1, 2, 4, 8\}$. Original intensity
profiles of $\mathcal{S}_{\mathrm{G}}$, $\mathcal{S}_{\mathrm{P}}$, $\mathcal{F}_{\mathrm{G}}$, and $\mathcal{F}_{\mathrm{P}}$ are
shown in gray, filtered profiles of the sources and filaments with $H{\,=\,}8$ and $16${\arcsec} are plotted in red and blue,
whereas profiles of $\mathcal{N}$ and $\mathcal{B}$ are colored in cyan and green, respectively. Large sliding windows
($W{\,>\,}2$) truncate Gaussian shapes so efficiently (cf. Table~\ref{truncation}) that some of the annotated curves become
invisible (fall entirely below the lower edge of the plots).
} 
\label{gausspower}
\end{figure*}

\begin{table*}
\caption
{ 
Truncation factors for median-filtered sources ($\mathcal{S}_{\mathrm{G}}$, $\mathcal{S}_{\mathrm{P}}$) and filaments
($\mathcal{F}_{\mathrm{G}}$, $\mathcal{F}_{\mathrm{P}}$) using sliding windows with different radii $W{\,=\,}R/H$, corresponding to
the images described in Sect.~\ref{sourceremoval} and their profiles plotted in Fig.~\ref{gausspower}. The factors are defined as
the ratio $f_{\mathrm{T}}{\,=\,}I_\mathrm{P}/I_\mathrm{F}$ of the original peak intensity to the filtered peak intensity. For the
images with background, $f_{\mathrm{T}}$ were computed in background-subtracted images. Some entries for the largest windows
present additional information with respect to the plots in Fig.~\ref{gausspower}.
} 
\begin{tabular}{rcccccccccccc}
\hline
\noalign{\smallskip}
$\!W\,\,\,$&
$\mathcal{S}_\mathrm{G}$&$\mathcal{S}_\mathrm{G}{+}\mathcal{N}$&$\mathcal{S}_\mathrm{G}{+}\mathcal{N}{+}\mathcal{B}$&
$\mathcal{S}_\mathrm{P}$&$\mathcal{S}_\mathrm{P}{+}\mathcal{N}$&$\mathcal{S}_\mathrm{P}{+}\mathcal{N}{+}\mathcal{B}$&
$\mathcal{F}_\mathrm{G}$&$\mathcal{F}_\mathrm{G}{+}\mathcal{N}$&$\mathcal{F}_\mathrm{G}{+}\mathcal{N}{+}\mathcal{B}$&
$\mathcal{F}_\mathrm{P}$&$\mathcal{F}_\mathrm{P}{+}\mathcal{N}$&$\mathcal{F}_\mathrm{P}{+}\mathcal{N}{+}\mathcal{B}$\\
\noalign{\smallskip}
\hline
\noalign{\smallskip}
\!$1$\,\,\,\,&$4$&$4$&$4$&$3$&$3$&$3$&$1.6$&$1.6$&$1.6$&$1.7$&$1.7$&$1.7$\\
\!$2$\,\,\,\,&$260$&$200$&$130$&$9$&$9$&$9$&$6.8$&$6.8$&$6.8$&$3.5$&$3.6$&$3.6$\\
\!$4$\,\,\,\,&$4{\times}10^9$&$2000$&$720$&$33$&$33$&$30$&$1000$&$240$&$160$&$11$&$11$&$11$\\
\!$8$\,\,\,\,&--&--&--&$130$&$110$&$84$&$10^{8}$&$710$&$220$&$44$&$43$&$40$\\
\!$16$\,\,\,\,&--&--&--&$510$&$340$&$250$&$5{\times}10^9$&$1700$&$350$&$170$&$140$&$55$\\
\noalign{\smallskip}
\hline
\end{tabular}
\label{truncation}
\end{table*}

Efficiently reducing noise and various artifacts, median filtering also erases other types of structures, such as sources and
filaments. The steeper their intensity distributions, the better they are removed with smaller sliding windows. This property of
median filtering is demonstrated below using simulated images of sources and linear filaments with Gaussian and power-law intensity
distributions: $\mathcal{S}_{\mathrm{G}}$, $\mathcal{S}_{\mathrm{P}}$, $\mathcal{F}_{\mathrm{G}}$, and $\mathcal{F}_{\mathrm{P}}$.
The sources and filaments have sizes $H{\,=\,}8$ and $16${\arcsec} (full-width at half-maximum, FWHM) and peak intensities
$I_{\mathrm{P}}{\,=\,}100$ (in arbitrary units), the filaments extend across the entire image. The simulated images have dimensions
of $803{\,\times\,}803$ with $0.67{\arcsec}$ pixels. Another variant of the images adds random Gaussian noise $\mathcal{N}$
convolved to a $2${\arcsec} resolution with a standard deviation $\sigma{\,=\,}0.33$ (zero mean). The third variant includes a
large-scale background $\mathcal{B}$ modeled as a wide Gaussian with a peak value of $50$ and a half-maximum width of
$512${\arcsec}.

\begin{figure*}                                                               
\centering
\centerline{\resizebox{0.33\hsize}{!}{\includegraphics{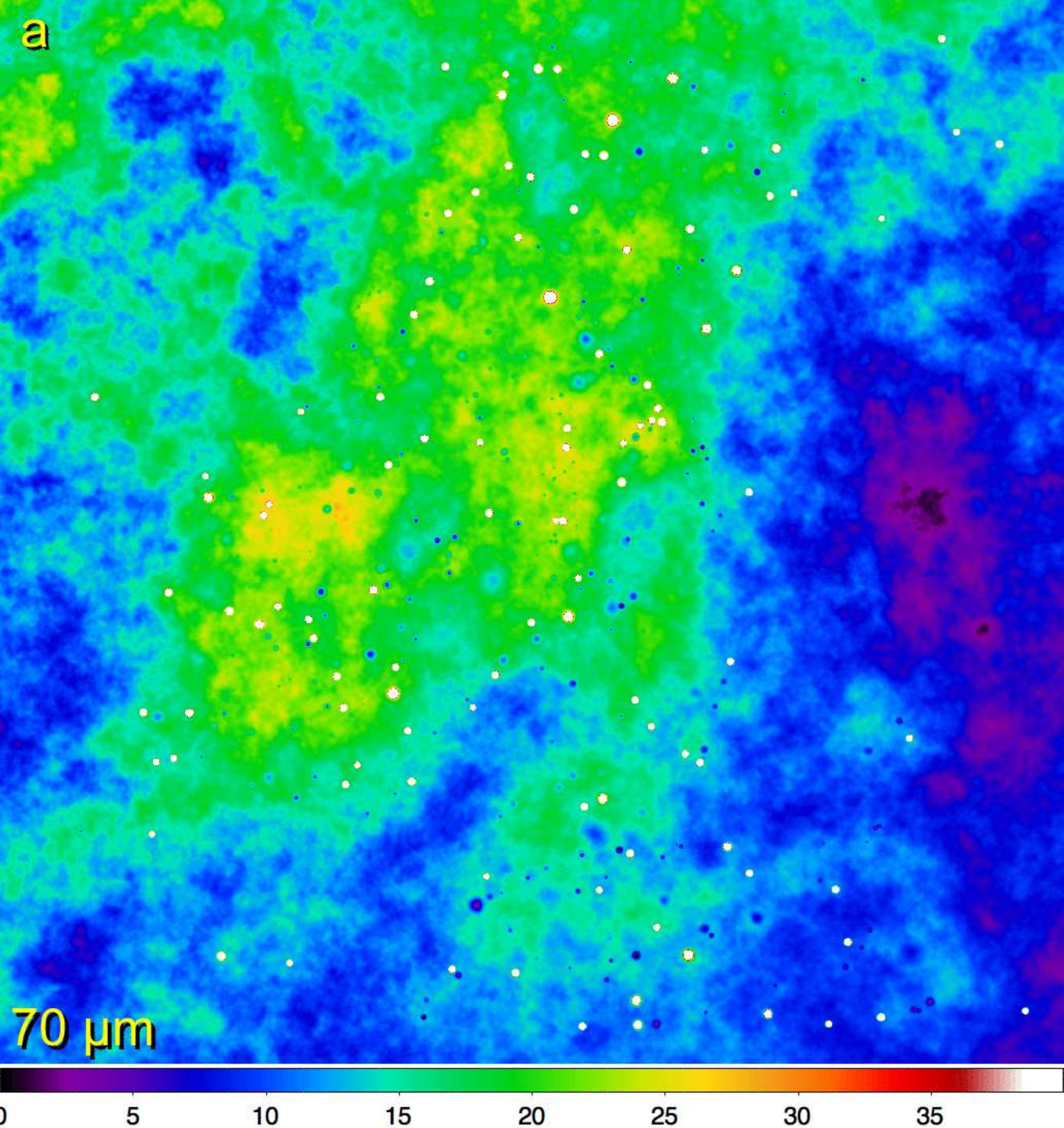}}
            \resizebox{0.33\hsize}{!}{\includegraphics{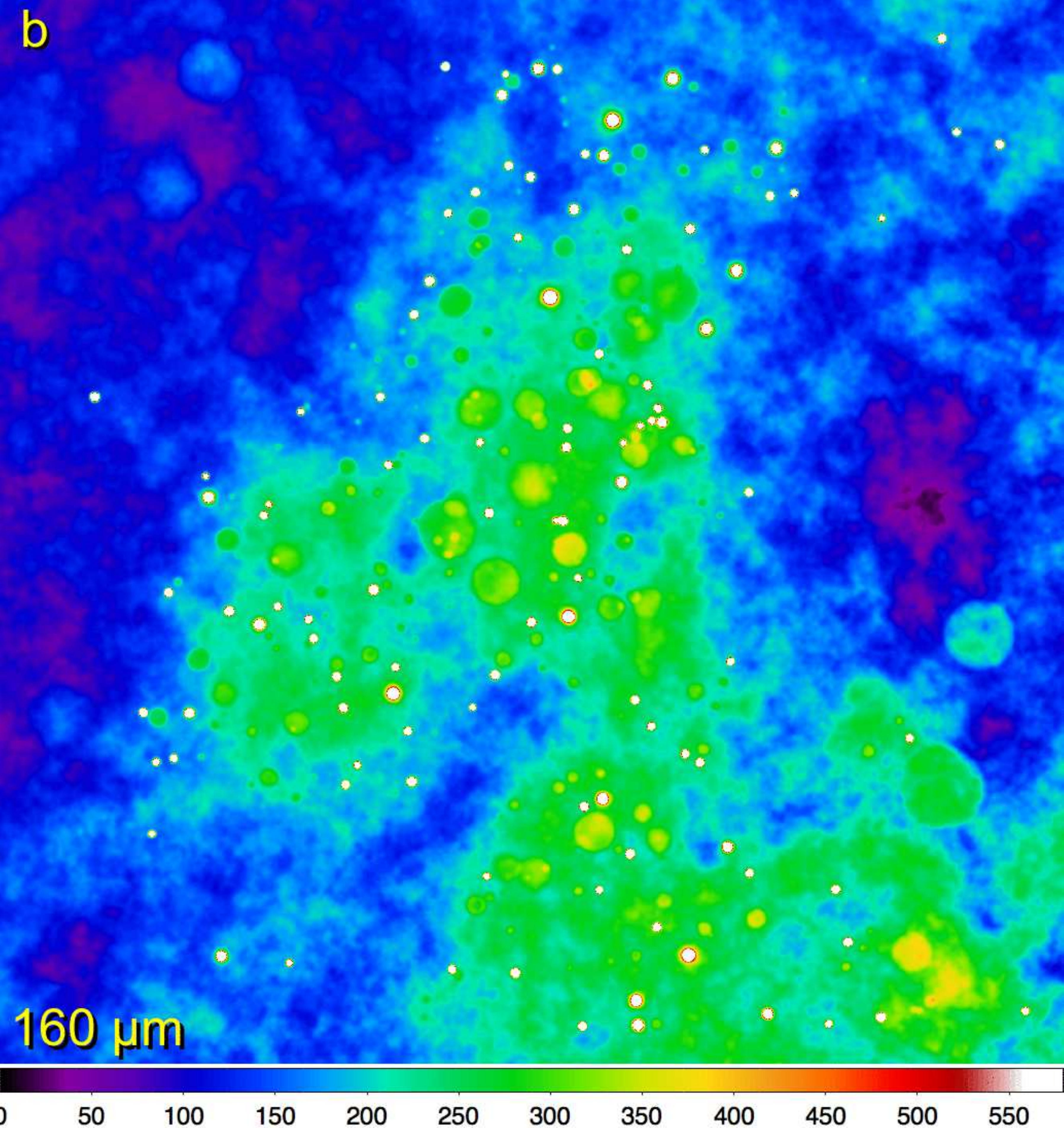}}
            \resizebox{0.33\hsize}{!}{\includegraphics{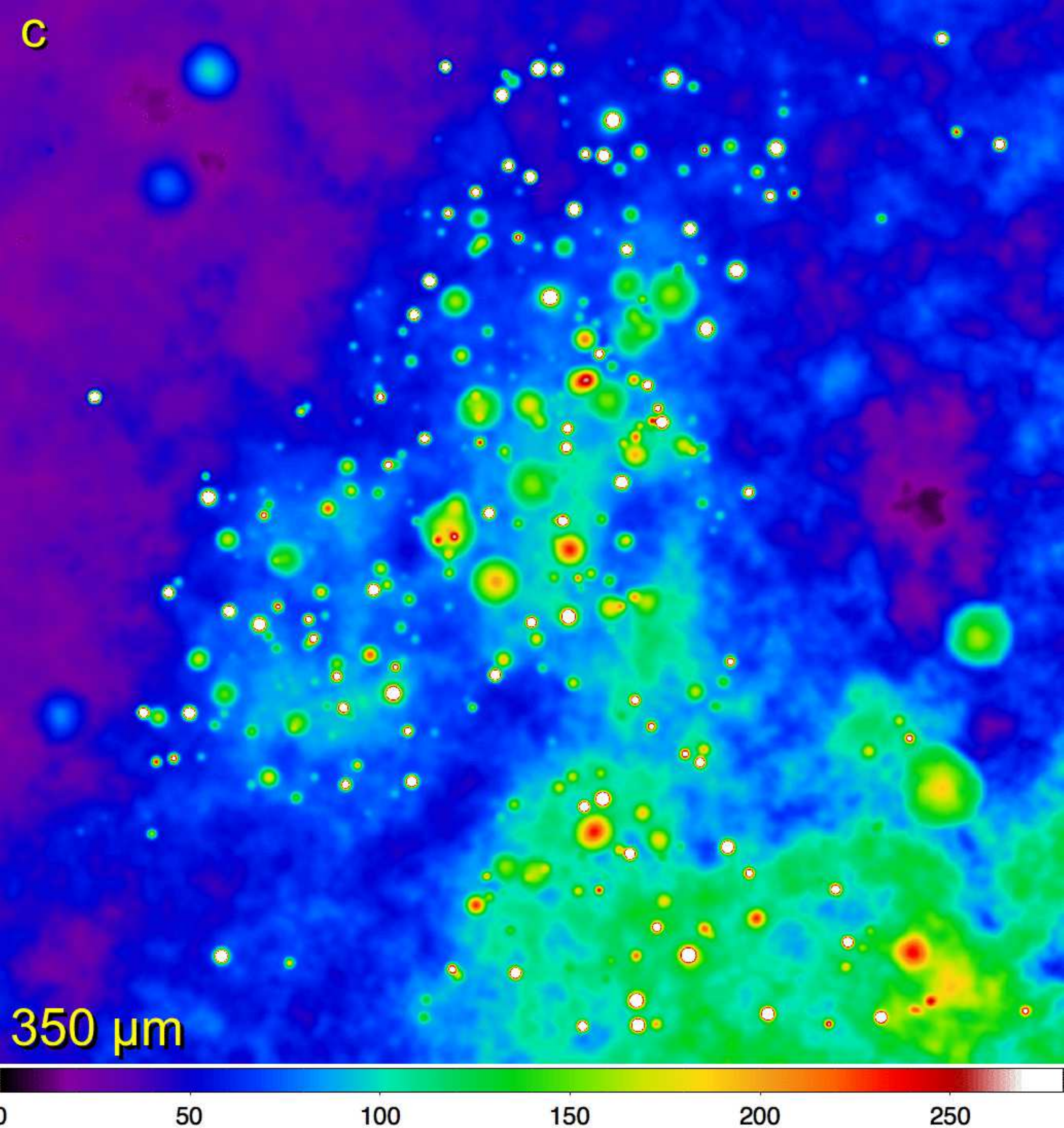}}}
\centerline{\resizebox{0.33\hsize}{!}{\includegraphics{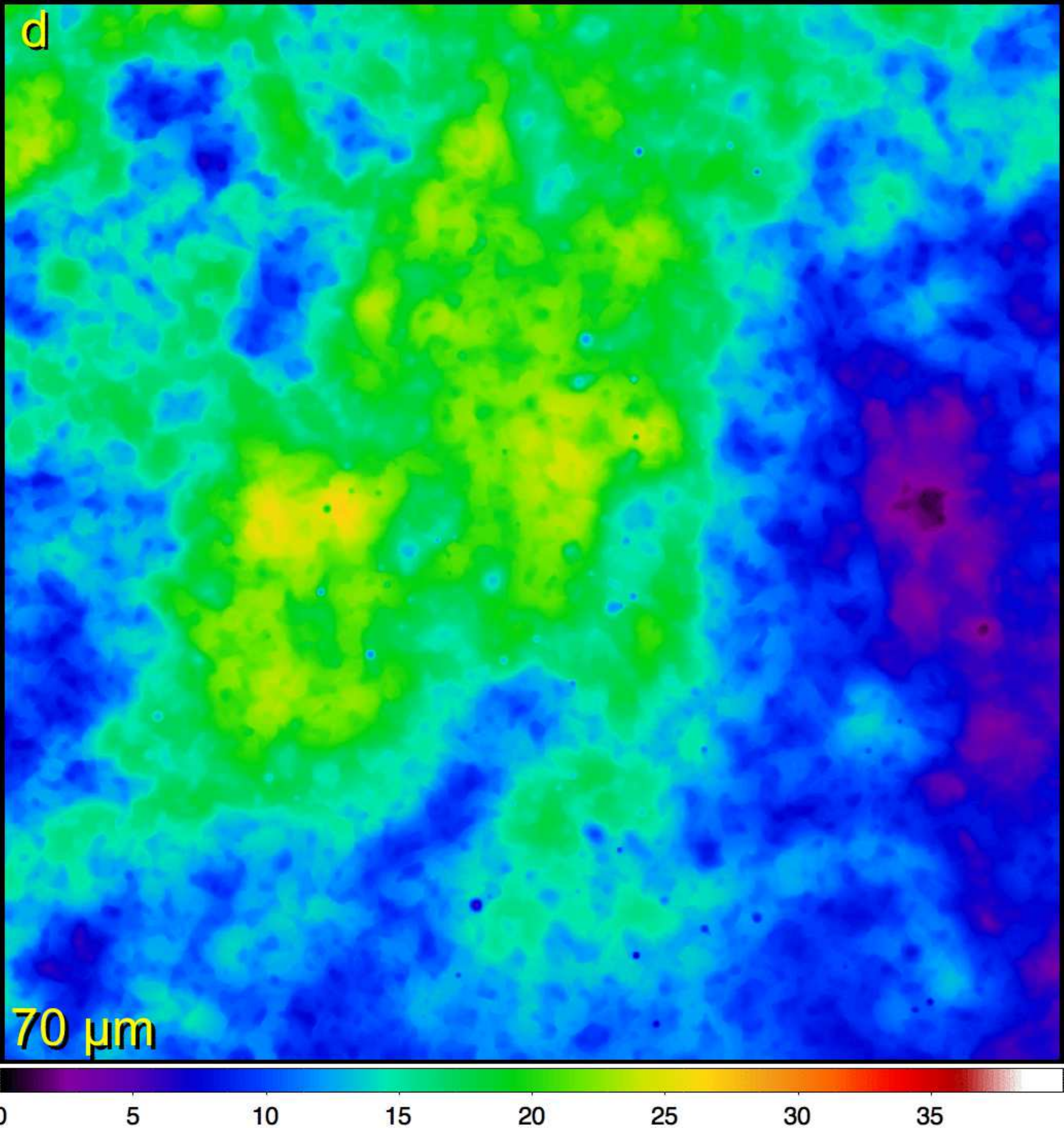}}
            \resizebox{0.33\hsize}{!}{\includegraphics{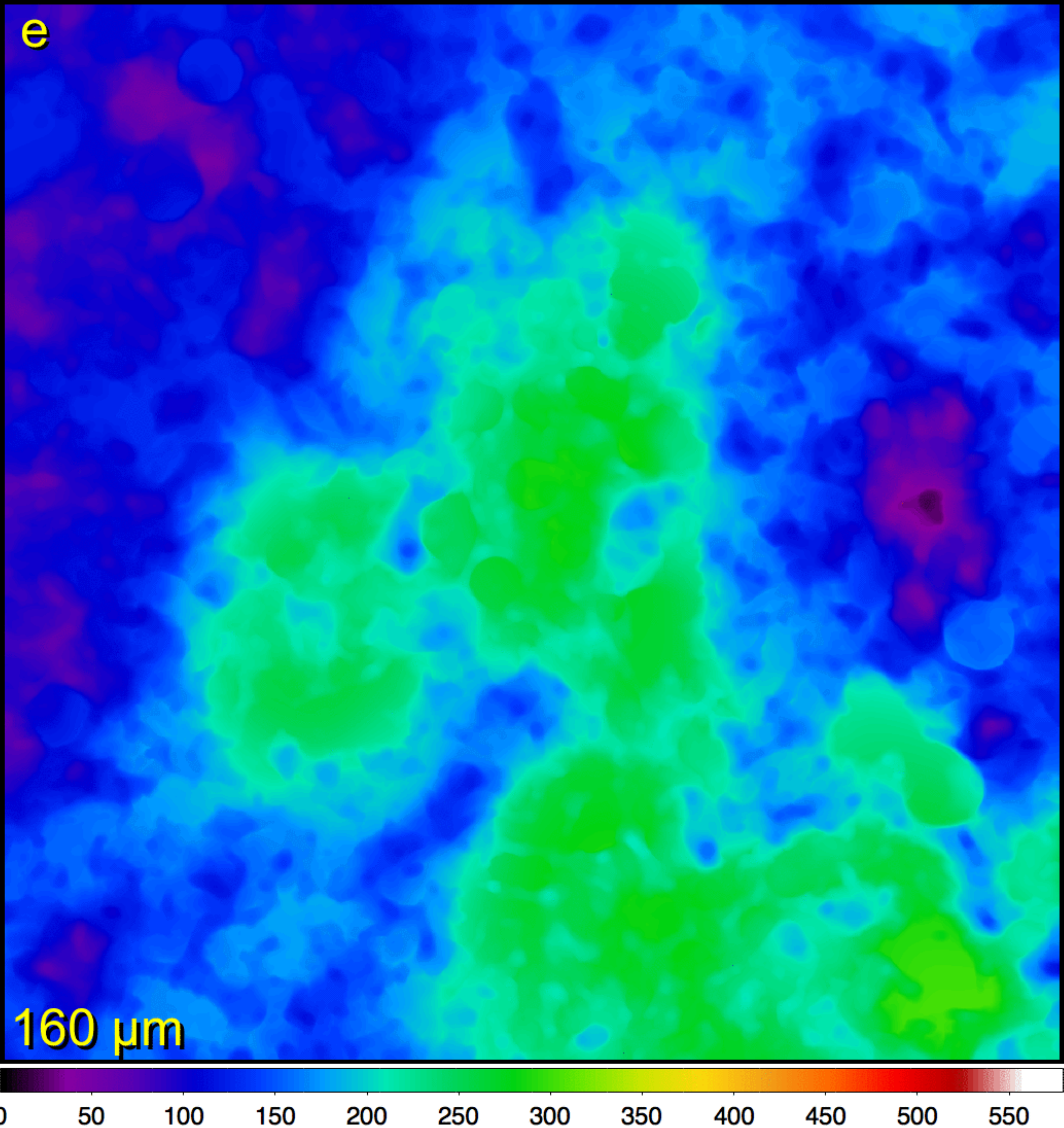}}
            \resizebox{0.33\hsize}{!}{\includegraphics{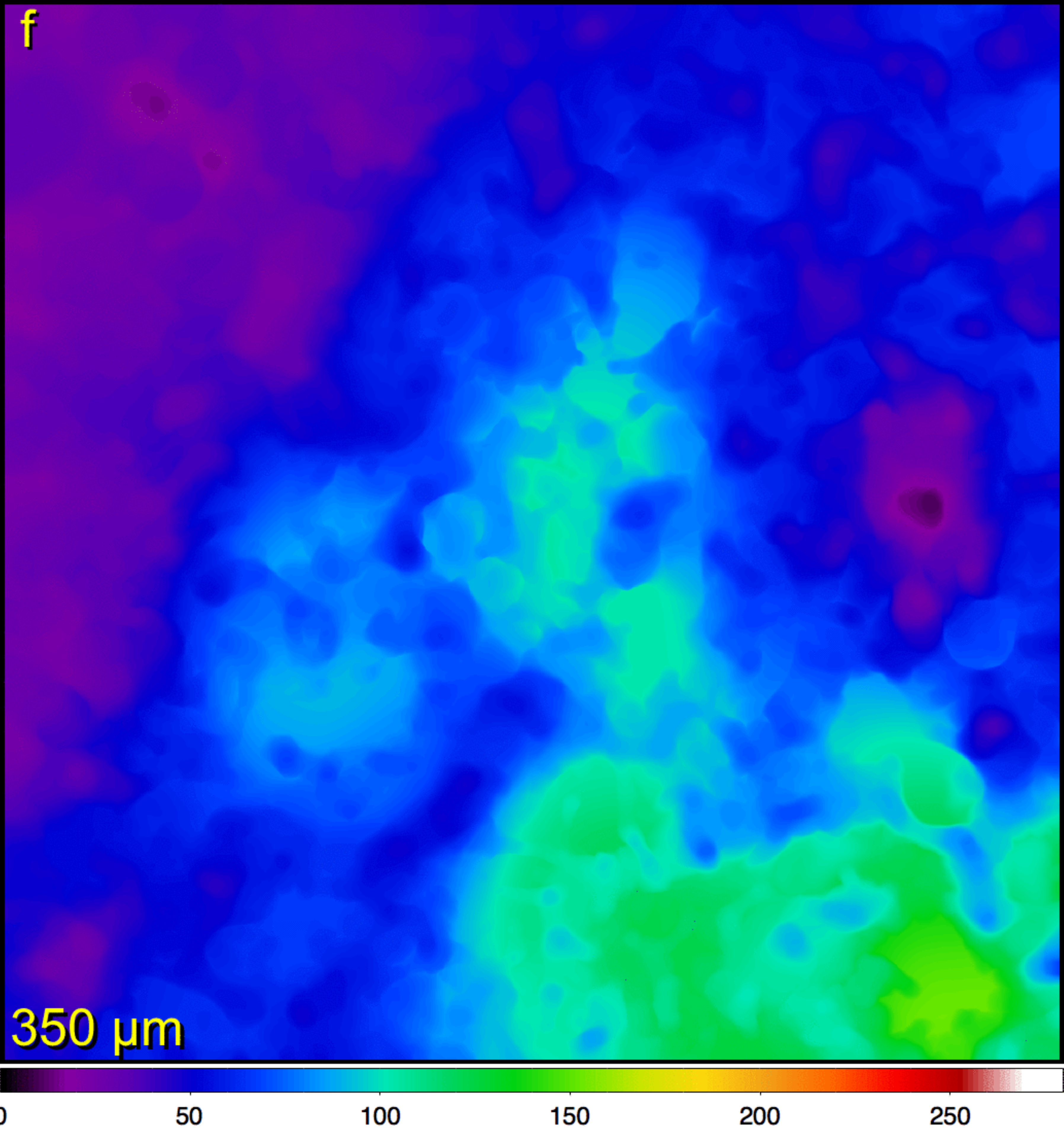}}}
\centerline{\resizebox{0.33\hsize}{!}{\includegraphics{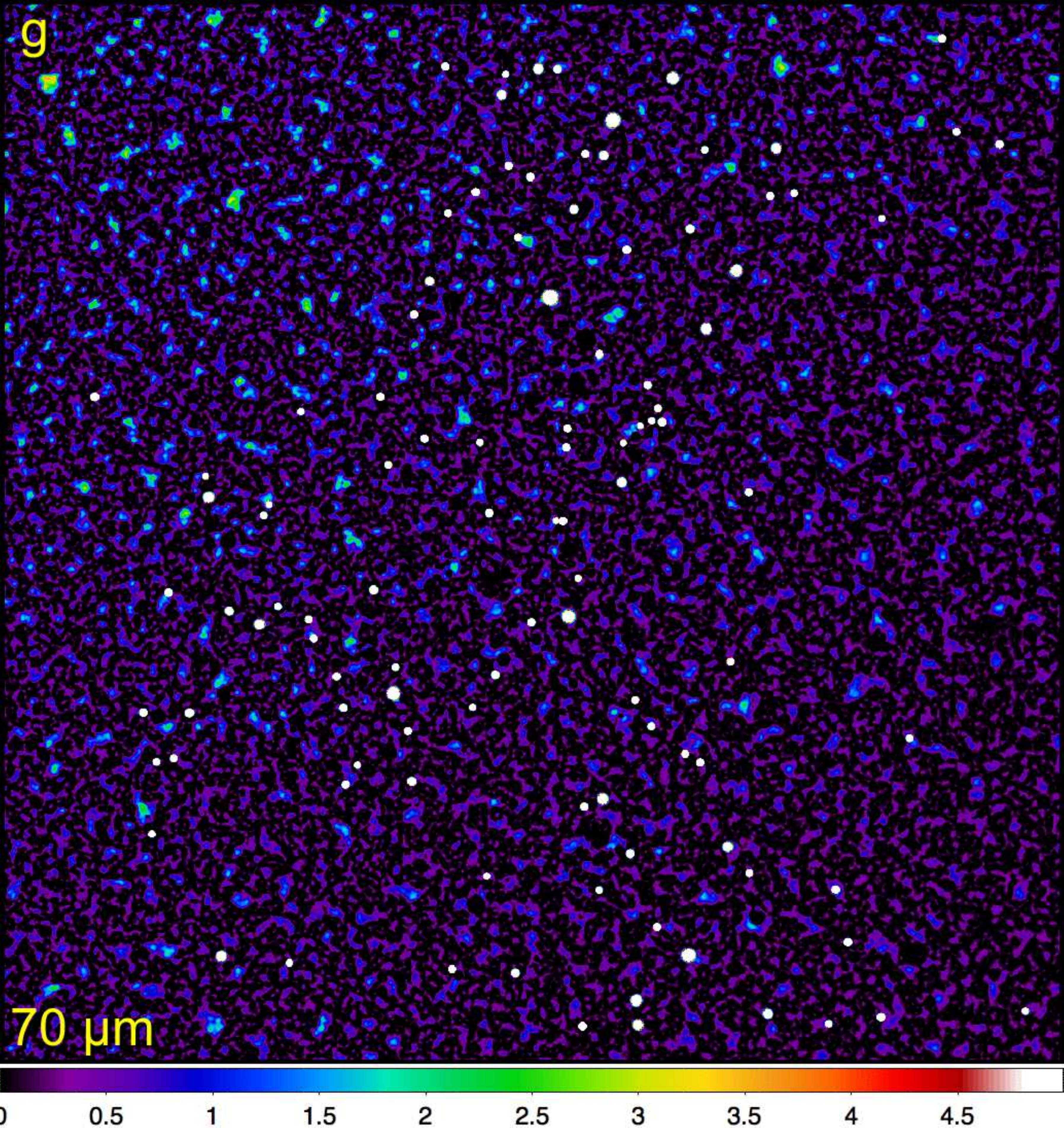}}
            \resizebox{0.33\hsize}{!}{\includegraphics{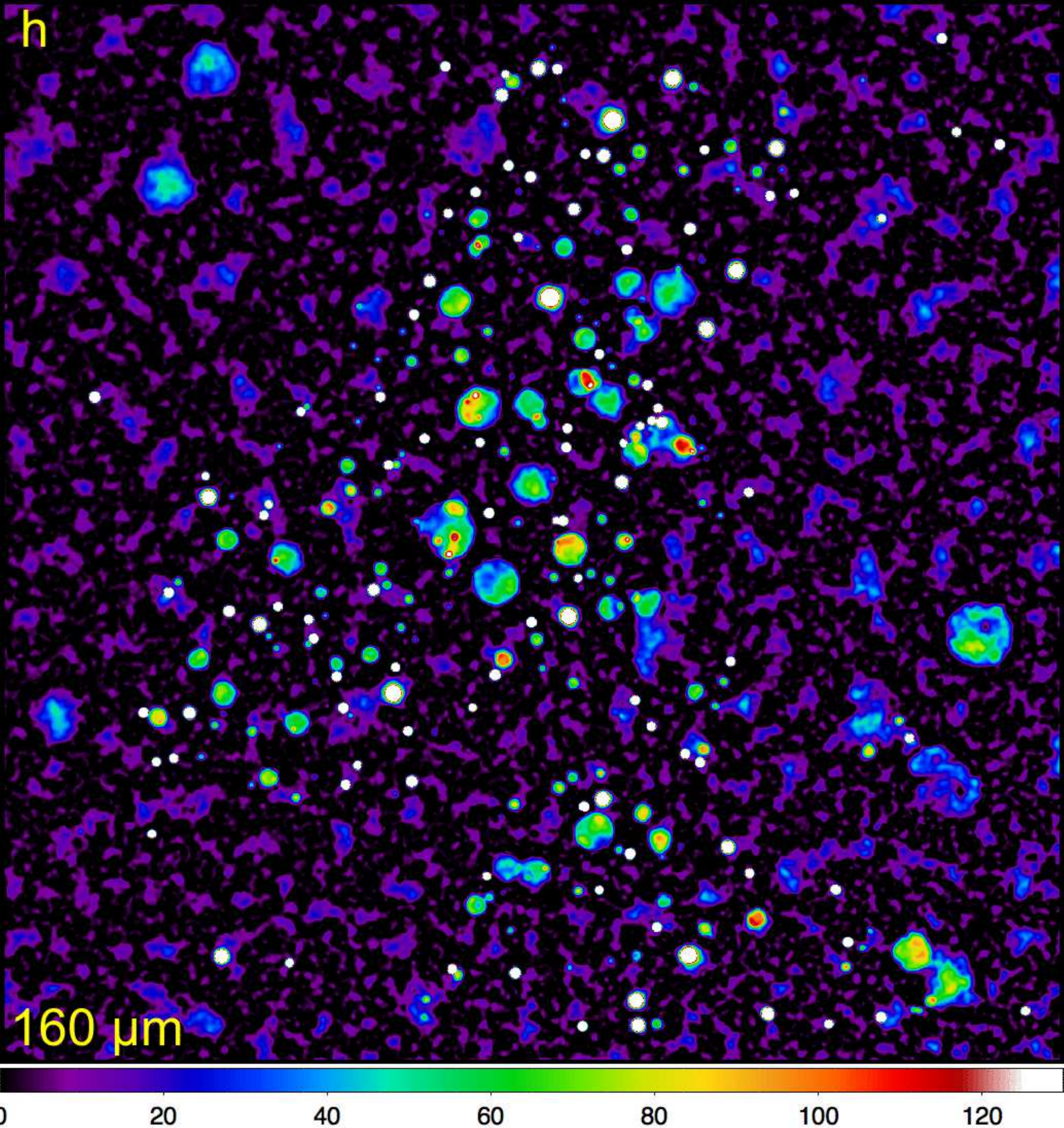}}
            \resizebox{0.33\hsize}{!}{\includegraphics{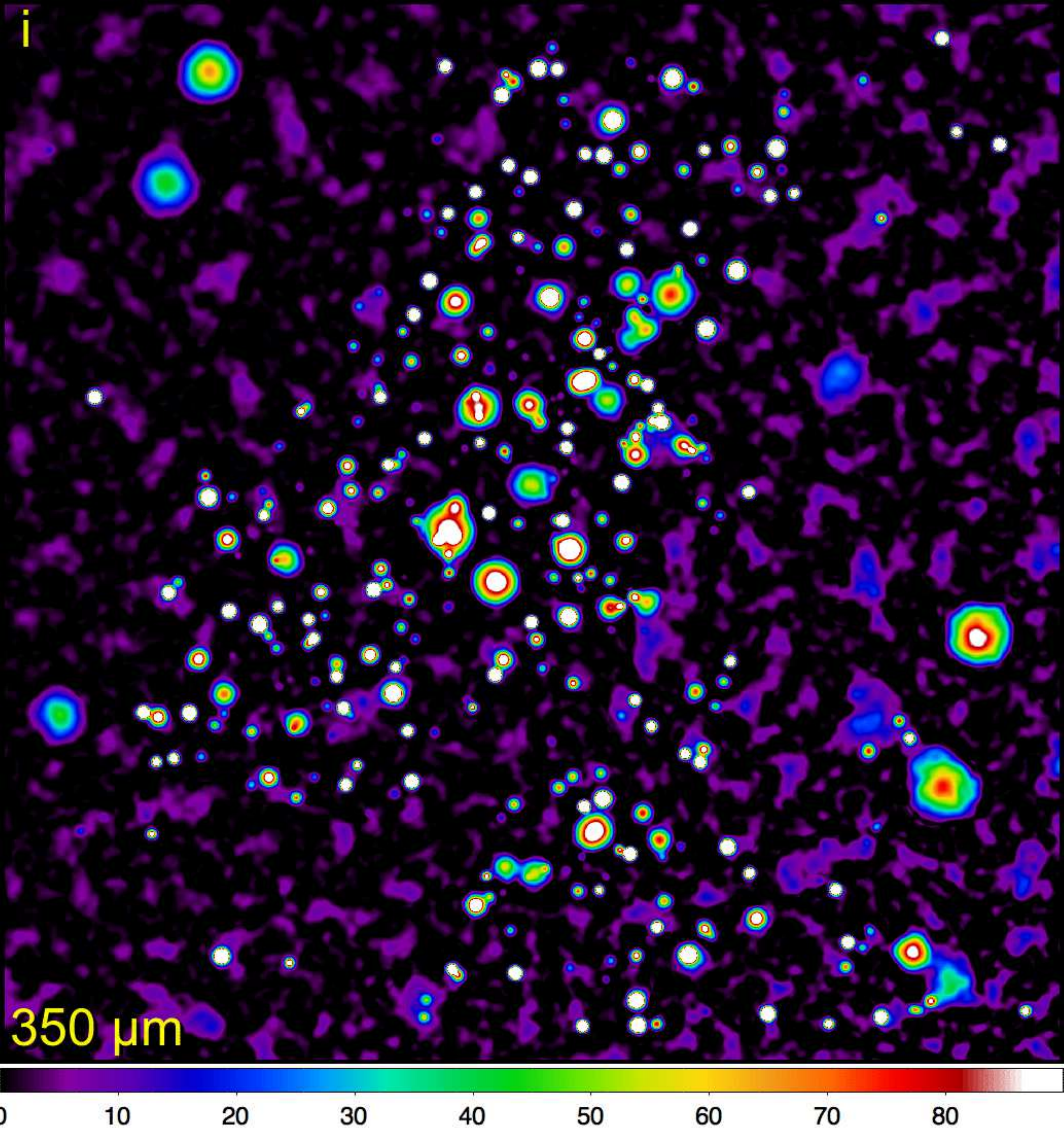}}}
\caption
{ 
Background derivation by median filtering. Shown are the images $\mathcal{I}_{\!\lambda}$ of a simulated star-forming region
(\emph{upper row}), median-filtered background $\tilde{\mathcal{B}}_{\lambda}$ (\emph{middle row}), and background-subtracted
images $\tilde{\mathcal{S}}_{\lambda}$ (\emph{bottom row}) at selected \emph{Herschel} wavelengths (after $M{\,=\,}30$ iterations).
The maximum source sizes $X_{\lambda}$ of $25$, $100$, and $100${\arcsec} for the \textsl{getimages} method were estimated directly
from $\mathcal{I}_{\!\lambda}$ (cf. Sect.~\ref{maxsize}). In panel \emph{a}, small holes are starless cores seen in absorption at
$70$\,{${\mu}$m}, and white emission peaks are protostellar sources. Intensities (in MJy/sr) are limited in range, and their color
scaling is linear.
} 
\label{bgsub}
\end{figure*}

\begin{figure*}                                                               
\centering
\centerline{\resizebox{0.33\hsize}{!}{\includegraphics{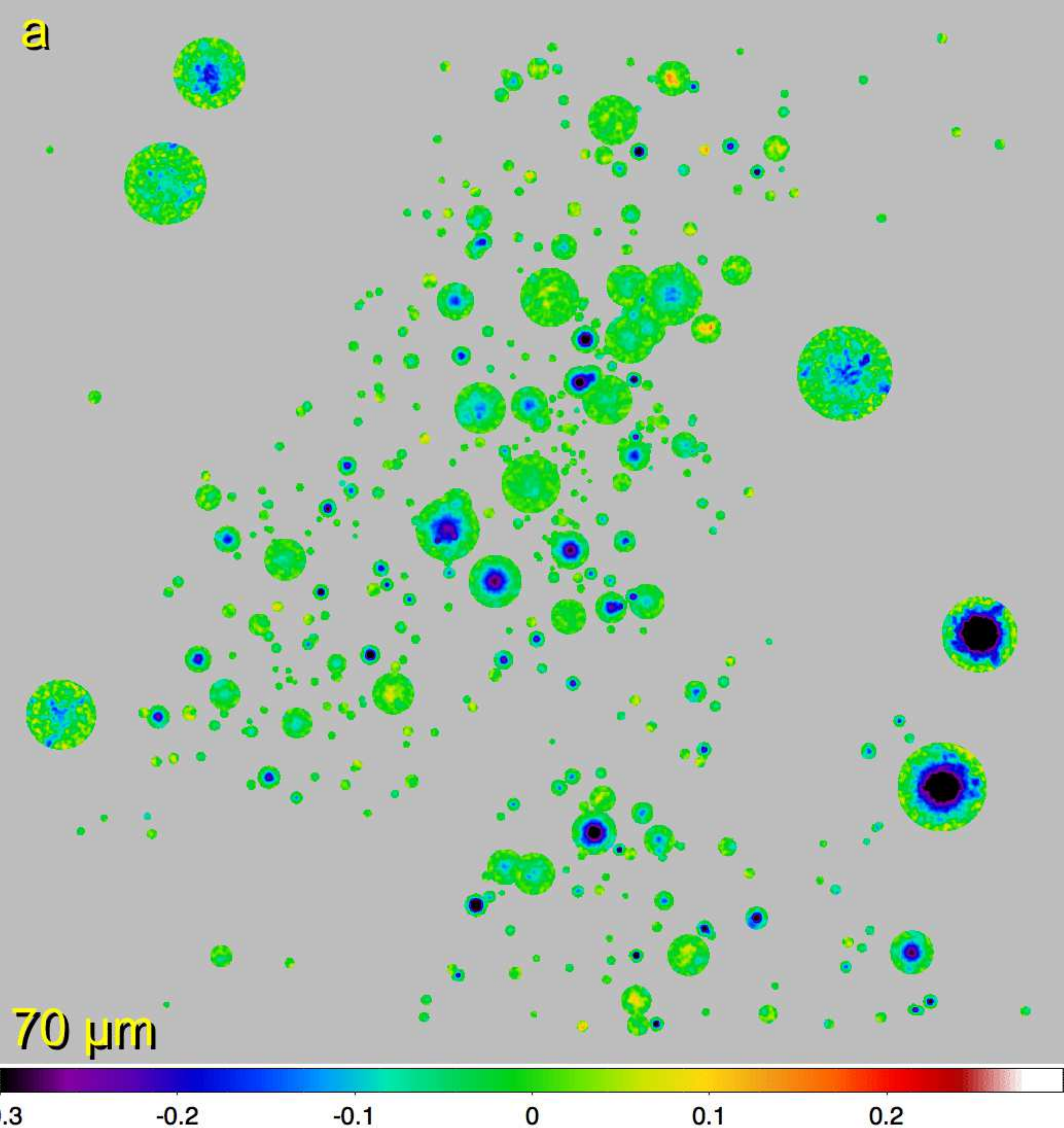}}
            \resizebox{0.33\hsize}{!}{\includegraphics{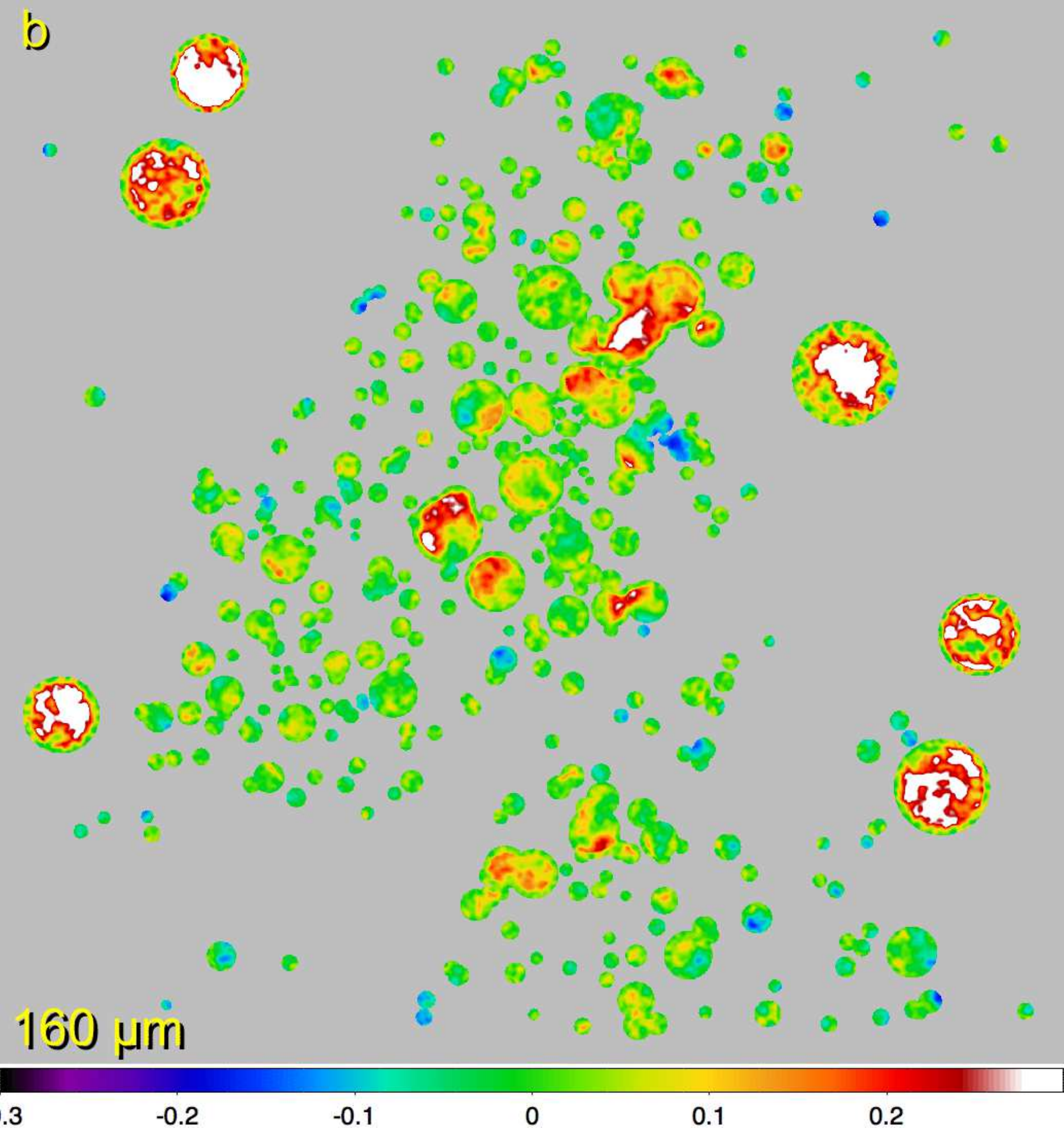}}
            \resizebox{0.33\hsize}{!}{\includegraphics{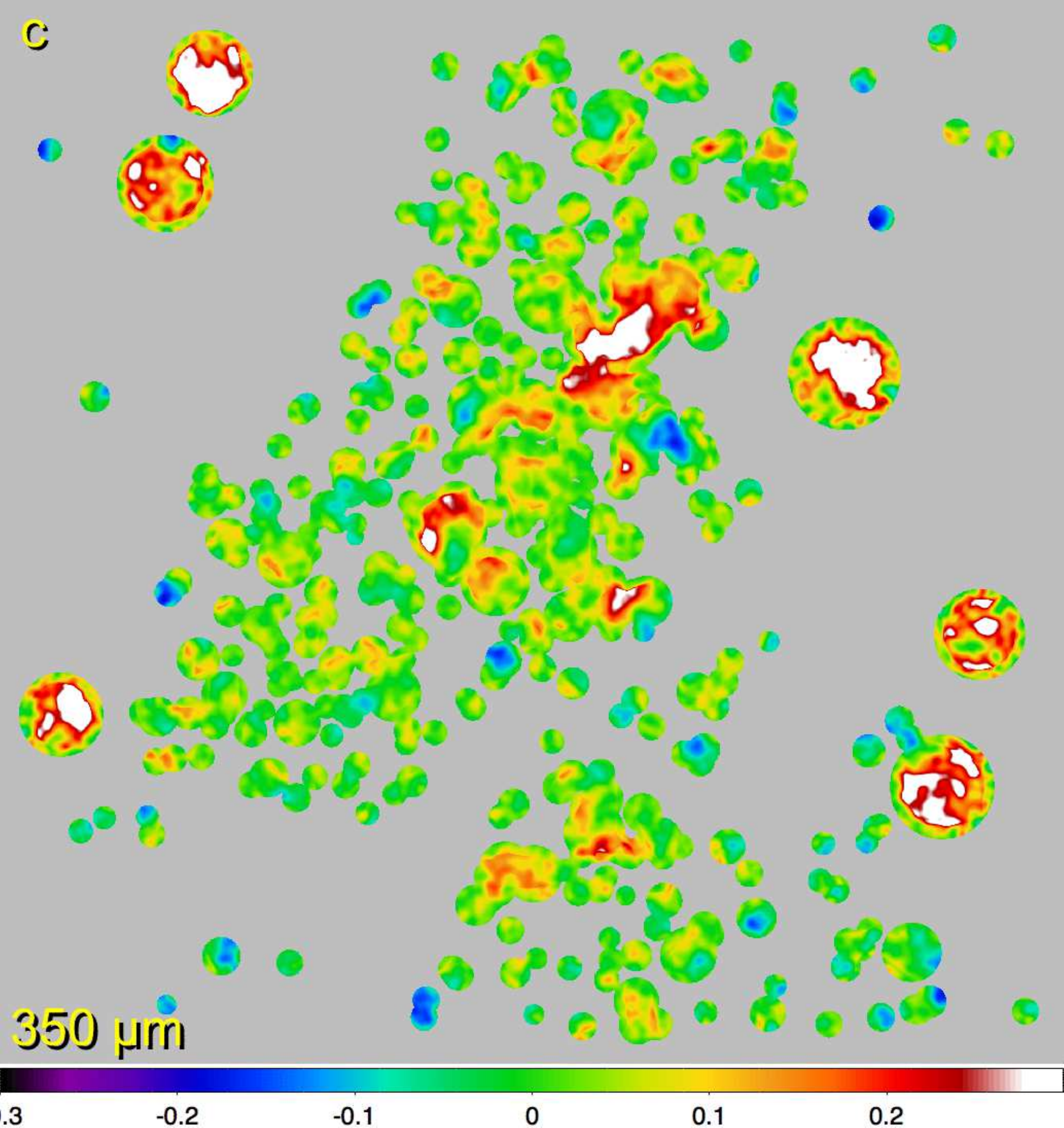}}}
\caption
{ 
Quality of the derived background presented in Fig.~\ref{bgsub}. Shown are the relative accuracies
$\tilde{\mathcal{B}}_{\lambda}/\mathcal{B}_{\lambda}{\,-\,}1$ of the background $\tilde{\mathcal{B}}_{\lambda}$ obtained by median
filtering with respect to the true background $\mathcal{B}_{\lambda}$. Only the relevant pixels within the model sources are shown
for values limited by the range $[-0.3, 0.3]$ with linear color coding. In panel \emph{a}, the minima visible in the centers of
some footprints are caused by the absorption of radiation in the central parts of starless cores; the two deepest minima are at
levels of $-0.7$ and $-0.6$. In panels \emph{b} and \emph{c}, the maxima are at levels of $0.8$ and $0.7$, respectively.
} 
\label{accbg}
\end{figure*}

Results of median filtering of all simulated structures are presented in Fig.~\ref{gausspower} by the intensity profiles through
their peaks (see Appendix~\ref{AppendixB} for the corresponding images). The resulting truncation factors, defined as the ratio
$f_{\mathrm{T}}{\,=\,}I_\mathrm{P}/I_\mathrm{F}$ of the original and filtered peak intensities, are summarized in
Table~\ref{truncation}. Median filtering was parameterized by $W{\,=\,}R/H$, the window size in units of the structure half-maximum
size, increasing by factor of two ($1$, $2$, $4$, etc.). The original structures of different sizes ($H{\,=\,}8$ and $16${\arcsec})
are truncated by the same factors depending only on $W$. For the same window, Gaussian shapes are erased much more efficiently than
power-law shapes and sources are truncated much deeper than filaments. This is a simple consequence of the fact that the more
extended or elongated structures fill sliding windows to a much higher degree. The presence of noise and background does not affect
the truncation factors until they increase to $f_{\mathrm{T}}{\,\sim\,}100$, whereas for larger windows, noise and background tend
to reduce the factors.

This demonstrates that median filtering efficiently removes Gaussian sources and filaments in small windows ($W{\,\approx\,}2$ and
$4$) with very low residuals of ${\la}1${\%}. More extended intensity distributions of the power-law sources and filaments require
larger windows ($W{\,\approx\,}8$ and $16$) for their removal to the same high accuracy. Median filtering with a smaller window
($W{\,=\,}4$) erases power-law sources and filaments to reasonably good levels of $3$ and $10${\%}, respectively
(Table~\ref{truncation}).

If the sources with size $H$ are cleanly erased, the filtered image closely approximates background $\mathcal{B}_H$ for those
sources and $\mathcal{I}{\,-\,}\mathcal{B}_H$ becomes their background-subtracted image. In the easiest case of isolated sources on
a \emph{simple} background, their accurate sizes and fluxes might be measured directly in $\mathcal{I}{\,-\,}\mathcal{B}_H$. The
problem is that in real-life images sources have different sizes and backgrounds are not simple.

\subsection{Background derivation by median filtering}
\label{bgestim}

Astrophysical backgrounds are highly filamentary and strongly variable on all spatial scales. Since median filtering truncates any
peaks, in addition to sources, it also removes some intensity peaks that belong to the filaments and complex backgrounds.
Backgrounds obtained by median filtering may therefore become underestimated in some places, hence the background-subtracted images
may contain contributions from the filaments and backgrounds. Direct measurements of source sizes and fluxes in such images are not
always feasible because they will be inaccurate for some sources. However, with drastically reduced contributions of filaments and
bright fluctuating backgrounds, the background-subtracted images are always much simpler than the originals. There are great
benefits in using them as detection images, in addition to the originals, which should be used as measurement images (cf. Papers I
and II).

Images of star-forming regions may be considered as superpositions of sources, backgrounds, filaments, and noise. To illustrate the
new algorithm, images of a simulated star-forming region were computed at \emph{Herschel} wavelengths, for simplicity, consisting
of only the sources and background: $\mathcal{I}_{\!\lambda}{\,=\,}\mathcal{S}_{\lambda}{\,+\,}\mathcal{B}_{\lambda}$. The images
are essentially identical to those described in Paper I (Appendix C) without instrumental noise; they have dimensions of
$1900{\,\times\,}1900$ with $2${\arcsec} pixels. The purely synthetic scale-free cirrus background images were made consistent
with a planar temperature gradient between $15$ and $20$\,K along one image diagonal (from lower right to upper left, as in
Sect.~2.8 of Paper I). Examples of the simulated images $\mathcal{I}_{\!\lambda}$ at wavelengths $\lambda$ of $70$, $160$, and
$350$\,{${\mu}$m}, convolved to the corresponding angular resolutions $O_{\lambda}$ of \emph{Herschel} ($8.4$,$13.5$, and
$24.9${\arcsec} FWHM), are shown in Fig.~\ref{bgsub} (upper panels).

Astronomical images observed with a limited resolution contain both unresolved and resolved sources whose sizes
$H_{\lambda}{\,\ge\,}O_{\lambda}$ are unknown before source extraction. Determining the background of such images using median
filtering with a single sliding window radius $R_{\lambda}$ may not be optimal, except when the goal is to extract sources of
similar sizes. For example, one might be interested in detecting unresolved or slightly resolved sources with
$H_{\lambda}{\,\cong\,}O_{\lambda}$ and use a sliding window with $R_{\lambda}{\,\approx\,}4\,O_{\lambda}$ (cf.
Sect.~\ref{sourceremoval}). When extracting sources with a wide range of sizes, it is conceivable to use a sliding window tailored
to the largest source. This may not always work, however, as median filtering with large windows spreads bright emission down to
the low-background areas, leading to an overestimated background and undetected faint sources in those areas of a
background-subtracted image.

The \textsl{getimages} method provides a simple and universal procedure to remove all structures of any width below an arbitrary
maximum size $X_{\lambda}$ (FWHM) and to derive background $\tilde{\mathcal{B}}_{\lambda}$ for the size range
$O_{\lambda}{\,\le\,}S_{\!\lambda}{\,\la\,}X_{\lambda}$. Given the maximum size $X_{\lambda}$, the algorithm defines $N$ sliding
windows with radii $R_{\lambda\,1}, R_{\lambda\,2},\dots, R_{\lambda\,N}$, such that $R_{\lambda\,1}{\,=\,}2\,O_{\lambda}$,
$R_{\lambda\,j}{\,=\,}f_\mathrm{W}\,R_{\lambda\,j-1}$, and $R_{\lambda\,N}{\,=\,}4\,X_{\lambda}$, where the factor
$f_\mathrm{W}{\,>\,}1$ must be small enough to sample the range of sizes. Median filtering of the original images
$\mathcal{I}_{\!\lambda}$ is repeated using $N$ sliding windows and minimizing the resulting set of images:
\begin{equation} 
\hat{\mathcal{B}}_{\lambda} = \min \left\{\,\mathrm{mf}_{R_{\lambda\,j}}(\mathcal{I}_{\!\lambda})\,\right\} 
\,\,\,({j = 1, 2,\dots, N}).
\label{bgfilter}
\end{equation} 
Finally, the median filtering with the largest window is repeated to improve the smoothness of the resulting image:
\begin{equation} 
\tilde{\mathcal{B}}_{\lambda} = \min \left\{\,\hat{\mathcal{B}}_{\lambda},\,
\mathrm{mf}_{R_{\lambda\,N}}(\hat{\mathcal{B}}_{\lambda})\,\right\}.
\label{bgfilter2}
\end{equation} 
The smallest window with $R_{\lambda\,1}{\,=\,}2\,O_{\lambda}$ completely removes Gaussian sources, and the largest window with
$R_{\lambda\,N}{\,=\,}4\,X_{\lambda}$ truncates Gaussian and power-law sources and filaments sufficiently well (cf.
Sect.~\ref{sourceremoval}, Table~\ref{truncation}). Practical details of the definition of $X_{\lambda}$ are discussed in
Sect.~\ref{maxsize}.

A reasonably low value $f_\mathrm{W}{\,=\,}2^{1/2}$ of the discretization factor is adopted in \textsl{getimages} by default.
Although lower values ($f_\mathrm{W}\,{\rightarrow}\,1$) ensure that the size range of the structures of interest is better
resolved, in practice, they slow down the procedure without any noticeable gain in accuracy. On the other hand, large factors
($f_\mathrm{W}{\,\ga\,}3$) lead to a poorly sampled size range and thus tend to give a less accurate background
$\tilde{\mathcal{B}}_{\lambda}$ for some structures.

This procedure ensures that the background obtained with small windows for the structures with
$H_{\lambda}{\,\approx\,}O_{\lambda}$ is preserved after median filtering with much larger windows that are suitable for
significantly wider sources or filaments ($H_{\lambda}{\,\gg\,}O_{\lambda}$). The resulting background image
$\tilde{\mathcal{B}}_{\lambda}$ becomes appropriate for structures within the entire size range
$O_{\lambda}{\,\le\,}S_{\!\lambda}{\,\la\,}X_{\lambda}$. The background-subtracted image is readily computed as
$\tilde{\mathcal{S}}_{\lambda}{\,=\,}\mathcal{I}_{\!\lambda}{\,-\,}\tilde{\mathcal{B}}_{\lambda}$, which is essentially the
original image of the structures $\mathcal{S}_{\lambda}$ containing a small differential contribution
$\mathcal{B}_{\lambda}{\,-\,}\tilde{\mathcal{B}}_{\lambda}$ induced by the inaccuracies in the estimated background.

To significantly improve $\tilde{\mathcal{B}}_{\lambda}$ and reduce contamination of $\tilde{\mathcal{S}}_{\lambda}$ by residual
background peaks, the \textsl{getimages} method employs iterations. In the iterative formulation, the median-filtering algorithm
described by Eqs.~(\ref{bgfilter}) and (\ref{bgfilter2}) remains the same, with two substitutions:
\begin{equation} 
\mathcal{I}_{\!\lambda}{\,\leftarrow\,}\tilde{\mathcal{S}}^{\,i-1}_{\lambda},\,\,\,
\tilde{\mathcal{B}}_{\lambda}{\,\leftarrow\,}\tilde{\mathcal{B}}^{\,i}_{\lambda}\,\,\,\,(i = 1, 2,\dots, M),
\label{iterations}
\end{equation} 
where $\tilde{\mathcal{S}}^{\,0}_{\lambda}{\,=\,}\mathcal{I}_{\!\lambda}$ and $M$ is the number of iterations. To construct
background-subtracted images exclusively for source detection, it would be sufficient to set $M{\,\approx\,}5{-}10$. To open the
possibility of using the images also for source measurements, more iterations ($M{\,\approx\,}20{-}30$) may be necessary. The final
improved images are computed as
\begin{equation} 
\tilde{\mathcal{B}}_{\lambda}{\,=\,}\sum_{i=1}^{M}{\tilde{\mathcal{B}}^{\,i}_{\lambda}},\,\,\,
\tilde{\mathcal{S}}_{\lambda}{\,=\,}\mathcal{I}_{\!\lambda}{\,-\,}\tilde{\mathcal{B}}_{\lambda}.
\label{iterations2}
\end{equation} 
Experience shows that the iterative scheme effectively reduces the differential contribution
$\mathcal{B}_{\lambda}{\,-\,}\tilde{\mathcal{B}}_{\lambda}$ of background peaks in $\tilde{\mathcal{S}}_{\lambda}$.

\begin{figure*}                                                               
\centering
\centerline{\resizebox{0.33\hsize}{!}{\includegraphics{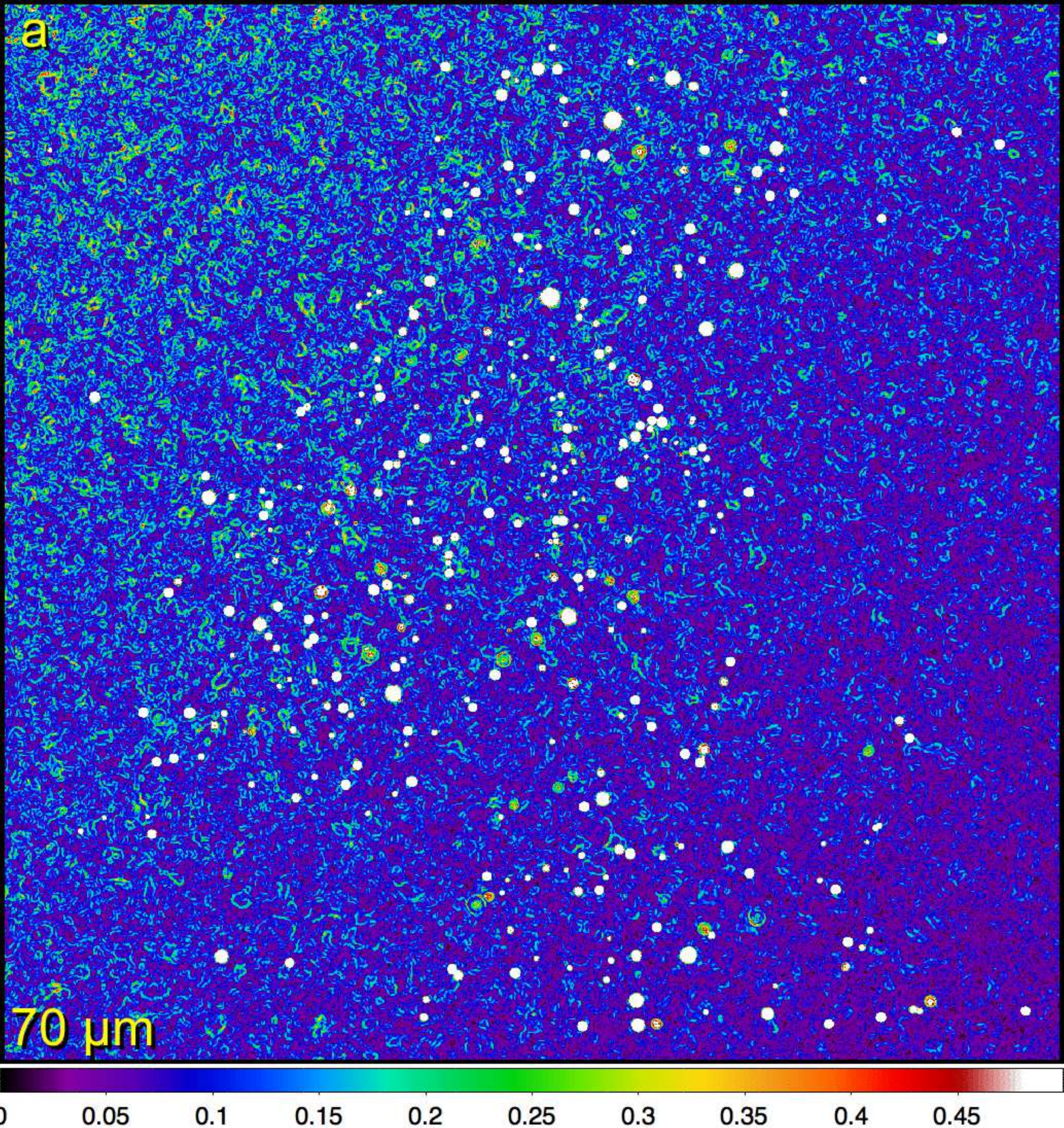}}
            \resizebox{0.33\hsize}{!}{\includegraphics{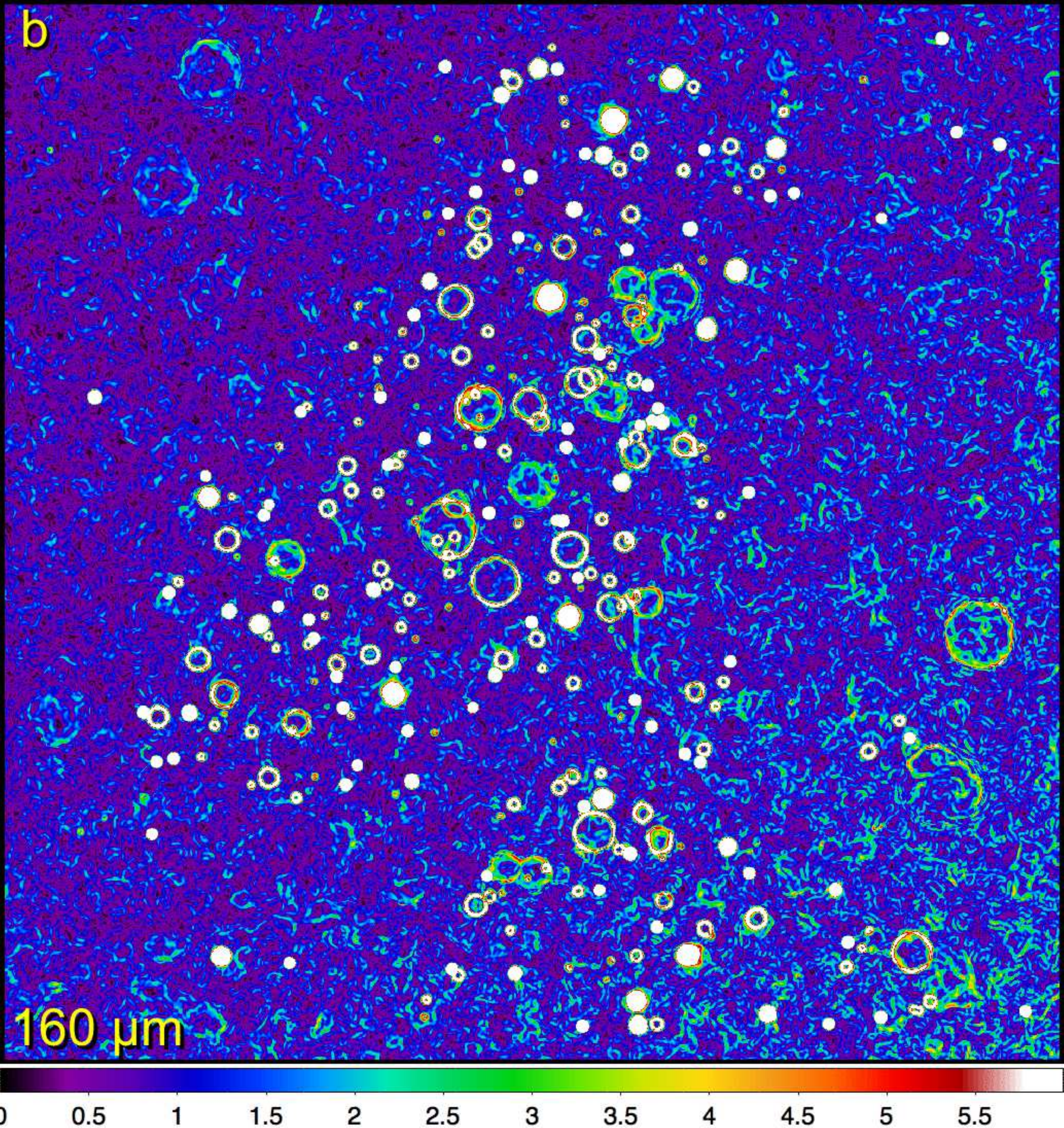}}
            \resizebox{0.33\hsize}{!}{\includegraphics{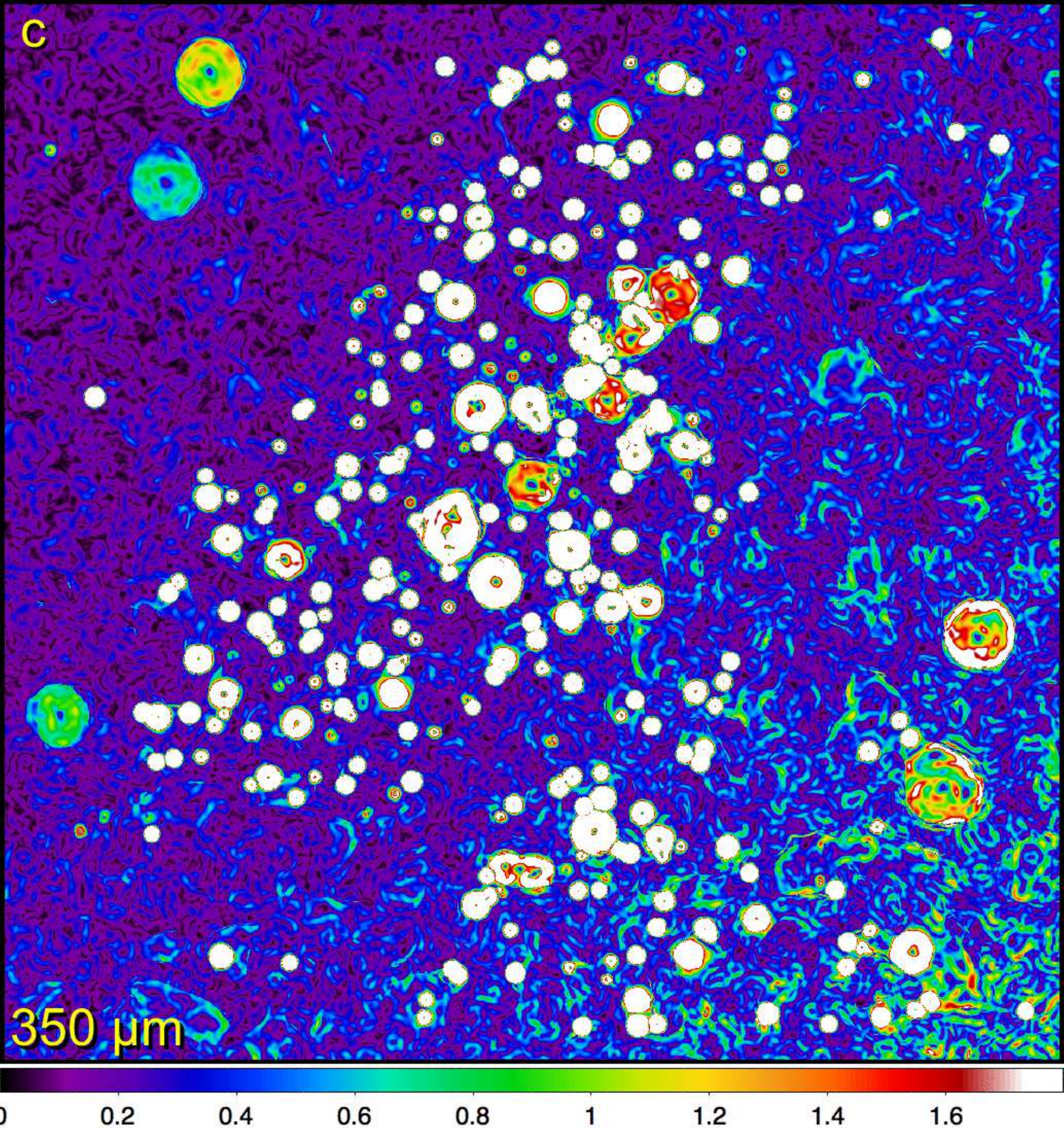}}}
\centerline{\resizebox{0.33\hsize}{!}{\includegraphics{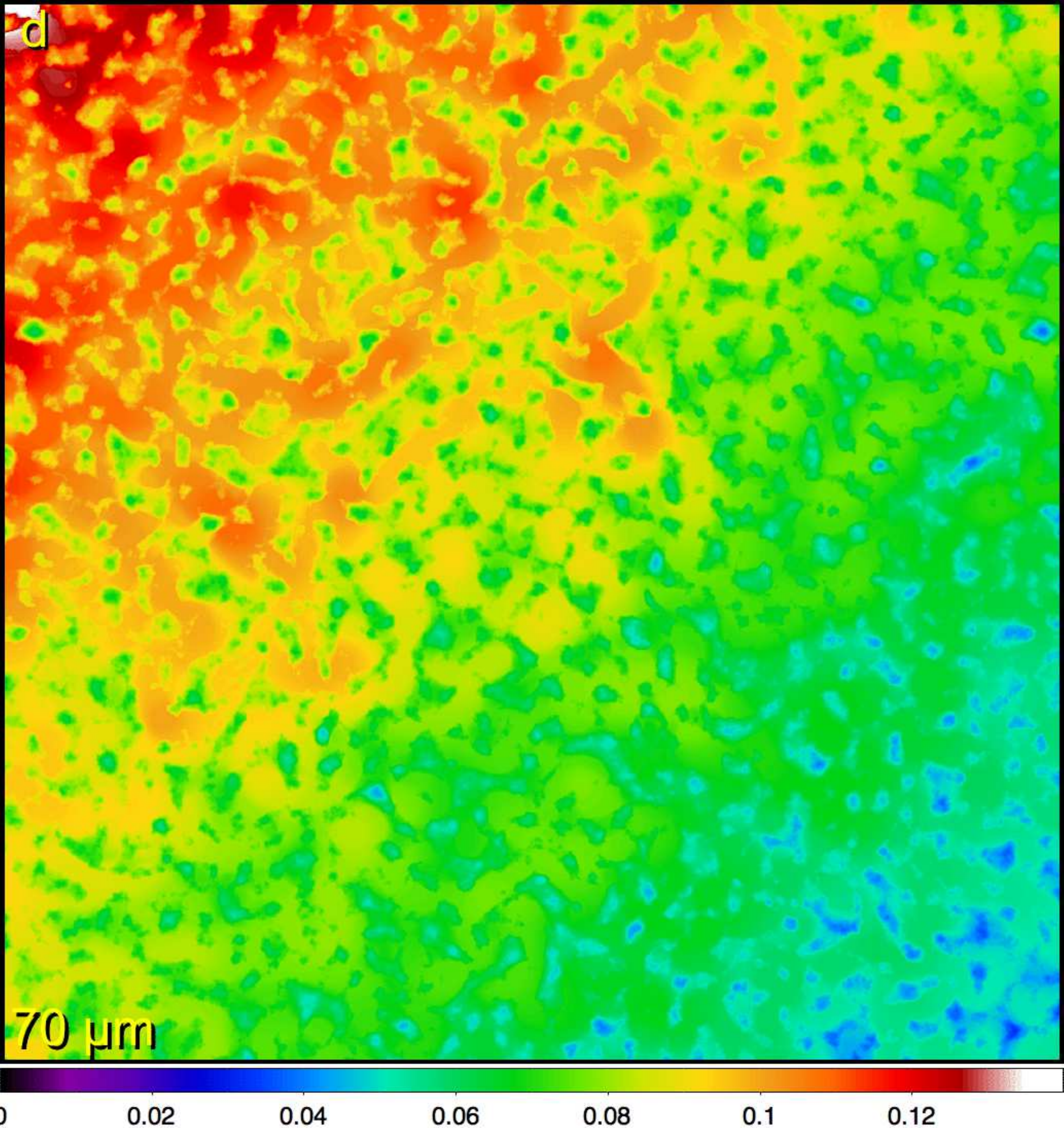}}
            \resizebox{0.33\hsize}{!}{\includegraphics{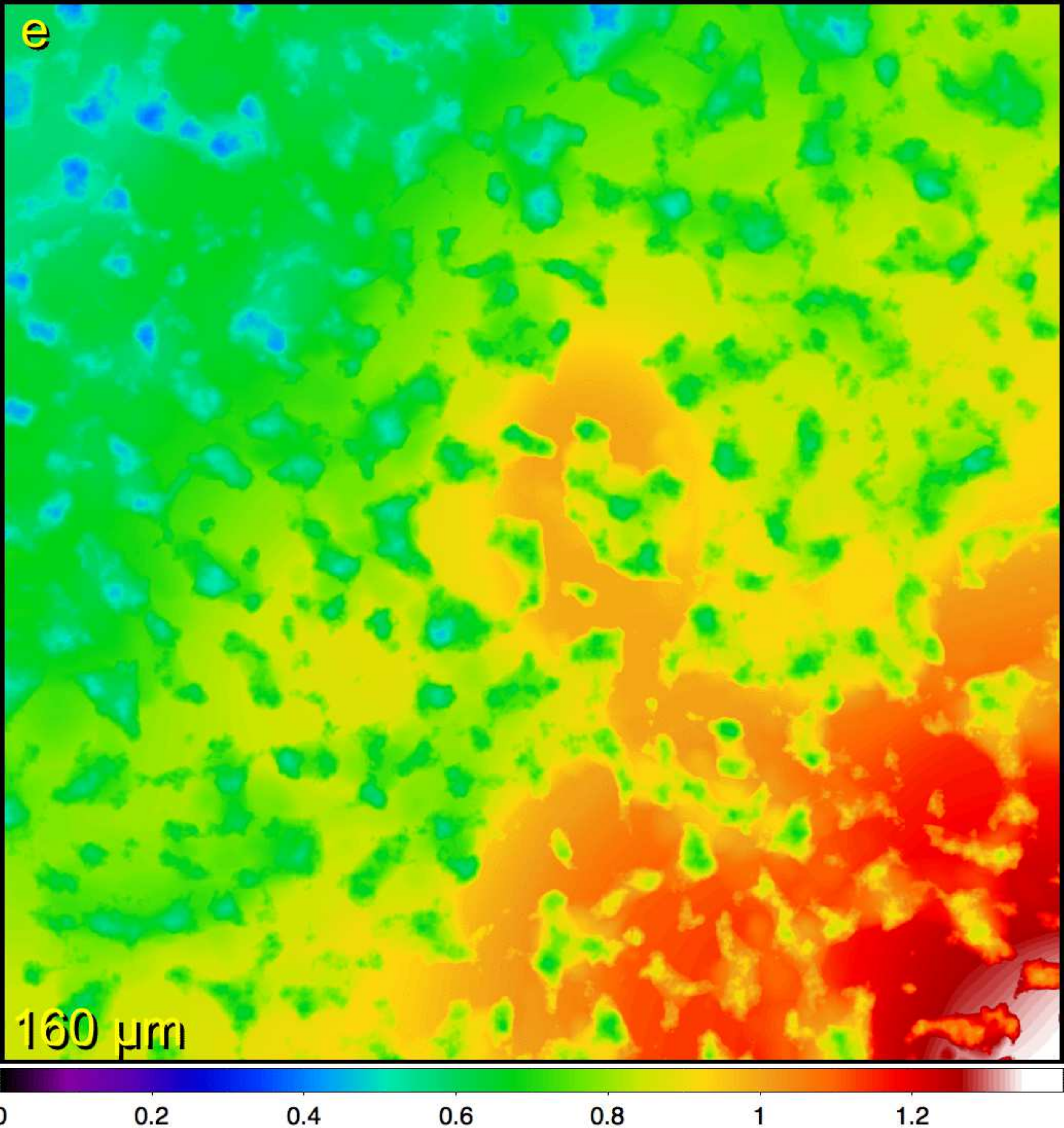}}
            \resizebox{0.33\hsize}{!}{\includegraphics{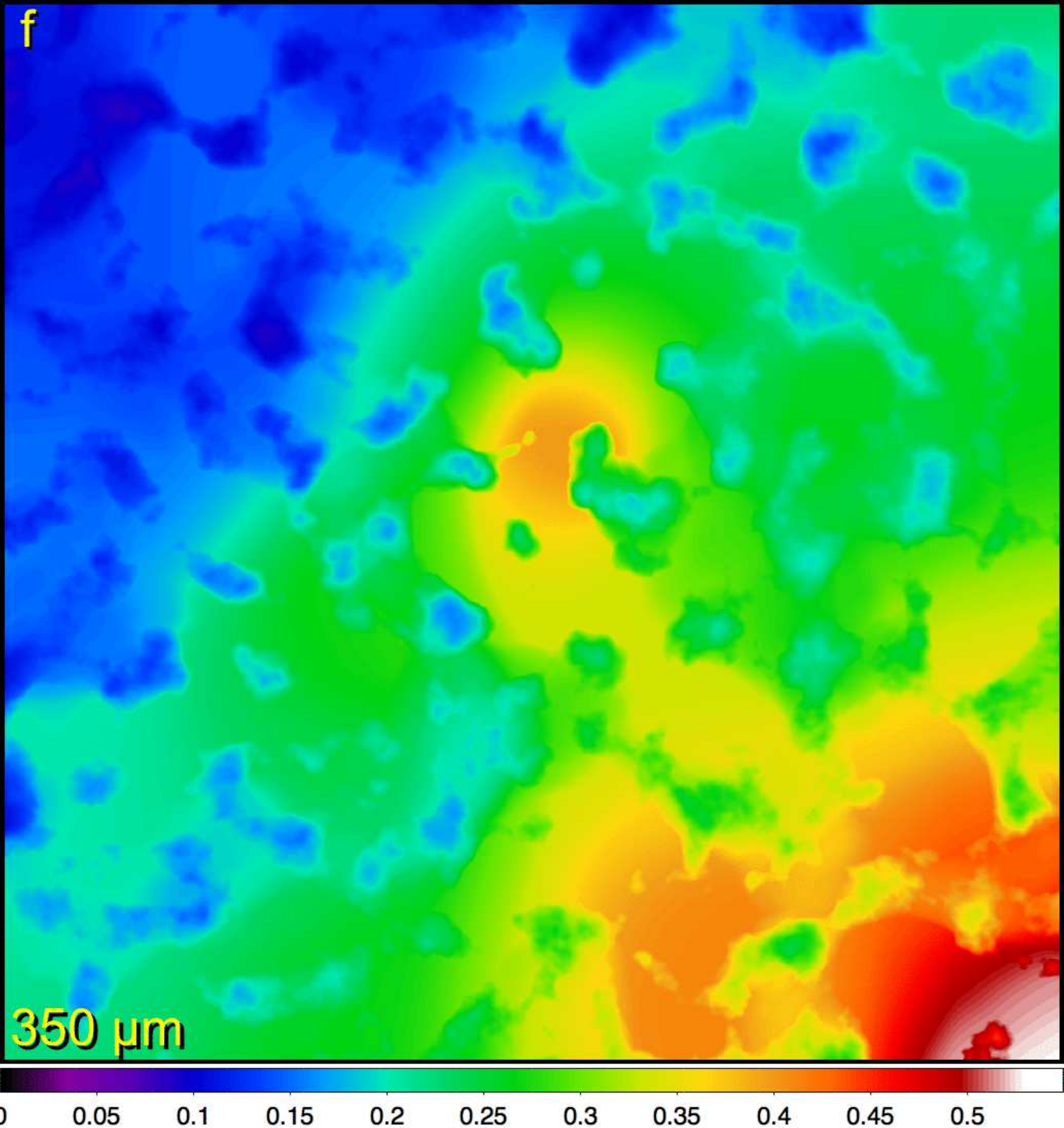}}}
\centerline{\resizebox{0.33\hsize}{!}{\includegraphics{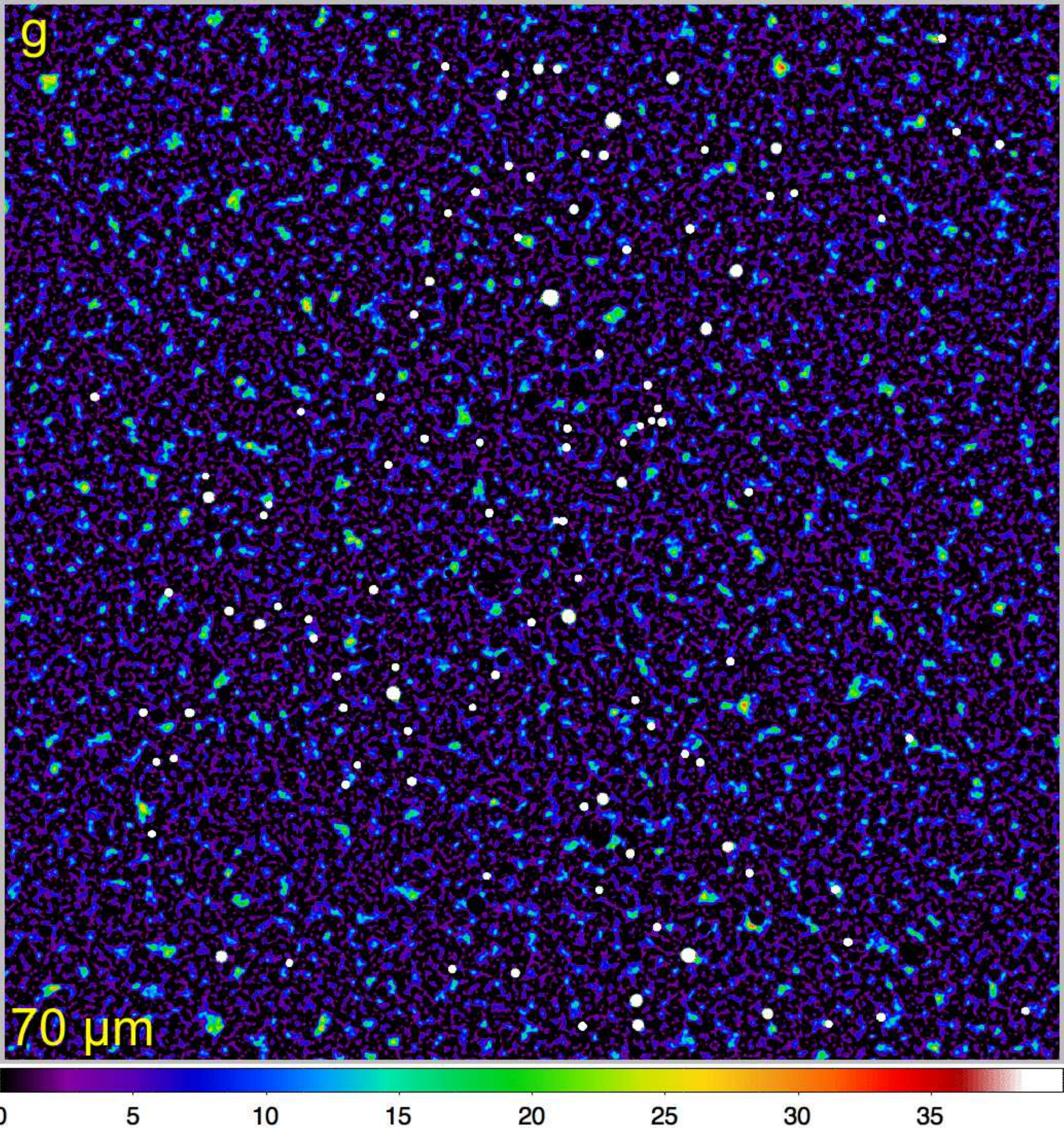}}
            \resizebox{0.33\hsize}{!}{\includegraphics{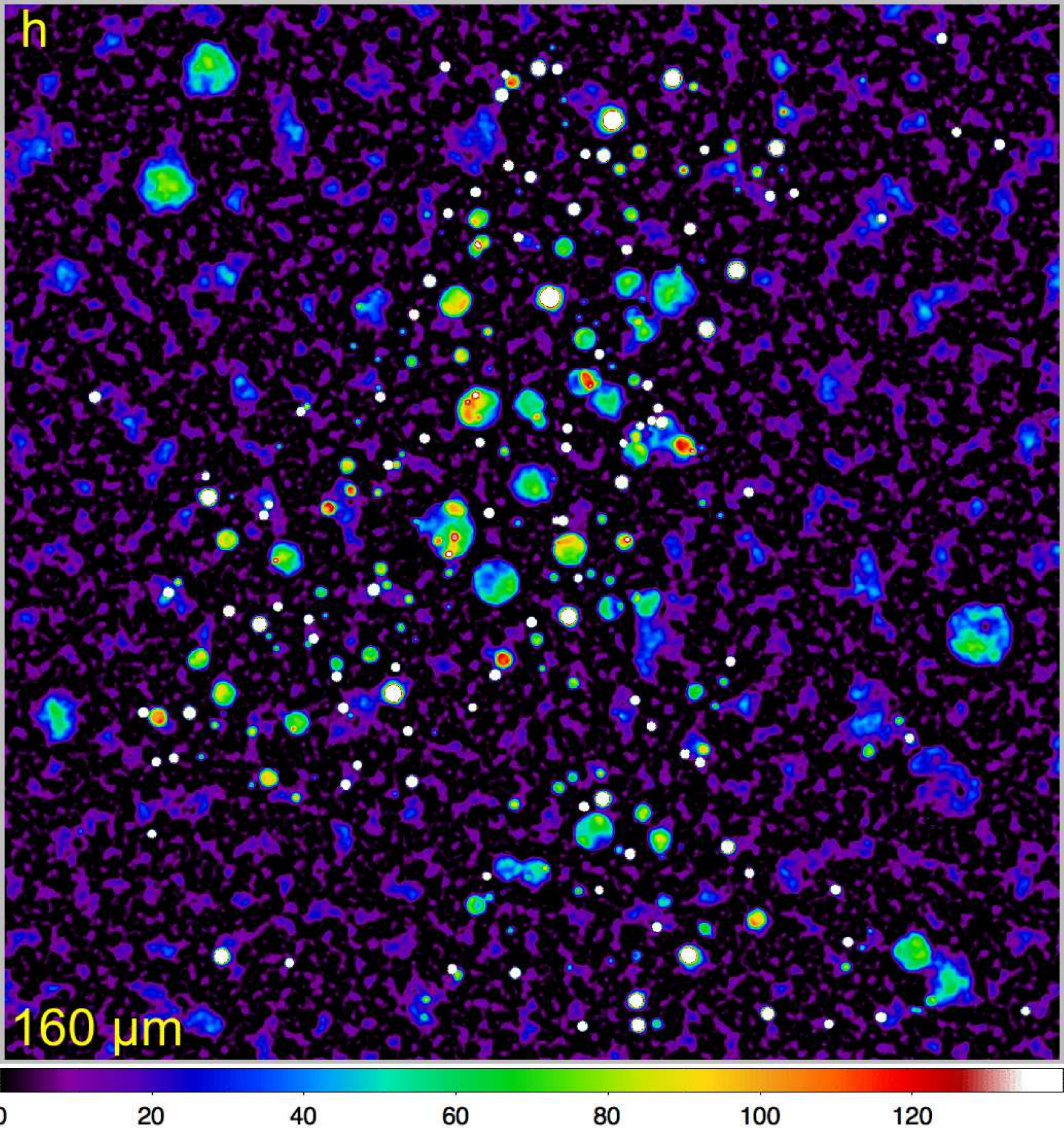}}
            \resizebox{0.33\hsize}{!}{\includegraphics{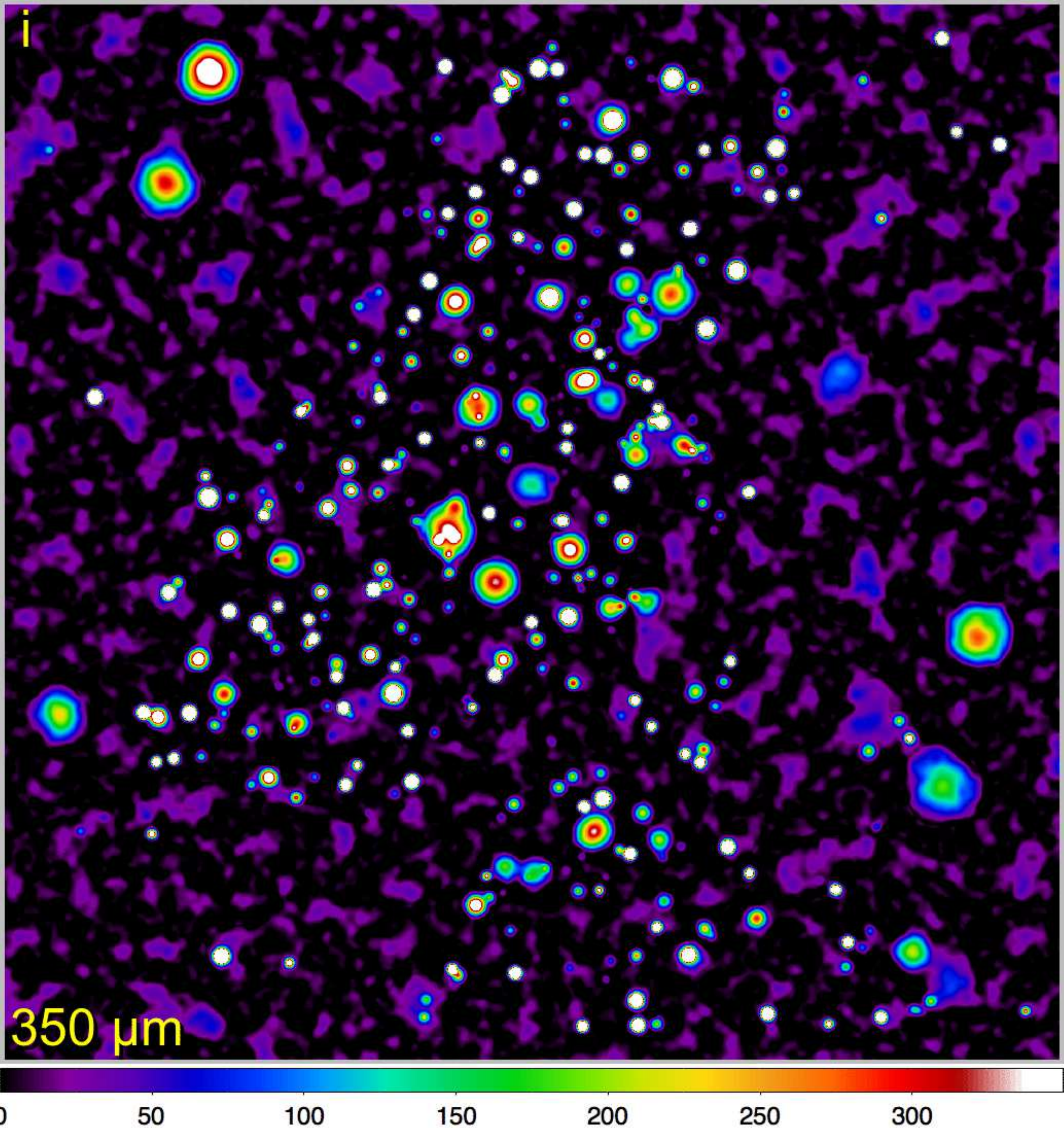}}}
\caption
{ 
Flattening background-subtracted images (Fig.~\ref{bgsub}). Shown are the standard deviations
$\mathcal{D}_{\lambda}{\,=\,}\mathrm{sd}_{9\,}(\tilde{\mathcal{S}_{\lambda}})$ of small-scale fluctuations in
$\tilde{\mathcal{S}_{\lambda}}$ (\emph{upper row}), median-filtered scaling (flattening) images $\tilde{\mathcal{F}}_{\!\lambda}$
(\emph{middle row}), and flattened detection images
$\mathcal{I}_{{\!\lambda}{\mathrm{D}}}{\,=\,}\tilde{\mathcal{S}}_{\lambda}{/}\tilde{\mathcal{F}}_{\!\lambda}$ (\emph{bottom row}) at
selected \emph{Herschel} wavelengths. In panel \emph{b}, thin ring-like structures reflect the off-center temperature peaks in
starless cores illuminated by the interstellar radiation field. The maximum sizes $X_{\lambda}$ of $25$, $100$, and $100${\arcsec}
are the same as those used for background derivation.
} 
\label{flat}
\end{figure*}

\begin{figure*}                                                               
\centering
\centerline{\resizebox{0.33\hsize}{!}{\includegraphics{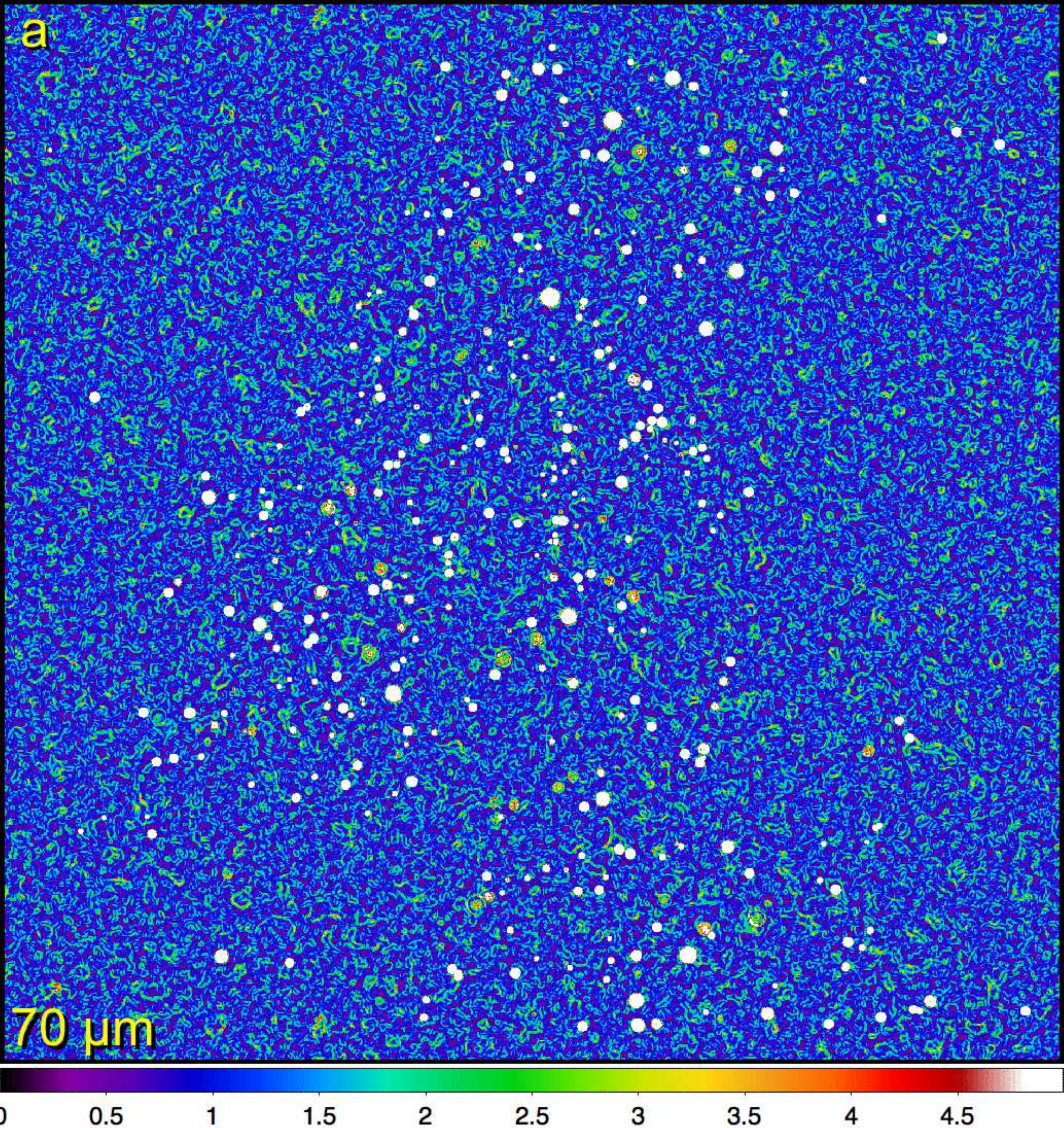}}
            \resizebox{0.33\hsize}{!}{\includegraphics{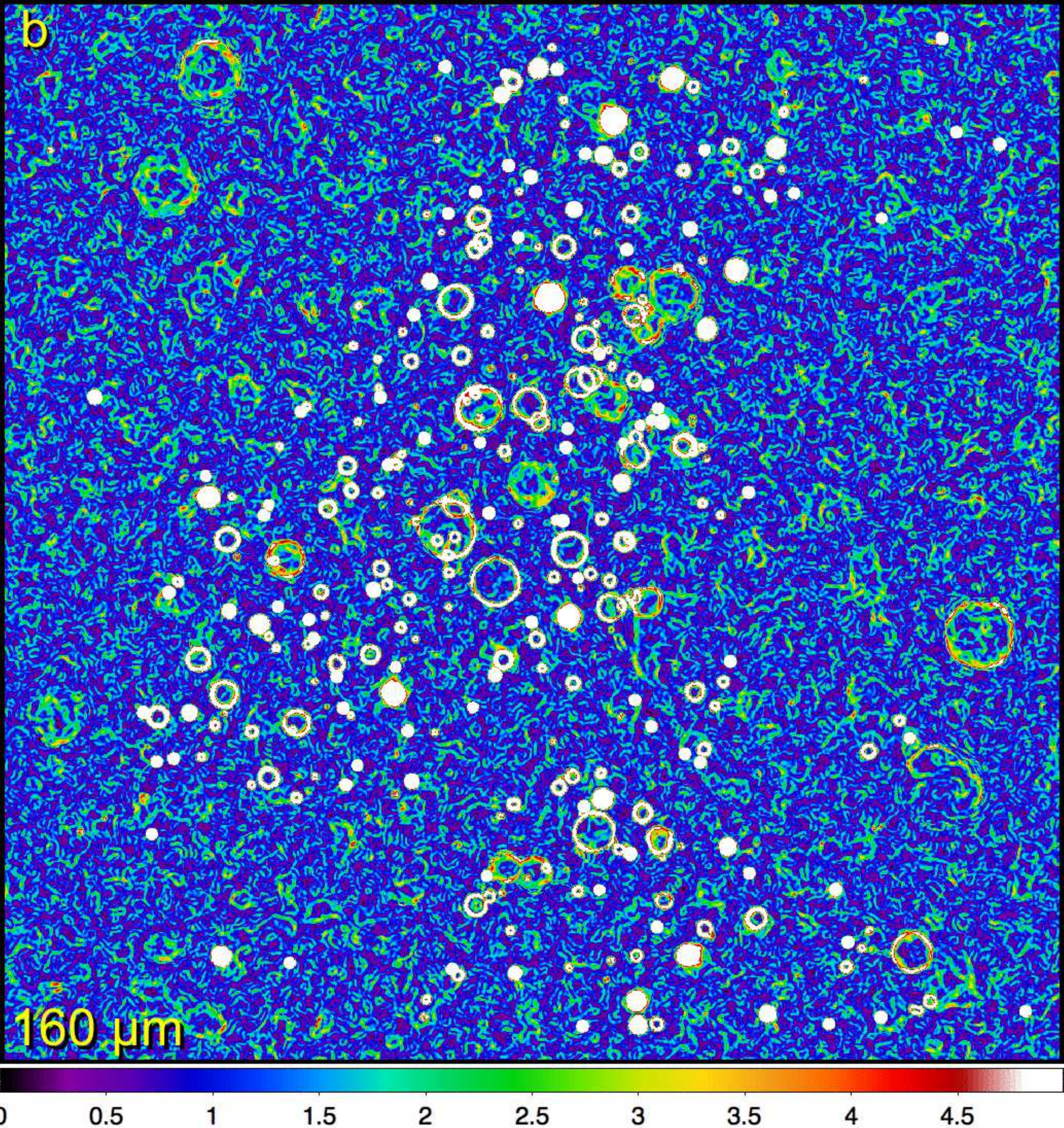}}
            \resizebox{0.33\hsize}{!}{\includegraphics{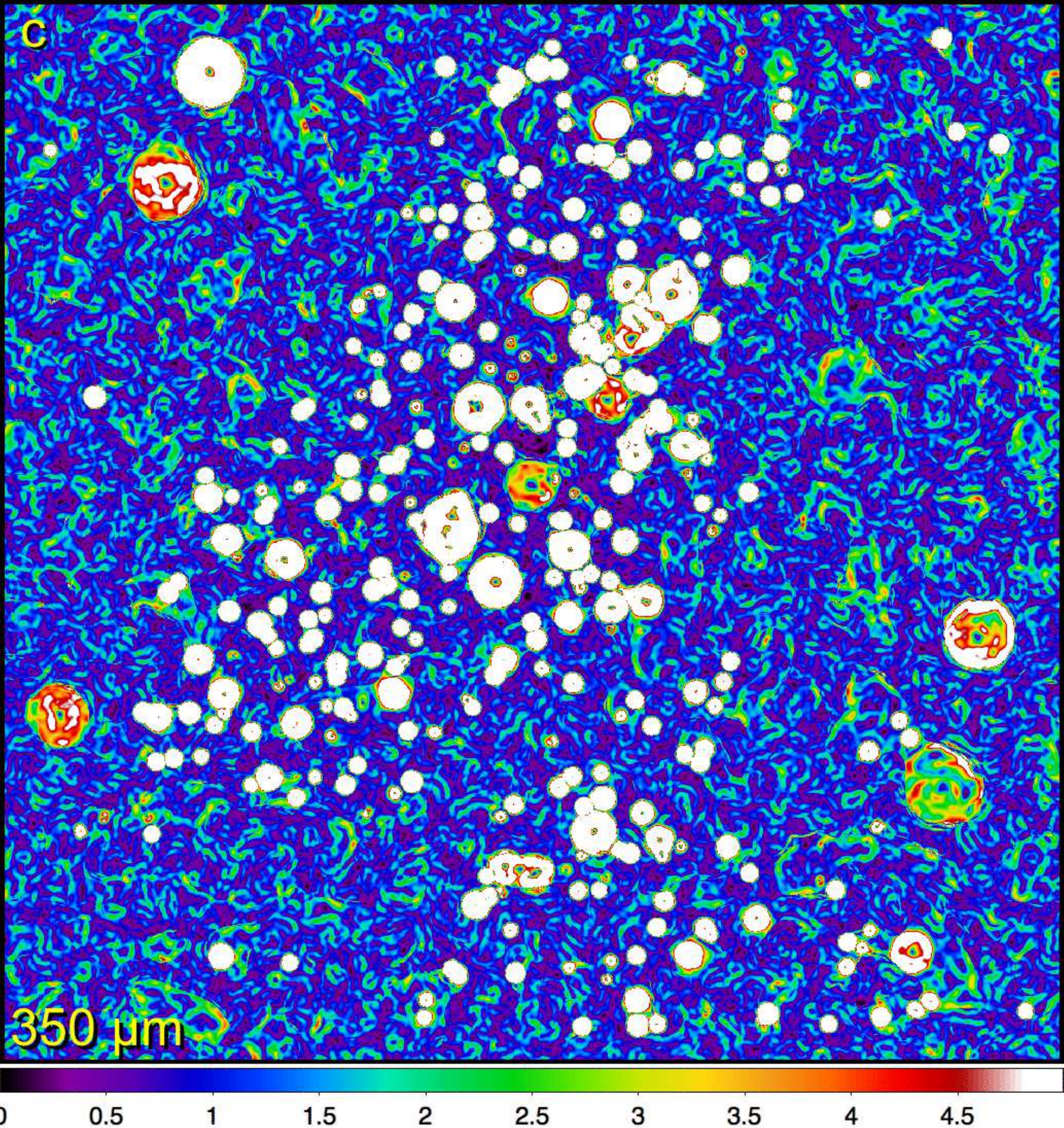}}}
\caption
{ 
Quality of the flattened images presented in Fig.~\ref{flat}. Shown are the standard deviations
$\mathrm{sd}_{9\,}(\mathcal{I}_{{\!\lambda}\mathrm{D}})$ of small-scale fluctuations in the flattened images
$\mathcal{I}_{{\!\lambda}\mathrm{D}}$. The fluctuations outside sources are very uniform, compared to
$\mathcal{D}_{\lambda}{\,=\,}\mathrm{sd}_{9\,}(\tilde{\mathcal{S}_{\lambda}})$ computed from the background-subtracted images.
} 
\label{accflat}
\end{figure*}

The background derivation algorithm is illustrated in Fig.~\ref{bgsub}. The adopted maximum source sizes $X_{\lambda}$ of $25$,
$100$, and $100${\arcsec} for the images at $70$, $160$, and $350$\,{${\mu}$m}, respectively, were estimated manually from
$\mathcal{I}_{\!\lambda}$ according to Eq.~(\ref{maxsizedef}) in Sect.~\ref{maxsize}. The median filtering procedure described by
Eqs.~(\ref{bgfilter}){--}(\ref{iterations2}) was applied $\mathcal{I}_{\!\lambda}$ and the background images
$\tilde{\mathcal{B}}_{\lambda}$ were obtained after $M{\,=\,}30$ iterations. With $X_{70}{\,=\,}25${\arcsec}, intensity variations
of $\tilde{\mathcal{B}}_{70}$ are preserved to smaller spatial scales in comparison to $\tilde{\mathcal{B}}_{160}$ and
$\tilde{\mathcal{B}}_{350}$ derived using larger sliding windows to accommodate sources up to $X_{\lambda}{\,=\,}100${\arcsec}.
Background-subtracted images $\tilde{\mathcal{S}}_{\lambda}$ are much simpler and flatter, but they preserve all structures with
sizes $H_{\lambda}{\,\la\,}X_{\lambda}$ that exist in $\mathcal{I}_{\!\lambda}$. Revealing the sources much more clearly than the
originals do, they contain a small additional contribution $\mathcal{B}_{\lambda}{\,-\,}\tilde{\mathcal{B}}_{\lambda}$ from the
median-filtered peaks of the true model background $\mathcal{B}_{\lambda}$. The background peaks remaining in
$\tilde{\mathcal{S}}_{\lambda}$ are narrower at $70$\,{${\mu}$m} than at $160$ and $350$\,{${\mu}$m}, according to the difference
in $X_{\lambda}$. The background residuals are the reason why, in general, $\tilde{\mathcal{S}}_{\lambda}$ can only be used for
detection, with the exception of simple and smooth extended backgrounds whose peaks have sizes $H_{\lambda}{\,\gg\,}X_{\lambda}$.

The relative accuracies of $\tilde{\mathcal{B}}_{\lambda}$ for all sources are presented in Fig.~\ref{accbg}, where only the
relevant pixels belonging to the model sources are displayed. The accuracy of derived background depends on the source size and its
position. For most of the sources, $\tilde{\mathcal{B}}_{\lambda}$ is estimated to within $10${\%}. Negative errors are found in
the places where the true fluctuating background has local peaks. Relatively large positive errors (red and white areas in
Fig.~\ref{accbg}) are observed only inside very extended isolated or overlapping sources, where wide areas of the true fluctuating
background $\mathcal{B}_{\lambda}$ may have deeper hollows. Such an overestimation is impossible to correct, as the true background
under real sources is very uncertain \citep[cf. Appendix B in][]{Men'shchikov2016}. When averaged over an entire source, the
background accuracy becomes much better. However, the accuracy of $\tilde{\mathcal{B}}_{\lambda}$ is irrelevant for detection
images that are not meant to be used for measurements. Detection images must only preserve all sources, their positions and sizes,
which is clearly the case for the results displayed in Fig.~\ref{bgsub}.

\subsection{Flattening background-subtracted images}
\label{flattening}

Observations demonstrate that background intensities and their fluctuations vary by orders of magnitude across images. They are
especially variable in the short-wavelength images that are affected by relatively strong dust temperature deviations induced by
the radiation from hot stars. Although background subtraction greatly simplifies $\mathcal{I}_{\!\lambda}$ (Sect.~\ref{bgestim}),
the removal of an average background $\tilde{\mathcal{B}}_{\lambda}$ on spatial scales larger than $X_{\lambda}$ does not strongly 
reduce variable fluctuations across $\tilde{\mathcal{S}}_{\lambda}$ on scales ${\la\,}X_{\lambda}$.

A flattening procedure introduced in \textsl{getsources} (Paper I) attempted to equalize the fluctuation levels in different parts
of detection images $\mathcal{I}_{{\!\lambda}{\mathrm{D}}}$ by dividing the latter with flattening images obtained from small-scale
background fluctuations. However, the original scheme required a complete preliminary source extraction in order to determine
background by cutting off the extracted sources. Depending on the extraction quality in the original complex images, the two-step
approach was not completely universal and hence not fully satisfactory. In contrast, the background derivation procedure introduced
in Sect.~\ref{bgestim} removes all sources automatically without any need of a prior iterative source extraction. The simple
median-filtering scheme of Eqs.~(\ref{bgfilter}) and (\ref{bgfilter2}) leads to a new straightforward and accurate flattening
procedure.

The standard deviations of
$\tilde{\mathcal{S}_{\lambda}}{\,=\,}\mathcal{S}_{\lambda}{\,+\,}(\mathcal{B}_{\lambda}{\,-\,}\tilde{\mathcal{B}_{\lambda}})$ are
computed in a small sliding window of $3{\,\times\,}3$ pixels\footnote{The small 9-pixel window ensures that fluctuations are
evaluated with the highest resolution, as locally as possible.}, and the operation is denoted as
$\mathcal{D}_{\lambda}{\,=\,}\mathrm{sd}_{9\,}(\tilde{\mathcal{S}_{\lambda}})$. For illustration, $\mathcal{D}_{\lambda}$ at
selected wavelengths of $70$, $160$, and $350$\,{${\mu}$m} is presented in Fig.~\ref{flat} (upper panels). Approximately, the
operation can be written as a sum of the individual components:
$\mathcal{D}_{\lambda}{\,\approx\,}\mathrm{sd}_{9\,}(\mathcal{S}_{\lambda}){\,+\,}\mathrm{sd}_{9\,}(\mathcal{B}
_{\lambda}{\,-\,}\tilde{\mathcal{B}_{\lambda}})$. This formulation is not precise (in pixels where
$\mathcal{S}_{\lambda}{\,\approx\,}\mathcal{B}_{\lambda}{\,-\,}\tilde{\mathcal{B}_{\lambda}}$) and used here only to highlight the
fact that $\mathcal{D}_{\lambda}$ contains contributions of the fluctuations induced by both sources and background residuals (in
general, also by filaments and noise). As a first step toward flattening, it is necessary to remove from $\mathcal{D}_{\lambda}$
the fluctuations produced by sources, that is, to determine its background
$\mathrm{sd}_{9\,}(\mathcal{B}_{\lambda}{\,-\,}\tilde{\mathcal{B}_{\lambda}})$.

The obvious similarity of the problems of deriving source-free images from $\mathcal{I}_{\lambda}$ and $\mathcal{D}_{\lambda}$
makes applying the median-filtering algorithm described in Sect.~\ref{bgestim} feasible. Using the same set of $N$ sliding windows
with radii $R_{\lambda\,1}, R_{\lambda\,2},\dots, R_{\lambda\,N}$, the algorithm median filters the standard deviations
$\mathcal{D}_{\lambda}$ and minimizes the resulting set of images:
\begin{equation} 
\hat{\mathcal{F}}_{\!\lambda} = \min \left\{\,\mathrm{mf}_{R_{\lambda\,j}}(\mathcal{D}_{\lambda})\,\right\}
\,\,\,({j = 1, 2,\dots, N}).
\label{stdfilter}
\end{equation} 
To conclude the procedure, the median filtering with the largest window is repeated (twice) to smooth the resulting image:
\begin{equation} 
\tilde{\mathcal{F}}_{\!\lambda} = \min \left\{\,\hat{\mathcal{F}}_{\!\lambda},\, 
\mathrm{mf}_{R_{\lambda\,N}}(\hat{\mathcal{F}}_{\!\lambda}),\,
\mathrm{mf}_{R_{\lambda\,N}}\bigl(\mathrm{mf}_{R_{\lambda\,N}}(\hat{\mathcal{F}}_{\!\lambda})\bigr)\,\right\}.
\label{stdfilter2}
\end{equation} 
The above procedure ensures that background fluctuations obtained with small windows for sources or filaments with
$H_{\lambda}{\,\approx\,}O_{\lambda}$ survive median filtering with much larger windows that are suitable for more extended
structures. As a consequence, contributions of all sources with sizes $O_{\lambda}{\,\le\,}S_{\!\lambda}{\,\la\,}X_{\lambda}$ are
removed from the image $\tilde{\mathcal{F}}_{\!\lambda}$ of background fluctuations.

A flattened detection image is obtained by dividing $\tilde{\mathcal{S}}_{\lambda}$ by the flattening (scaling) image:
$\mathcal{I}_{{\!\lambda}\mathrm{D}}{\,=\,}\tilde{\mathcal{S}}_{\lambda}/\tilde{\mathcal{F}}_{\!\lambda}$. This procedure
effectively equalizes fluctuation levels across the entire image $\mathcal{I}_{{\!\lambda}\mathrm{D}}$, while preserving the
intensity distributions of sources.

The flattening algorithm is illustrated in Fig.~\ref{flat}. The adopted maximum sizes $X_{\lambda}$ of $25$, $100$, and
$100${\arcsec} for the images at $70$, $160$, and $350$\,{${\mu}$m}, respectively, are the same as those used in
Sect.~\ref{bgestim} to derive background $\tilde{\mathcal{B}}_{\lambda}$. The standard deviations $\mathcal{D}_{\lambda}$ of the
background-subtracted $\tilde{\mathcal{S}}_{\lambda}$, computed in a 9-pixel sliding window, clearly show the sources and that the
residual background fluctuations increase along one diagonal. The amplifying fluctuations are induced by the planar temperature
gradient adopted in the synthetic background (Sect.~\ref{bgestim}). Such images are not well suited for a complete and reliable
source extraction. It is easy to see that global thresholding methods are bound to either produce spurious sources or leave some
faint sources undetected. The intensity gradient of fluctuations is also visible in the flattening images
$\tilde{\mathcal{F}}_{\!\lambda}$ obtained using the median-filtering procedure described by Eqs.~(\ref{stdfilter}) and
(\ref{stdfilter2}). By construction, the images $\mathcal{I}_{{\!\lambda}\mathrm{D}}$ in Fig.~\ref{flat} must have uniform
fluctuation levels and they do appear much flatter than $\tilde{\mathcal{S}}_{\lambda}$ (Fig.~\ref{bgsub}).

\begin{figure*}                                                               
\centering
\centerline{\resizebox{0.33\hsize}{!}{\includegraphics{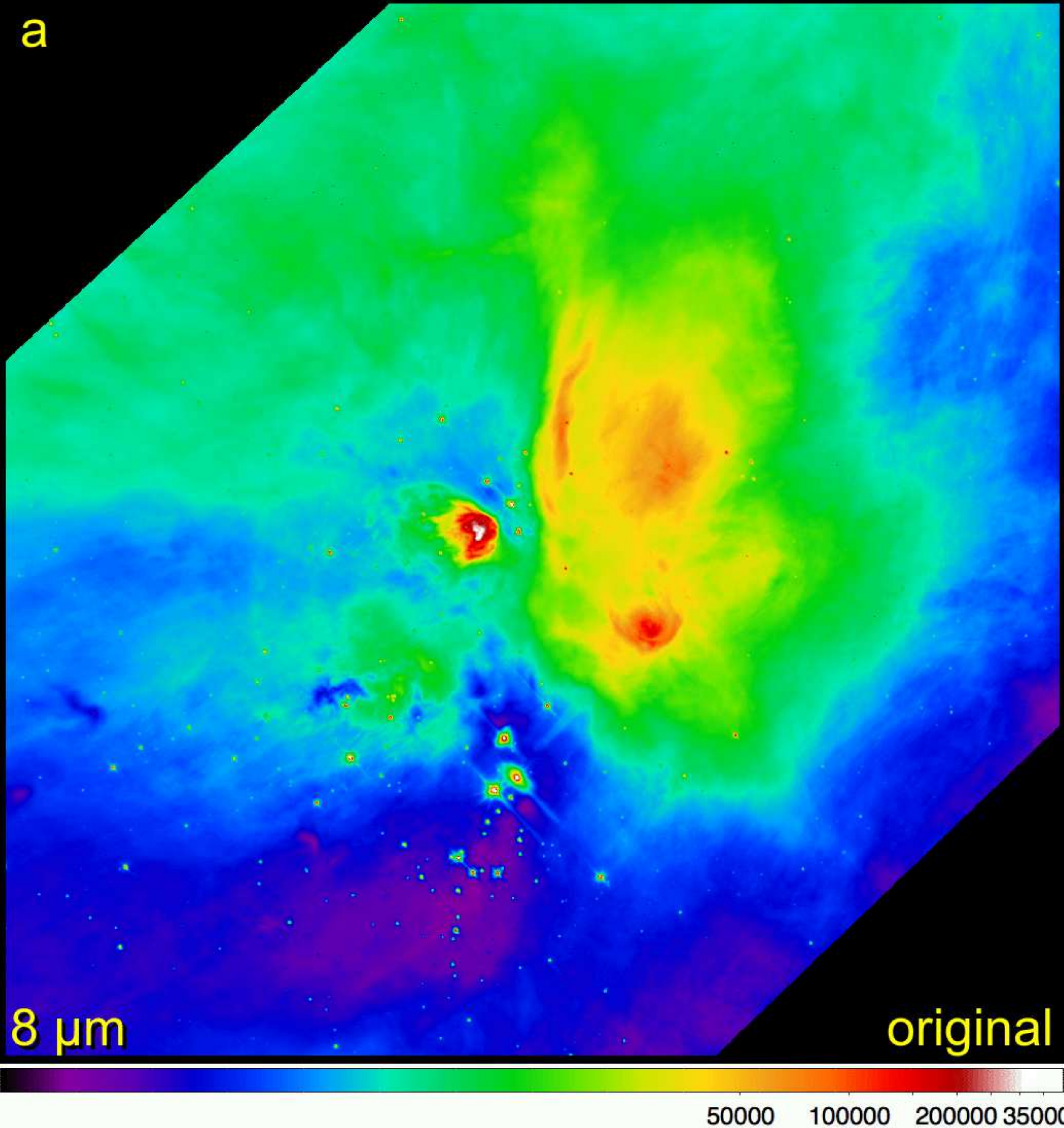}}
            \resizebox{0.33\hsize}{!}{\includegraphics{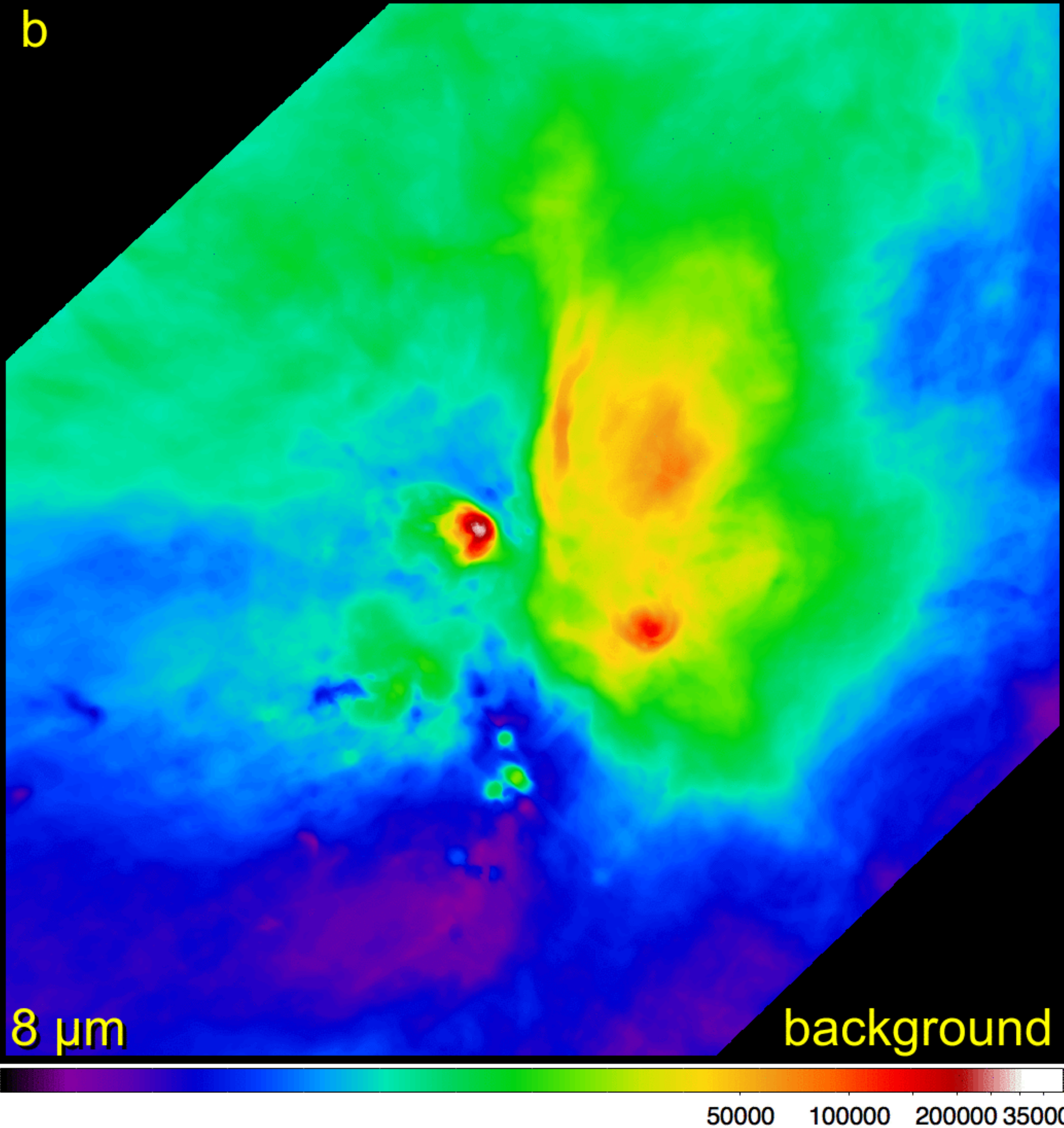}}
            \resizebox{0.33\hsize}{!}{\includegraphics{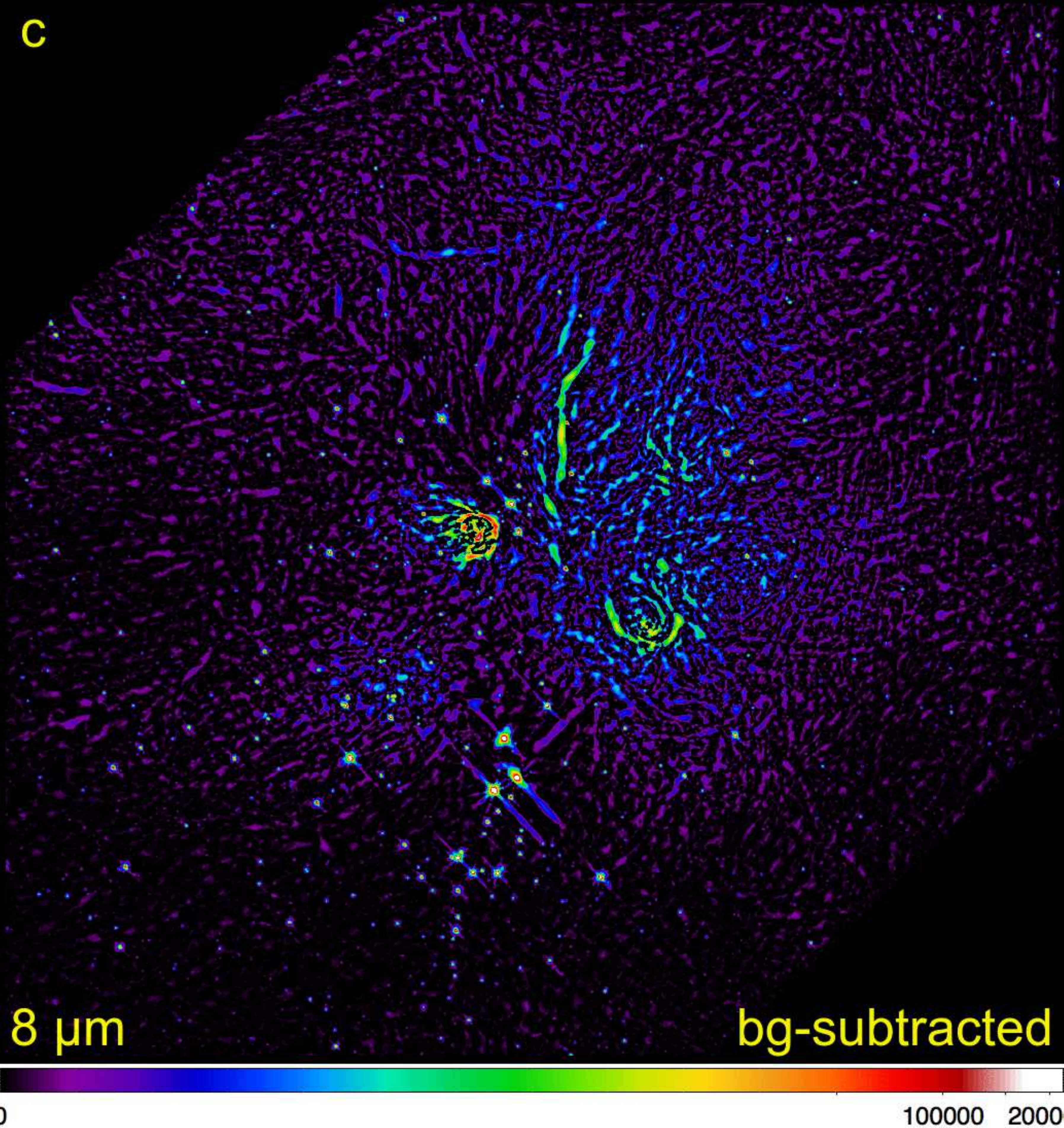}}}
\centerline{\resizebox{0.33\hsize}{!}{\includegraphics{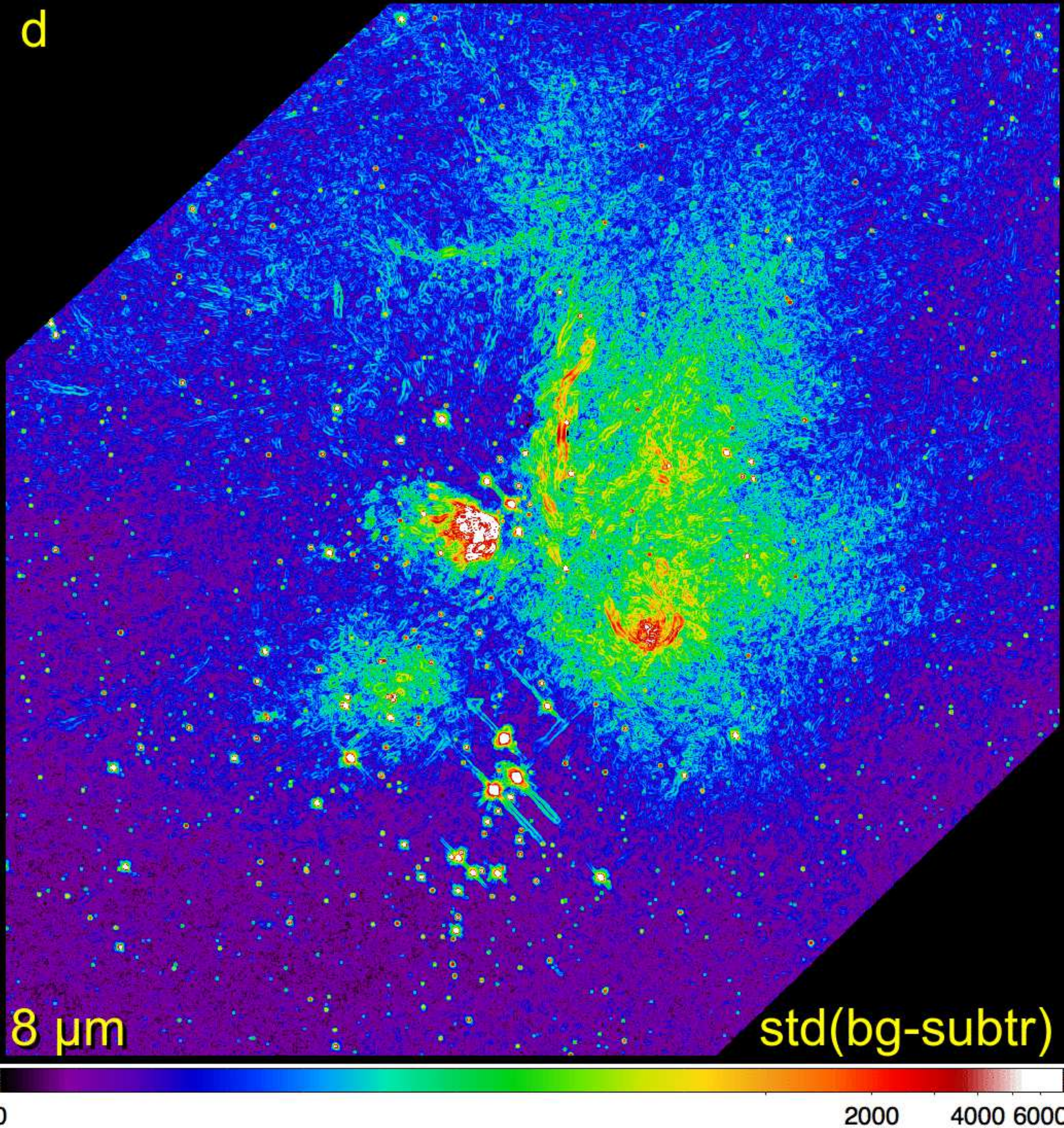}}
            \resizebox{0.33\hsize}{!}{\includegraphics{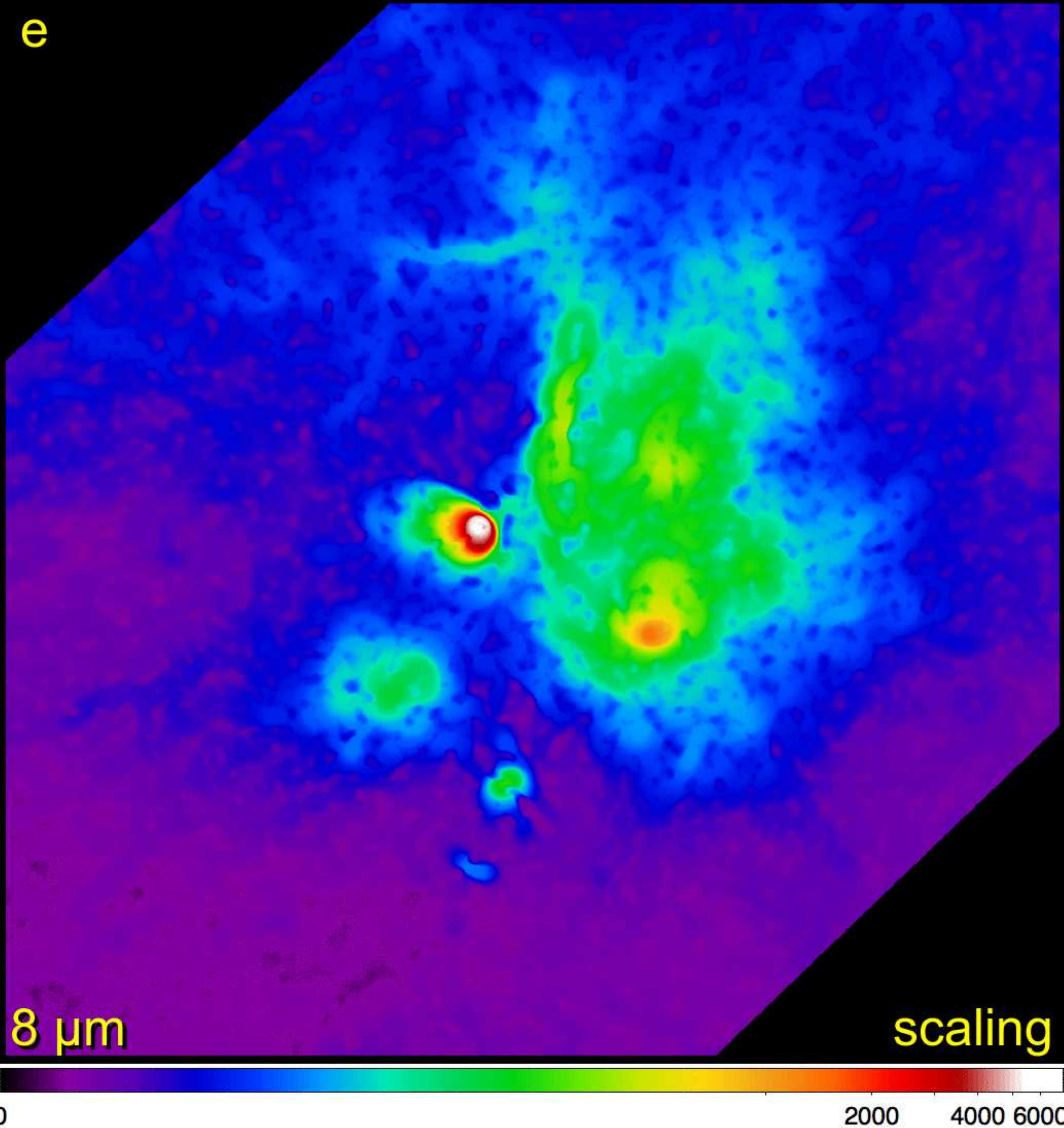}}
            \resizebox{0.33\hsize}{!}{\includegraphics{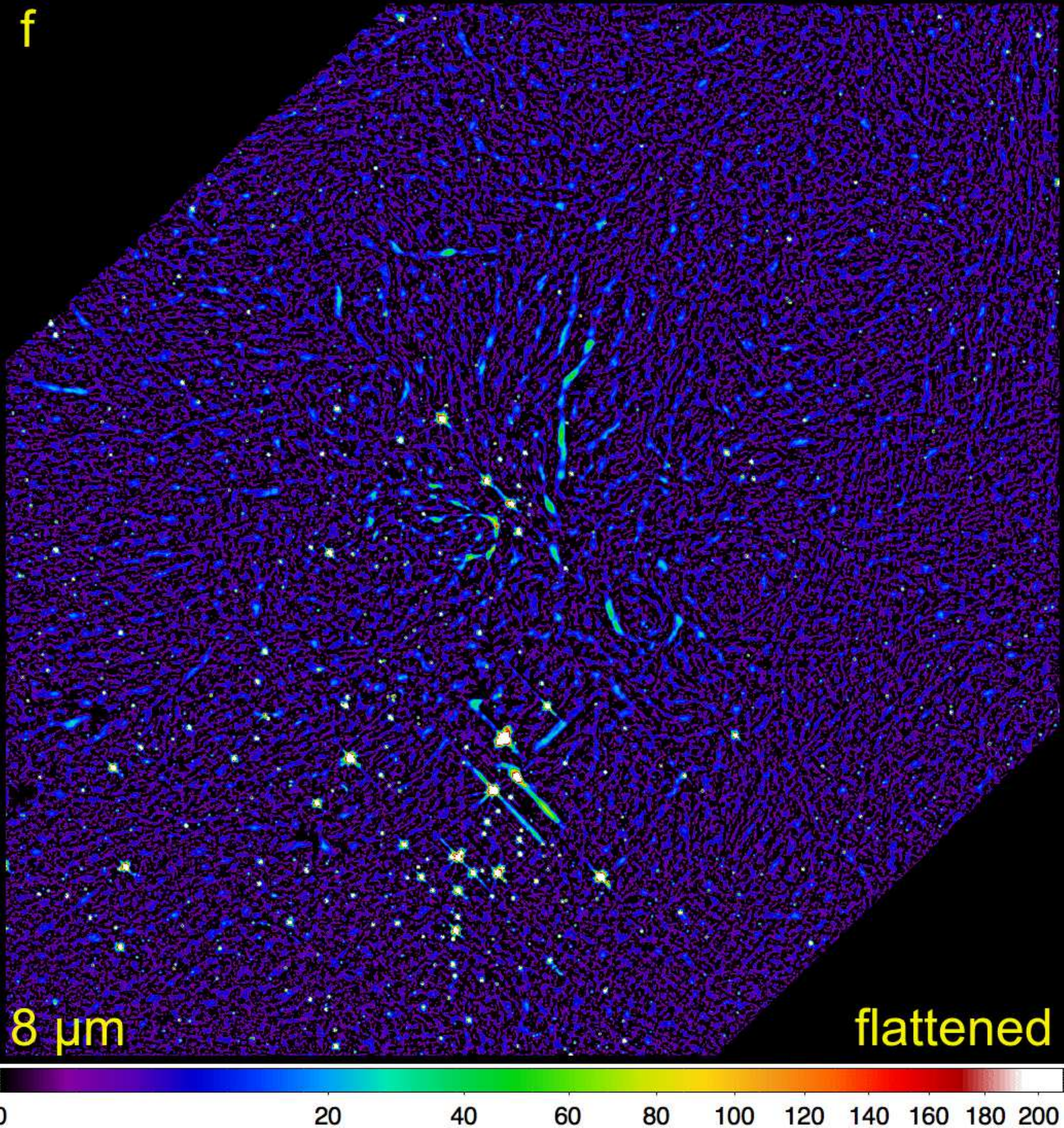}}}
\centerline{\resizebox{0.33\hsize}{!}{\includegraphics{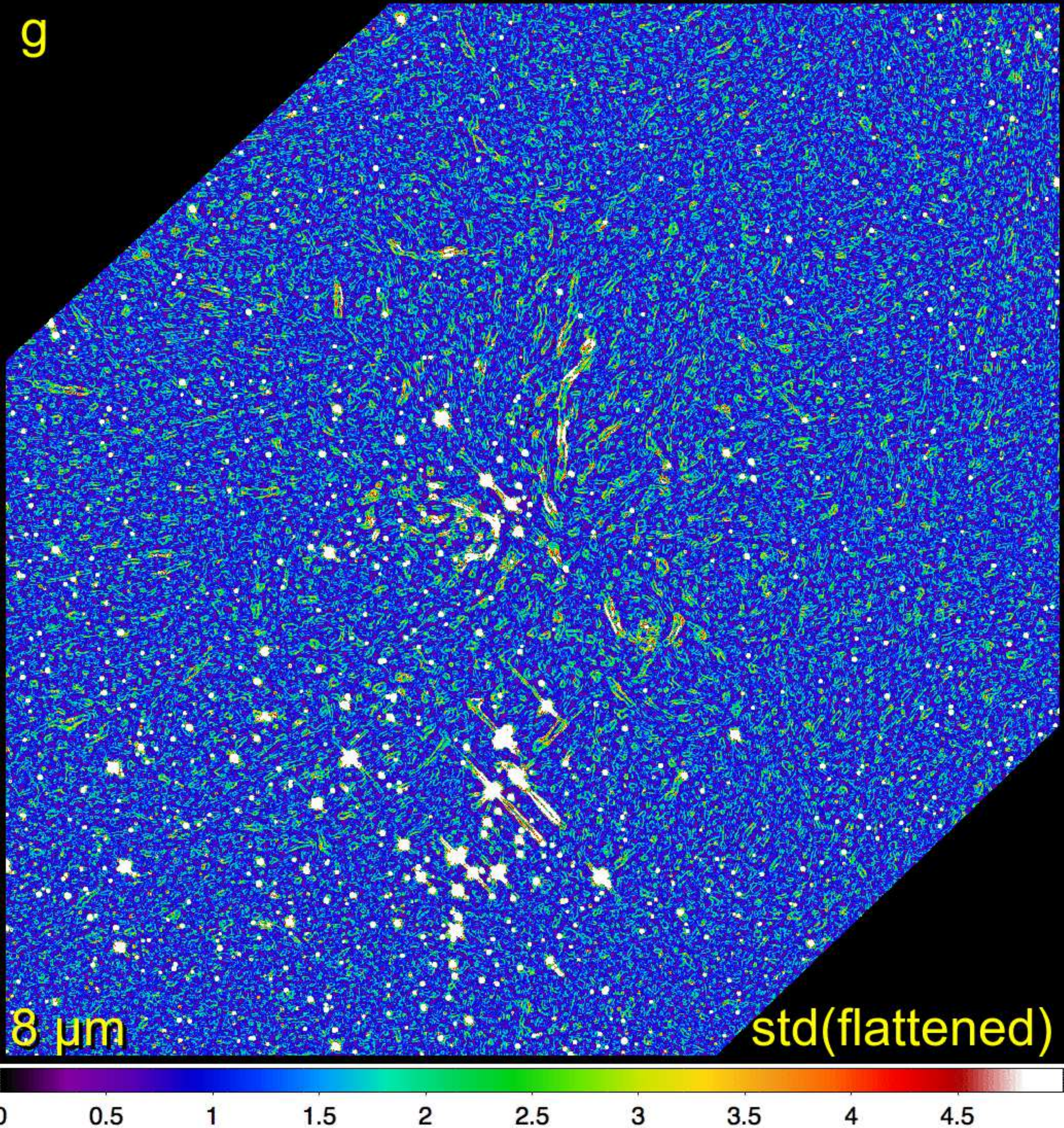}}
            \resizebox{0.33\hsize}{!}{\includegraphics{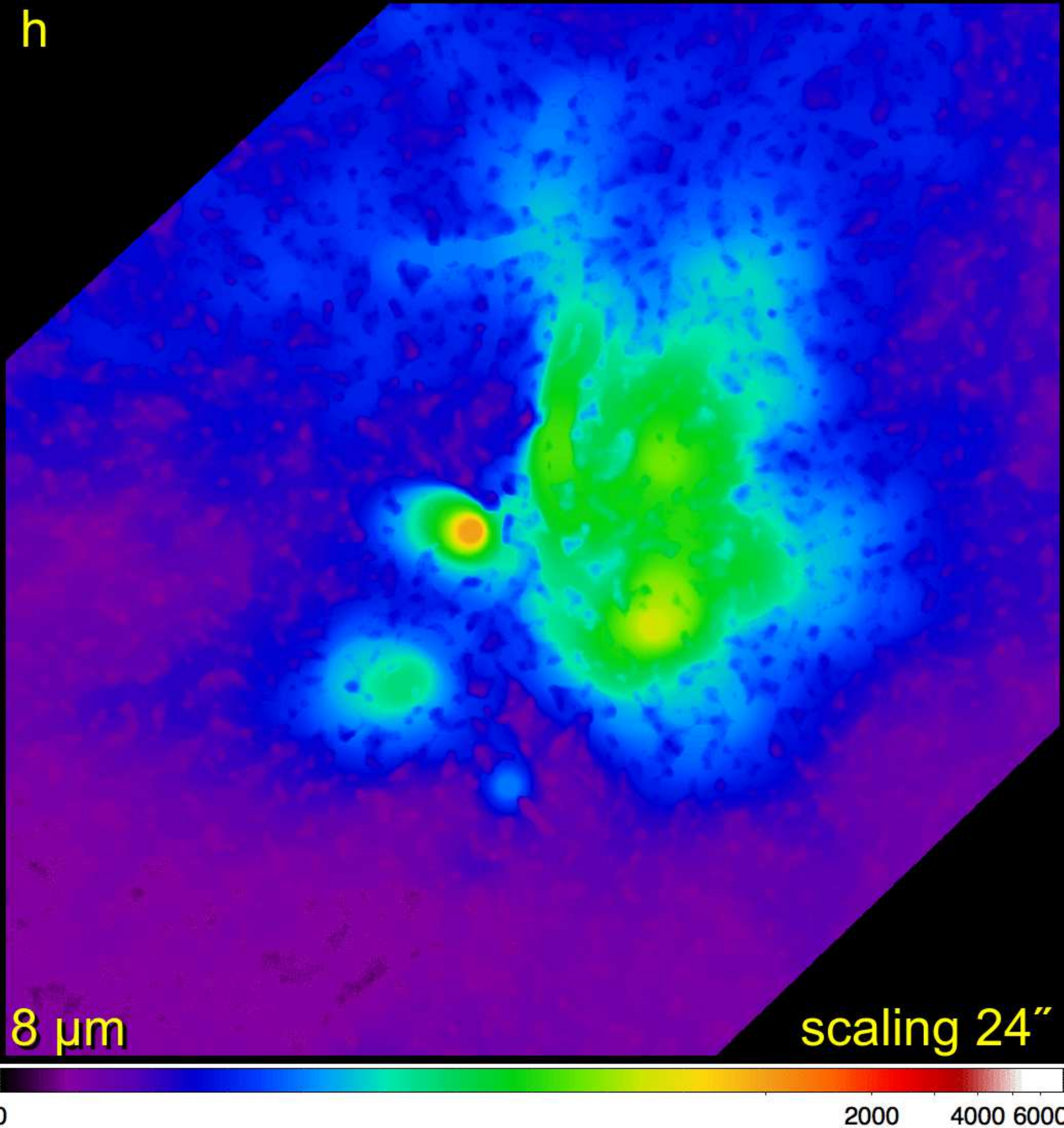}}
            \resizebox{0.33\hsize}{!}{\includegraphics{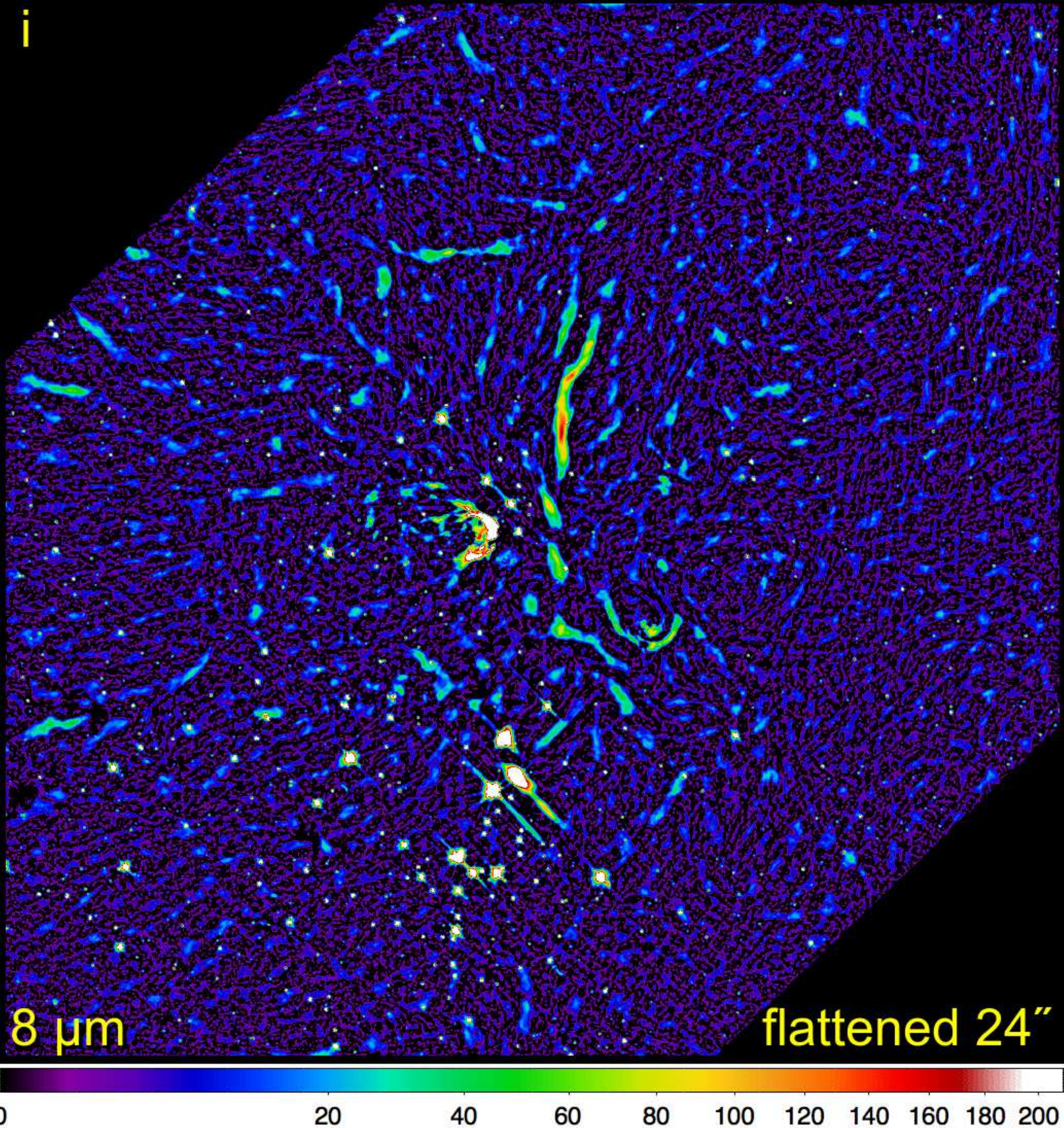}}}
\caption
{ 
Application of \textsl{getimages} (with $X_{\lambda}{\,=\,}12${\arcsec} and $M{\,=\,}30$) to real-life (\emph{Spitzer}
$\lambda{\,=\,}8$\,{${\mu}$m}) observations of the Ophiuchus \object{L\,1688} star-forming region: (\emph{a}) original image
$\mathcal{I}_{\!\lambda}$, (\emph{b}) estimated background $\tilde{\mathcal{B}}_{\lambda}$, (\emph{c}) background-subtracted image
$\tilde{\mathcal{S}}_{\lambda}$, (\emph{d}) standard deviations $\mathcal{D}_{\lambda}$, (\emph{e}) scaling image
$\tilde{\mathcal{F}}_{\!\lambda}$, (\emph{f}) flattened detection image $\mathcal{I}_{{\!\lambda}\mathrm{D}}$, and (\emph{g})
standard deviations $\mathrm{sd}_{9\,}(\mathcal{I}_{{\!\lambda}\mathrm{D}})$. Selected results for $X_{\lambda}{\,=\,}24${\arcsec}
are also shown: (\emph{h}) scaling image $\tilde{\mathcal{F}}_{\!\lambda}$ and (\emph{i}) flattened image
$\mathcal{I}_{{\!\lambda}{\mathrm{D}}}$. Intensities (in MJy/sr) are somewhat limited in range and their color scaling is
logarithmic, except for panels \emph{f} and \emph{i} where it is the square root of intensity, and \emph{g}, where it is linear.
} 
\label{oph-l1688}
\end{figure*}

\begin{figure*}                                                               
\centering
\centerline{\resizebox{0.33\hsize}{!}{\includegraphics{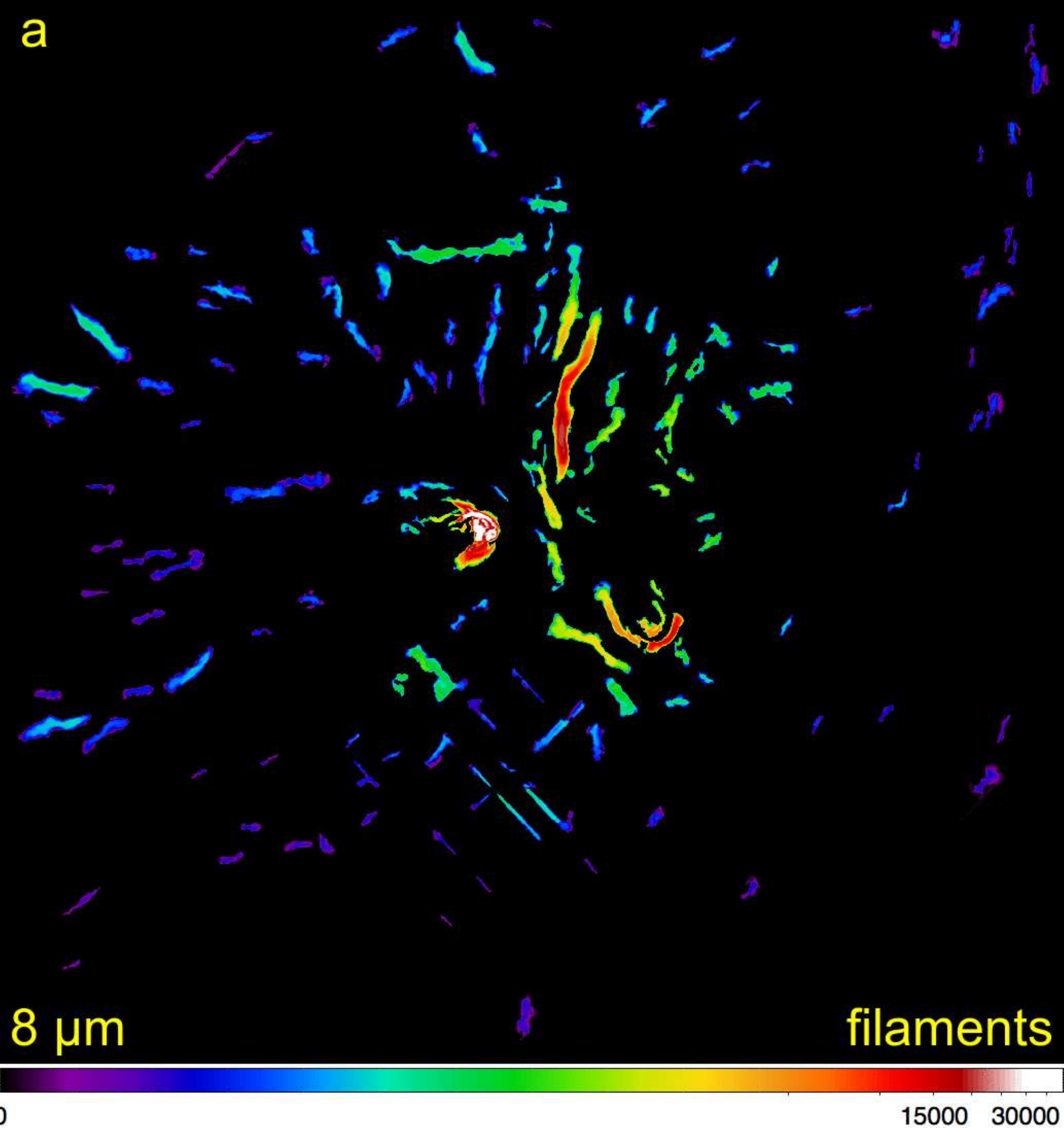}}
            \resizebox{0.33\hsize}{!}{\includegraphics{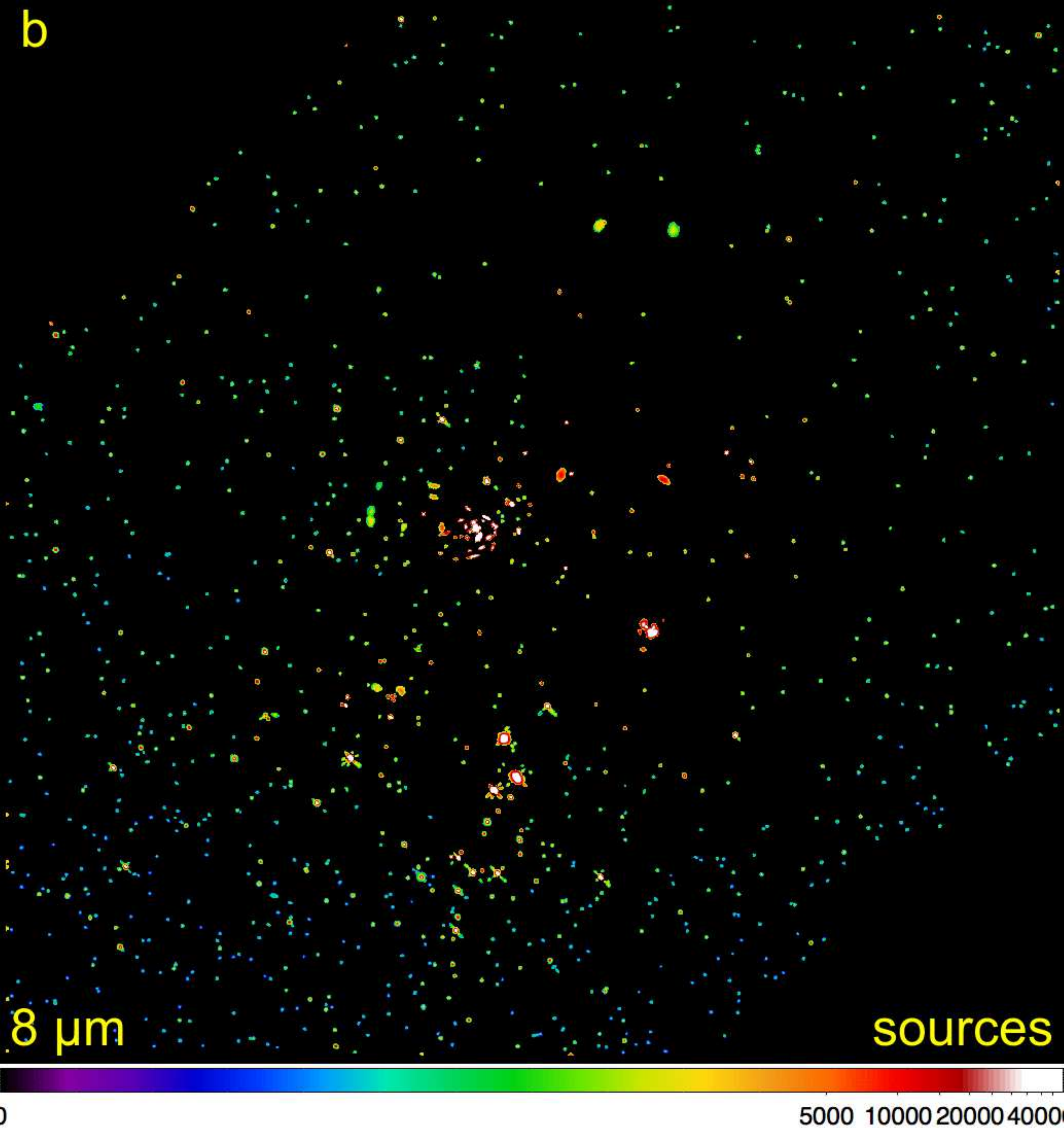}}
            \resizebox{0.33\hsize}{!}{\includegraphics{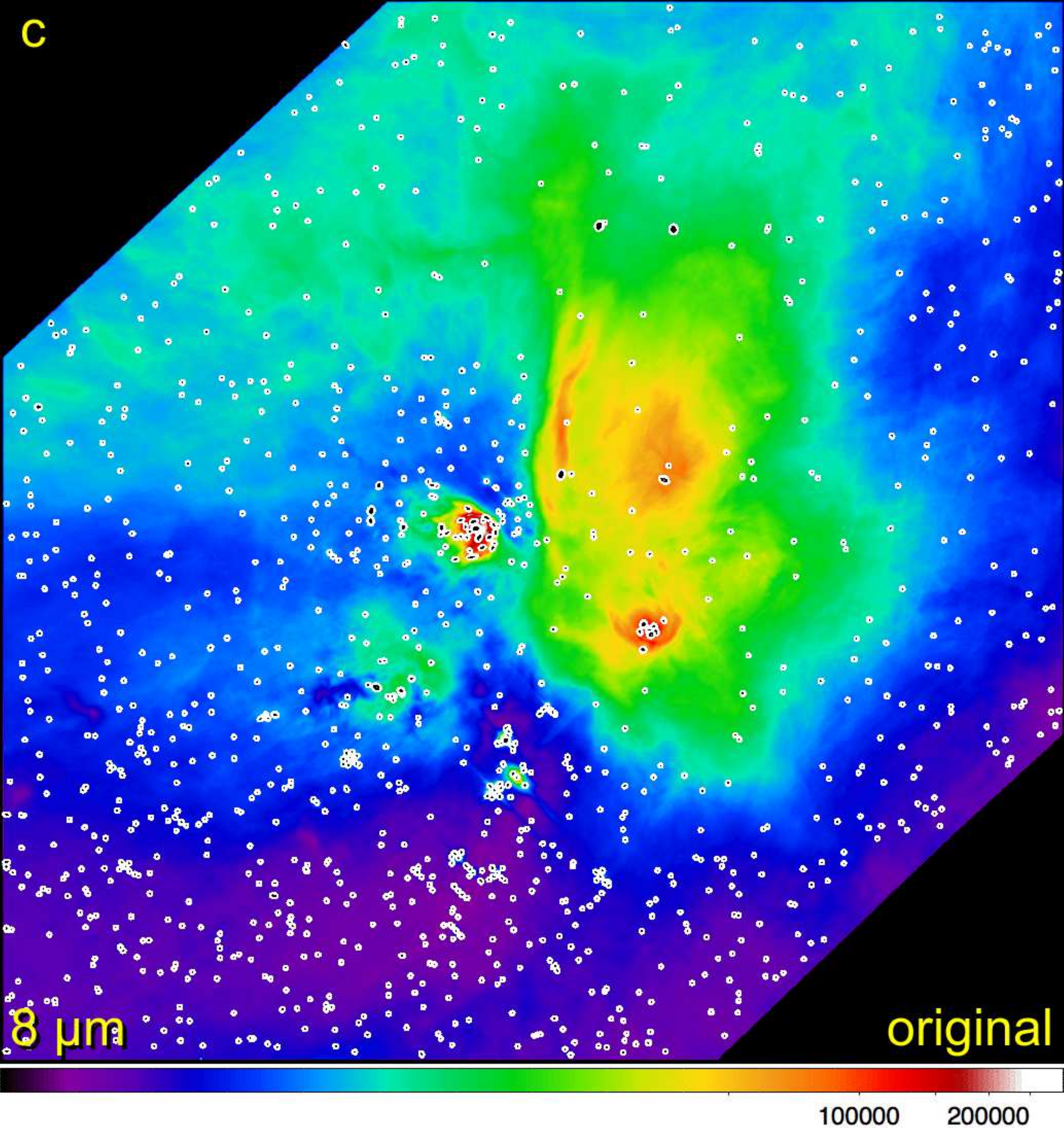}}}
\caption
{ 
Application of \textsl{getsources} and \textsl{getfilaments} to the flattened detection images
$\mathcal{I}_{{\!\lambda}{\mathrm{D}}}$ shown in Fig.~\ref{oph-l1688}{\;\!\emph{i}}. Different structural components are separated
as independent images of filaments (\emph{a}) and sources (\emph{b}). For reference, the extraction ellipses for all detected
sources (most of them may be background galaxies) are also overlaid on the original observed image $\mathcal{I}_{\!\lambda}$
(\emph{c}). Intensities (in MJy/sr) are plotted with logarithmic color scaling.
} 
\label{extraction}
\end{figure*}

The quality of the flattening algorithm is visualized in Fig.~\ref{accflat} by the image of
$\mathrm{sd}_{9\,}(\mathcal{I}_{{\!\lambda}\mathrm{D}})$ computed from the flat detection images. A comparison with
$\mathcal{D}_{\lambda}{\,=\,}\mathrm{sd}_{9\,}(\tilde{\mathcal{S}}_{\lambda})$ makes it quite clear that
$\mathcal{I}_{{\!\lambda}\mathrm{D}}$ have remarkably uniform fluctuations (outside the sources). Such flat images are optimal for
detecting sources (with $H_{\lambda}{\,\la\,}X_{\lambda}$) using extraction methods that employ global thresholding.

\section{Discussion}
\label{discussion}

The background estimation and flattening method \textsl{getimages} was validated in Sect.~\ref{bgsflat} using a simulated
star-forming region. To demonstrate its performance on very complex observed images, it was applied to a \emph{Spitzer}
image\footnote{The $8$\,{${\mu}$m} image was retrieved from the \emph{Spitzer} Heritage Archive (\emph{c2d} Legacy Program, PI Neal
Evans II) by Bilal Ladjelate.} of the star-forming region \object{L\,1688} in Ophiuchus\footnote{The method has been validated on a
dozen regions observed in the \emph{Herschel} Gould Belt \citep{Andre_etal2010} and HOBYS \citep{Motte_etal2010} surveys.}. The
only aim of this section is to clarify various aspects of the new method and any astrophysical analysis based on the image is
beyond the scope of this paper.

\subsection{Application to \object{L\,1688} in Ophiuchus}
\label{applic}

The \emph{Spitzer} $8$\,{${\mu}$m} image of \object{L\,1688} ($1{\,\times\,}1${\degr} with a $2${\arcsec} resolution,
Fig.~\ref{oph-l1688}{\;\!\emph{a}}) displays very complex intensity distribution $\mathcal{I}_{\!\lambda}$, with the background
varying on different spatial scales by more than two orders of magnitude. In addition to many compact sources in both faint and
bright background areas, it also shows some filamentary structures, the most prominent one running up from the middle of the image.
Sources blended with such complex and variable background are very difficult to extract reliably and measure accurately using
automated methods. To apply the \textsl{getimages} method, it is necessary to specify a maximum size $X_{\lambda}$ for the
structures of interest (cf. Sect.~\ref{maxsize}). In this application to \object{L\,1688}, $X_{\lambda}{\,=\,}12{\arcsec}$ was
adopted, and as a variation, a higher value of $24${\arcsec} was also used in order to demonstrate the effects of this parameter.
The results obtained after $M{\,=\,}30$ iterations are presented below.

The background $\tilde{\mathcal{B}}_{\lambda}$ of the \object{L\,1688} image, derived by \textsl{getimages}, is displayed in
Fig.~\ref{oph-l1688}{\;\!\emph{b}}. The median-filtering algorithm described by Eqs.~(\ref{bgfilter}){--}(\ref{iterations2}) has
cleanly erased all sources of various sizes, perhaps with the exception of two large and very bright sources below the image
center, which appear to be incompletely removed. The two residual maxima in $\tilde{\mathcal{B}}_{\lambda}$ may indeed be parts of
the extended power-law sources, or they might also be unrelated background structures. Similarly, the prominent vertical filament
seems to have left some residuals in $\tilde{\mathcal{B}}_{\lambda}$. Nevertheless, the main part of the structures was adequately
truncated and small residuals at that level are unimportant for detection images.

The background-subtracted image $\tilde{\mathcal{S}}_{\lambda}{\,=\,}\mathcal{I}_{\!\lambda}{\,-\,}\tilde{\mathcal{B}}_{\lambda}$
of \object{L\,1688} is presented in Fig.~\ref{oph-l1688}{\;\!\emph{c}}. Without the dominating bright and strongly variable
background, the image becomes substantially simpler, displaying all sources and structures with widths
$X_{\lambda}{\,\la\,}12${\arcsec} much more clearly. Obviously, the brightness remains highly fluctuating across
$\tilde{\mathcal{S}}_{\lambda}$, and subtraction of a large-scale background cannot improve the image to ensure a problem-free
source extraction.

The fluctuations are quantified by the image of standard deviations
$\mathcal{D}_{\lambda}{\,=\,}\mathrm{sd}_{9\,}(\tilde{\mathcal{S}}_{\lambda})$ in \object{L\,1688} shown in
Fig.~\ref{oph-l1688}{\;\!\emph{d}}. Pixel values spanning several orders of magnitude highlight the difficulties that are
encountered when using thresholds in terms of the global standard deviations to separate structures from background and noise. For
such images, a solution would be either to measure local fluctuations around each structure or to equalize background and noise
fluctuations across the entire image $\tilde{\mathcal{S}}_{\lambda}$. The former approach is very problematic, as it would require
an accurate prior source extraction, whereas the latter is the flattening approach adopted by \textsl{getimages}.

To render the fluctuations uniform, the method computes a median-filtered background $\tilde{\mathcal{F}}_{\!\lambda}$ of the
standard deviations $\mathcal{D}_{\lambda}$, according to the algorithm of Eqs.~(\ref{stdfilter}) and (\ref{stdfilter2}). The
scaling (flattening) image $\tilde{\mathcal{F}}_{\!\lambda}$ shown in Fig.~\ref{oph-l1688}{\;\!\emph{e}} represents background and
noise fluctuations in \object{L\,1688}, excluding all structures of widths $X_{\lambda}{\,\la\,}12${\arcsec}. Similarly to
$\tilde{\mathcal{B}}_{\lambda}$, which approximates the local background for all sources and filaments,
$\tilde{\mathcal{F}}_{\!\lambda}$ evaluates their local fluctuation levels. To equalize the latter over the entire image, it is
sufficient to divide $\tilde{\mathcal{S}}_{\lambda}$ by the scaling image $\tilde{\mathcal{F}}_{\!\lambda}$.

The flattened detection image
$\mathcal{I}_{{\!\lambda}\mathrm{D}}{\,=\,}\tilde{\mathcal{S}}_{\lambda}/\tilde{\mathcal{F}}_{\!\lambda}$ for \object{L\,1688} is
presented in Fig.~\ref{oph-l1688}{\;\!\emph{f}}. Compared with the background-subtracted image $\tilde{\mathcal{S}}_{\lambda}$
(Fig.~\ref{oph-l1688}{\;\!\emph{c}}), the scaled image of \object{L\,1688} appears remarkably flat. This visual impression is
further quantified by the image of standard deviations $\mathrm{sd}_{9\,}(\mathcal{I}_{{\!\lambda}\mathrm{D}})$ in
Fig.~\ref{oph-l1688}{\;\!\emph{g}}, and it is confirmed by comparison with the image $\mathcal{D}_{\lambda}$ computed before the
flattening (Fig.~\ref{oph-l1688}{\;\!\emph{d}}). Such greatly simplified (background-subtracted and flat) detection images are
highly beneficial for extracting sources and filamentary structures.

Possible imperfections of the results produced by \textsl{getimages} are related to an inadequate choice of $X_{\lambda}$ and they
must become clearly visible in $\tilde{\mathcal{B}}_{\lambda}$ and $\tilde{\mathcal{F}}_{\!\lambda}$. Indeed,
Figs.~\ref{oph-l1688}{\;\!\emph{b}} and \ref{oph-l1688}{\;\!\emph{e}} show some residuals in places of the vertical filament and of
the two bright sources, above and below the center of the images, respectively. Such levels of inaccuracies in
$\mathcal{I}_{{\!\lambda}\mathrm{D}}$ are not important for source extraction with \textsl{getsources} (Paper I) because the source
measurements are made in the original image $\mathcal{I}_{\!\lambda}$. One might be interested, however, in a more complete
reconstruction and accurate profiling of the prominent filament seen in Fig.~\ref{oph-l1688}. Filament extraction with
\textsl{getfilaments} (Paper II) is made in detection images $\mathcal{I}_{{\!\lambda}\mathrm{D}}$, using the scaling images as
well to obtain correct intensities. To minimize the residuals in $\tilde{\mathcal{B}}_{\lambda}$ and
$\tilde{\mathcal{F}}_{\!\lambda}$, it is sufficient to simply rerun \textsl{getimages} with an increased value of $X_{\lambda}$.

An illustration of the effect of a twice higher value $X_{\lambda}{\,=\,}24${\arcsec} on the resulting images for \object{L\,1688}
is presented in Figs.~\ref{oph-l1688}{\;\!\emph{h}} and \ref{oph-l1688}{\;\!\emph{i}}. An inspection of the images shows that the
larger maximum size leads to lower residuals in $\tilde{\mathcal{F}}_{\!\lambda}$. The corresponding flattened image
$\mathcal{I}_{{\!\lambda}\mathrm{D}}$ displays structures that are brighter and roughly twice as wide (at zero level) than those in
Fig.~\ref{oph-l1688}{\;\!\emph{f}} derived with $X_{\lambda}{\,=\,}12${\arcsec}. With the higher parameter value, the vertical
filament is extracted more completely, and its profile can be determined more accurately. 

Figure~\ref{extraction} shows an example of the source and filament extraction in the \object{L\,1688} field using a flattened
detection image produced by \textsl{getimages} (Fig.~\ref{oph-l1688}{\;\!\emph{i}}). The \textsl{getfilaments} method extracts
filamentary structures in the image and subtracts them before extracting compact sources with \textsl{getsources}. The three
methods carefully separate blended structural components of the original image $\mathcal{I}_{\!\lambda}$ (background, filaments,
and sources) from each other. The linear observational artifacts seen as orthogonal spikes next to bright sources are automatically
identified as ``filaments'' (Fig.~\ref{extraction}{\;\!\emph{a}}) and therefore removed before the source extraction. Owing to the
radically simplified detection image, most of the extracted $1193$ sources (Fig.~\ref{extraction}{\;\!\emph{b}}) in the
\object{L\,1688} field are real. A small number of the remaining spurious sources are located around the unresolved bright peaks
that are due to the artifacts in the observed image, which are induced by the complex shape of the observational beam.

\subsection{Practical definition of the maximum size \textit{X}$_{\lambda}$}
\label{maxsize}

The maximum structure size is the only free parameter of \textsl{getimages}. Before applying the method to an image
$\mathcal{I}_{\!\lambda}$, it is necessary to evaluate $X_{\lambda}$ for the sources and filaments of interest in that image. Since
human eyes are good at identifying sources even on very complex backgrounds, the maximum size can easily be estimated directly from
the observed $\mathcal{I}_{\!\lambda}$ or, alternatively, assumed on the basis of additional considerations. The parameter
$X_{\lambda}$ controls the quality of the derived background, therefore it must be related to the size of a zero-level
footprint, not to the half-maximum size $H_{\lambda}$ of the structures of interest.

Following Papers I and II, footprints are defined here as areas of connected pixels that make a non-negligible contribution to the
total fluxes of sources or filaments. The footprint size $Z_{\lambda}$ is measured in $\mathcal{I}_{\!\lambda}$ at a radial
distance from the peak or crest, where the structure intensity completely merges with background and noise fluctuations (``zero
level''). Therefore, $Z_{\lambda}$ has the meaning of the major axis of an elliptical source or the largest width of a filamentary
structure at zero intensity level after background subtraction. The relationship between $H_{\lambda}$ and $Z_{\lambda}$ varies for
different structures, depending on their intensity distribution, signal-to-noise ratio S/N, and background properties. To define
the half-maximum $X_{\lambda}$ as a proxy to zero-level $Z_{\lambda}$, it is necessary to eliminate the dependence on unknown
intensity profiles.

\begin{figure}
\centering
\centerline{\resizebox{0.85\hsize}{!}{\includegraphics{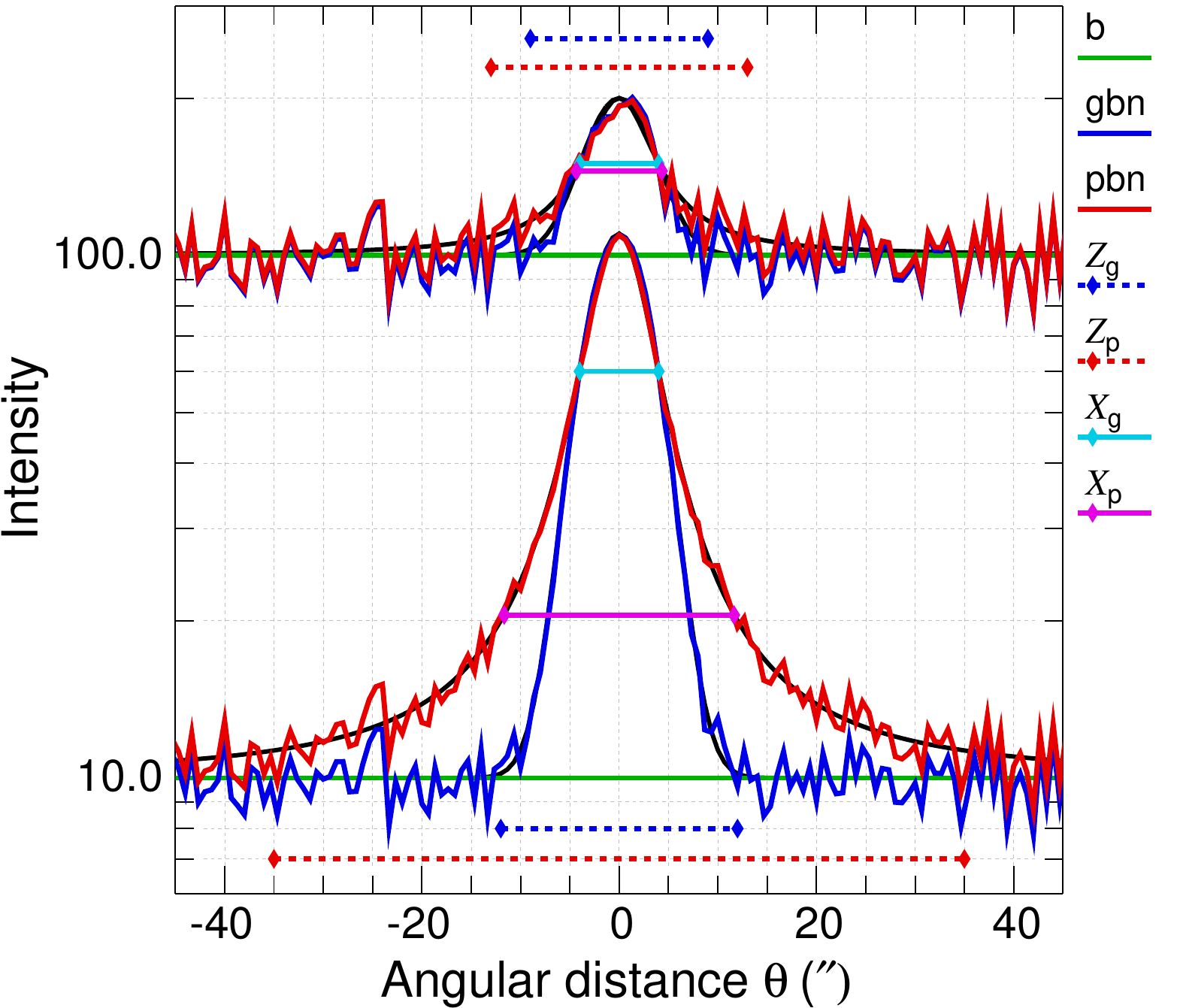}}}
\caption
{ 
Definition of the maximum size $X$. The two sets of curves represent faint (\emph{upper}) and bright (\emph{lower}) sources, with
an S/N of $10$ and $100$. Shown are profiles of Gaussian (\emph{blue}) and power-law (\emph{red}) sources with a peak intensity
$I_{\mathrm{P}}{\,=\,}100$ and an FWHM size $H{\,=\,}8${\arcsec} on a flat background (\emph{green}), affected by random Gaussian
noise (with $\sigma{\,=\,}10$ and $1$). The noise-free profiles are shown by black curves. Dashed horizontal lines (blue and red)
visualize zero-level sizes $Z$ that would be estimated for the two sources and the solid horizontal lines (cyan and magenta) show
the corresponding $X$ values (assuming $\eta{\,=\,}3$). In the high-S/N case, $Z{\,\approx\,}24$ and $70${\arcsec}, whereas
$X{\,=\,}8$ and $23${\arcsec}. In the low-S/N case, $Z{\,\approx\,}18$ and $26${\arcsec}, whereas $X{\,=\,}8$ and $8.7${\arcsec}.
} 
\label{maxsizes}
\end{figure}

It makes sense to convert $Z_{\lambda}$ into $X_{\lambda}$ assuming an equivalent Gaussian intensity profile and determine the
\textsl{getimages} parameter as $X_{\lambda}{\,=\,}\eta^{-1} Z_{\lambda}$, where $\eta$ is a scaling factor within the range of
$2{-}3$. To create $\mathcal{I}_{{\!\lambda}\mathrm{D}}$ exclusively for detection, it is sufficient to adopt $\eta{\,=\,}3$ (see
Fig.~\ref{maxsizes}). To subtract the background more accurately and open the possibility of using
$\mathcal{I}_{{\!\lambda}\mathrm{D}}$ for measurements as well, it is advisable to adopt $\eta{\,=\,}2$, which translates
$Z_{\lambda}$ into larger $X_{\lambda}$.

Incorporating angular resolution $O_{\lambda}$ as a lower limit, \textsl{getimages} defines the parameter as
\begin{equation} 
X_{\lambda} = \max \left\{\,\eta^{-1} Z_{\lambda},\,O_{\lambda}\,\right\}.
\label{maxsizedef}
\end{equation} 
This formulation also gives reasonable values of $X_{\lambda}$ for faint unresolved sources whose $Z_{\lambda}$ values obtained
from the observed images are likely to be underestimated. An obvious upper limit for $X_{\lambda}$ is a reasonably small fraction
(${\la\,}5${\%}) of the image size to prevent the largest windows from sliding beyond the image edges. In practice, it is
recommended to restrict the parameter $X_{\lambda}$ to the lowest values possible, as median filtering with unnecessarily large
windows affects much greater areas of images and hence might make the results less local.

The above definition is illustrated in Fig.~\ref{maxsizes}, which shows model examples of Gaussian and power-law sources with an
FWHM size $H{\,=\,}8${\arcsec} and peak intensity $I_{\mathrm{P}}{\,=\,}100$ (in arbitrary units) on a simple planar background
with additional random noise. The background brightness ($1$ and $10$) and noise fluctuations ($\sigma{\,=\,}1$ and $10$) were
chosen to simulate both bright and faint sources. The bright sources have S/N${\,=\,}100$ (lower curves), whereas the relatively
faint sources correspond to S/N${\,=\,}10$ (upper curves).

An estimation of the footprint size would give values $Z{\,\approx\,}24$ and $70${\arcsec} for the bright Gaussian and power-law
sources, respectively (Fig.~\ref{maxsizes}). According to the definition in Eq.~(\ref{maxsizedef}), the values correspond to
$X{\,=\,}8$ and $23${\arcsec} (for $\eta{\,=\,}3$), and the maximum size for the steep Gaussian intensity distribution equals the
source size $H$. For much fainter sources, the low-intensity outskirts of the source intensity distributions dissolve into much
stronger background and noise fluctuations (Fig.~\ref{maxsizes}). An evaluation of the zero-level size would therefore
underestimate the true source values, giving $Z{\,\approx\,}18$ and $26${\arcsec}, corresponding to $X{\,=\,}8$ and $8.7${\arcsec}.
For the faint sources, $Z$ is underestimated and $X{\,=\,}H$ for the Gaussian source because of the presence of $O_{\lambda}$ in
Eq.~(\ref{maxsizedef}).

From the definition of the maximum size, $X_{\lambda}{\,>\,}H_{\lambda}$ for power-law sources and hence
$R_{\lambda\,N}{\,=\,}4\,X_{\lambda}{\,>\,}4\,H_{\lambda}$ (Sects.~\ref{bgestim} and \ref{flattening}). The model values of
truncation factors $f_{\mathrm{T}}$ presented in Table~\ref{truncation} (Sect.~\ref{sourceremoval}) as a function of $R/H$ are
therefore only lower limits for power-law sources. The actual truncation factors and hence the quality of source removal by median
filtering in \textsl{getimages} are much better. For example, images of the power-law sources profiled in Fig.~\ref{maxsizes} show
that $f_{\mathrm{T}}$ for $R/X{\,=\,}4$ must increase by a factor of $5.6$ in comparison with the $R/H{\,=\,}4$ values listed in
Table~\ref{truncation}.

\section{Conclusions}
\label{conclusions}

This paper described \textsl{getimages}, a new general method of background estimation and image flattening, which solves the two
problems in a multiscale median-filtering approach. The method does not need any preliminary source (or filament) extraction or
any prior information about the location and sizes of sources. A single free parameter of \textsl{getimages}, the maximum size
$X_{\lambda}$ of the structures of interest, can easily be evaluated directly from the observed images $\mathcal{I}_{\!\lambda}$
(Sect.~\ref{maxsize}). The method produces background-subtracted and flattened images that are radically simplified by the removal
of huge intensity variations of large-scale backgrounds (Sect.~\ref{bgestim}) and global equalization of the small-scale
fluctuation levels (Sect.~\ref{flattening}), for all structures with sizes $H_{\lambda}{\,\la\,}X_{\lambda}$. The resulting flat
images $\mathcal{I}_{{\!\lambda}\mathrm{D}}$ are highly advantageous for detecting sources and other structures, whereas
measurements of the physical quantities are usually carried out in the original observed images $\mathcal{I}_{\!\lambda}$. When
accurately derived, the background-subtracted images $\tilde{\mathcal{S}}_{\lambda}$ can also be used for measurements.

Any source or filament extraction method that clearly distinguishes between the detection and measurement images could greatly
benefit from using the flat detection images produced by \textsl{getimages}. Background subtraction and flattening are especially
important for the methods that employ global thresholding to detect sources. Specifically, the \textsl{getsources} and
\textsl{getfilaments} source and filament extraction methods (Papers I and II) are greatly simplified and improved when using the
flat detection images $\mathcal{I}_{{\!\lambda}\mathrm{D}}$. Instead of the two (initial and final) extractions of the original
approach, a single extraction is now sufficient (see Appendix~\ref{AppendixC} for details). Most importantly, the new flattening
algorithm of \textsl{getimages} is much more reliable and universal than the original algorithm.

Observational and map-making artifacts are also eliminated or greatly reduced in the flattened detection images. For example,
mosaicking of several independently observed images of the same field often produces significantly deviating noise levels in the
tiles of a resulting large image. As a bonus, \textsl{getimages} automatically equalizes the levels of noise fluctuations between
the different tiles, which makes the composite detection image seamless.


\begin{acknowledgements} 
This study employed \textsl{SAOImage DS9} (by William Joye) developed at the Smithsonian Astrophysical Observatory (USA), the
\textsl{CFITSIO} library (by William D. Pence) developed at HEASARC NASA (USA), and \textsl{SWarp} (by Emmanuel Bertin) developed
at Institut d'Astrophysique de Paris (France). The \textsl{plot} utility and \textsl{ps12d} library used in this work to draw
figures directly in the \textsl{PostScript} language were written by the author using the \textsl{PSPLOT} library (by Kevin E.
Kohler) developed at Nova Southeastern University Oceanographic Center (USA) and the plotting subroutines from the \textsl{AZEuS}
MHD code (by David A. Clarke and the author) developed at Saint Mary's University (Canada). This work used observations made with
the \emph{Spitzer} Space Telescope, which is operated by the Jet Propulsion Laboratory, California Institute of Technology, under a
contract with NASA. The author appreciates collaborative work within the \emph{Herschel} Gould Belt and HOBYS surveys key projects. 
Useful comments on a draft made by Pierre Didelon helped improve this paper.
\end{acknowledgements} 


\begin{appendix}

\section{Model intensity distributions}
\label{AppendixA}

\begin{figure*}
\centering
\centerline{\resizebox{0.3644\hsize}{!}{\includegraphics{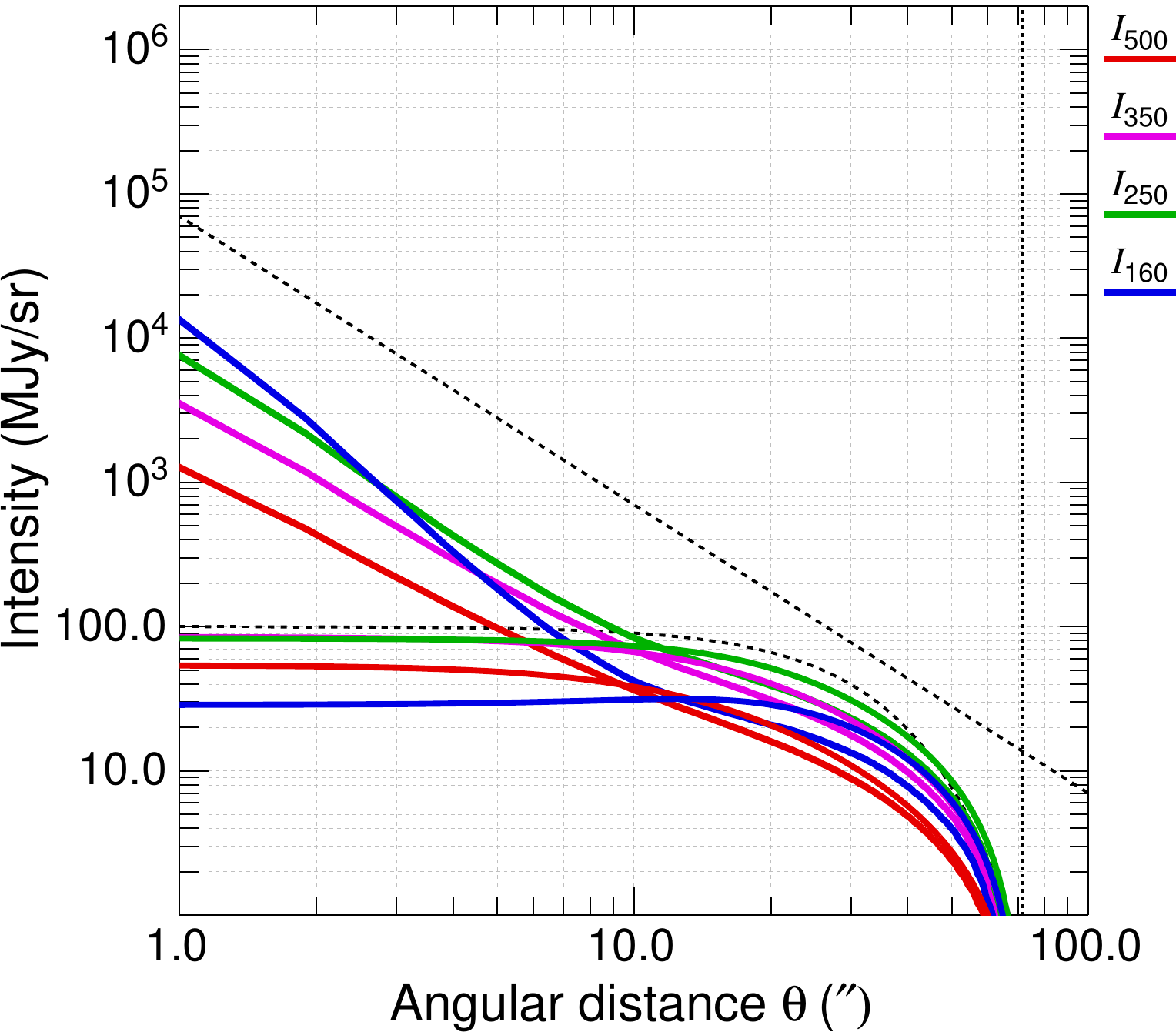}}
            \resizebox{0.3181\hsize}{!}{\includegraphics{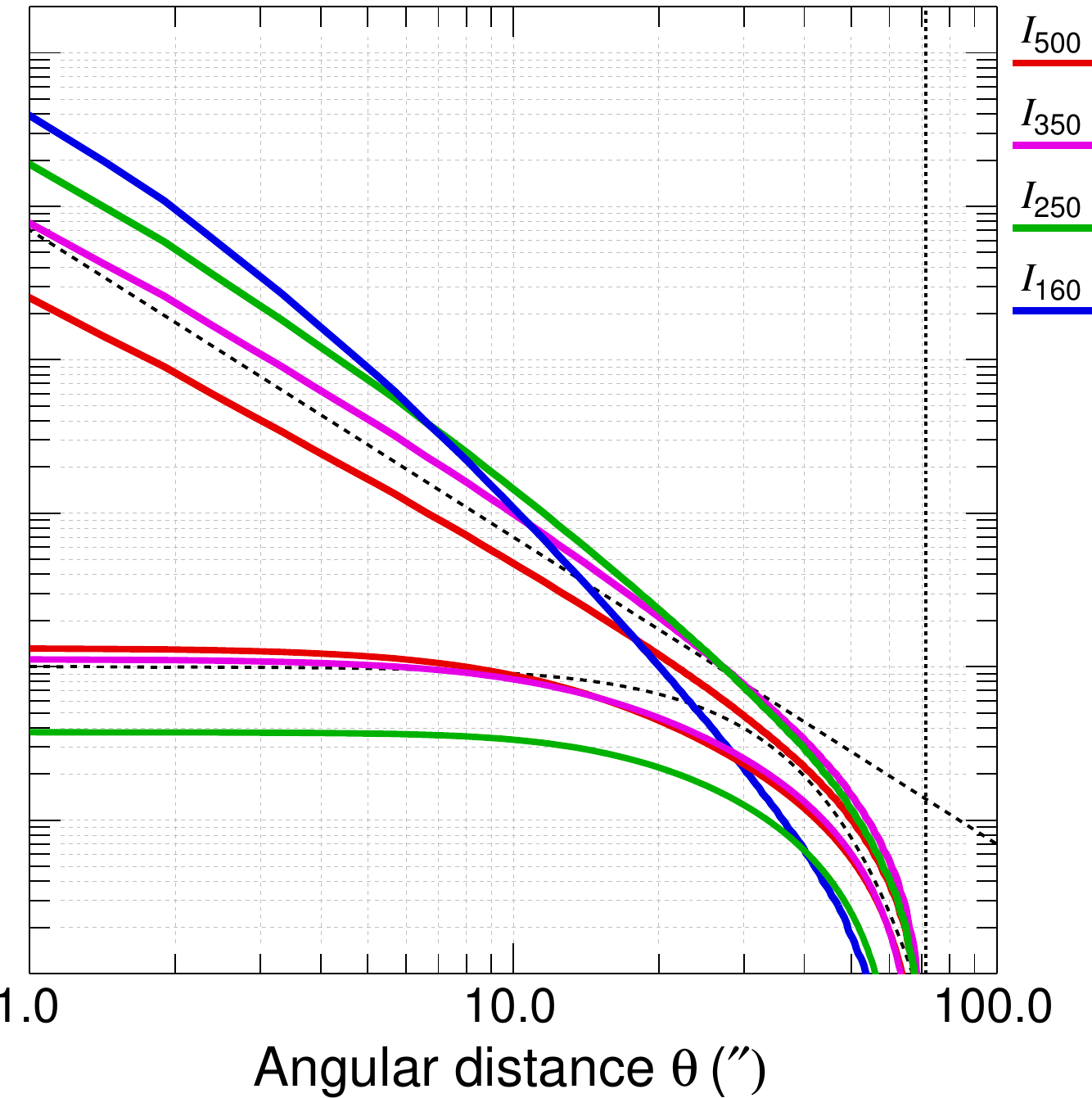}}
            \resizebox{0.3181\hsize}{!}{\includegraphics{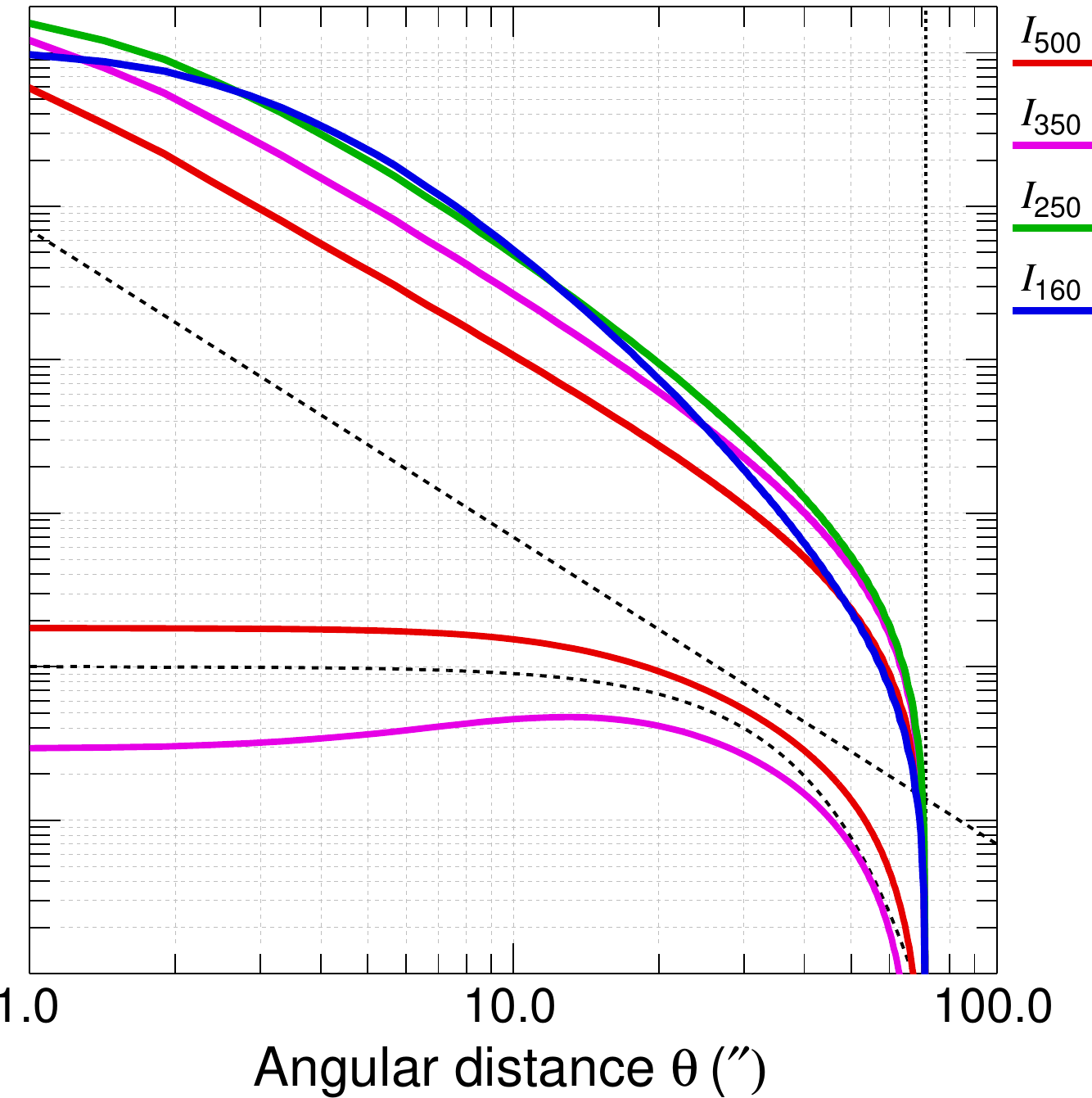}}}
\caption
{ 
Intensity profiles of radiative transfer models. The three panels from left to right display results for masses of $0.3$, $3$, and
$30\,M_{\sun}$. In each panel, the lower and upper sets of curves show background-subtracted profiles of starless cores and
protostellar envelopes, respectively, for selected \emph{Herschel} wavelengths of $160$, $250$, $350$, and $500$\,{${\mu}$m}. For
reference, the black dashed curves visualize a Gaussian profile and a power-law slope $I_{\lambda}{\,\propto\,}\theta^{\;\!-2}$,
whereas the dotted vertical line indicates the outer radius of the models.
} 
\label{profiles}
\end{figure*}

This section employs radiative transfer modeling to justify the choice of simple intensity shapes in Sect.~\ref{sourceremoval}. It
presents intensity profiles in selected \emph{Herschel} bands for several radiative transfer models of spherical starless cores and
protostellar envelopes with masses $0.3$, $3$, and $30\,M_{\sun}$, calculated and described in detail by \cite{Men'shchikov2016}.
The starless cores have a flat-topped density structure of an isothermal Bonnor-Ebert sphere \citep{Bonnor1956}. The protostellar
envelopes have a power-law radial density distribution $\rho(r){\,\propto\,}r^{\;\!-2}$ and accretion luminosity with the numerical
value (in $L_{\sun}$) of the model mass. The outer boundary of the models was placed at a radial distance of $10^{4}$\,AU, beyond
which an embedding constant-density cloud was assumed. At the adopted distance of $140$ pc, the outer model radius corresponds to
$71${\arcsec}. For more details of the radiative transfer models, see Sect.\,2 in \cite{Men'shchikov2016}.

Figure~\ref{profiles} displays intensity profiles for both starless cores and protostellar envelopes after subtracting the
background emission produced by the embedding cloud. For simplicity, obvious angular resolution effects are ignored, therefore the
curves show the true distribution of the model intensities (with a pixel size of $0.67${\arcsec}). The intensity profiles of the
starless cores resemble Gaussians and those of the protostellar envelopes can be approximated by a power law
$I_{\lambda}{\,\propto\,}\theta^{\;\!-2}$. Although the profiles of radiative transfer models are wavelength-dependent and more
complex than the simple Gaussian and power-law shapes, the latter are reasonable choices to illustrate the source removal by median
filtering in Sect.~\ref{sourceremoval}.

\section{Images of simulated structures}
\label{AppendixB}

\begin{figure*}                                                               
\centering
\centerline{\resizebox{0.246\hsize}{!}{\includegraphics{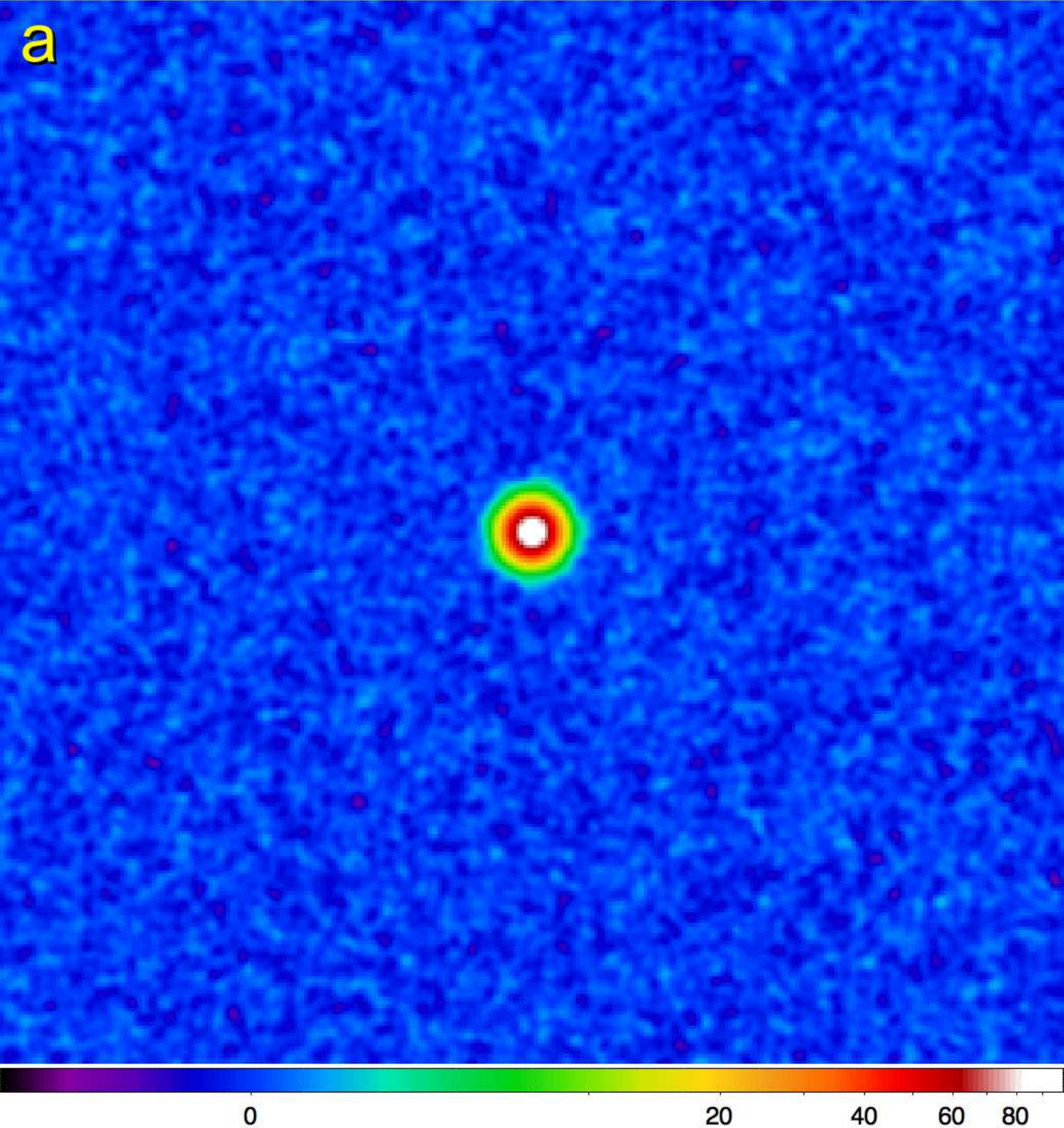}}
            \resizebox{0.246\hsize}{!}{\includegraphics{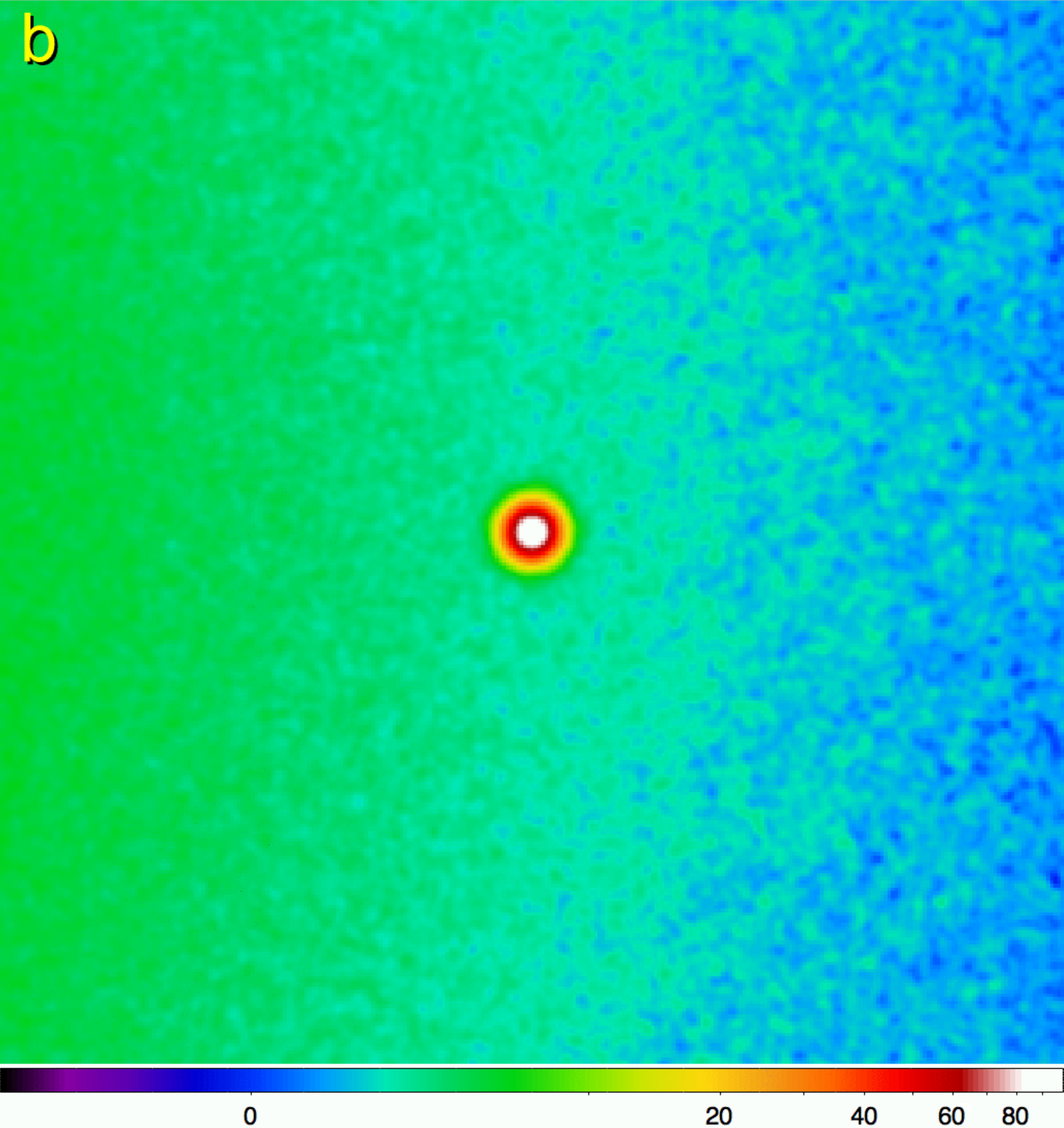}}
            \resizebox{0.246\hsize}{!}{\includegraphics{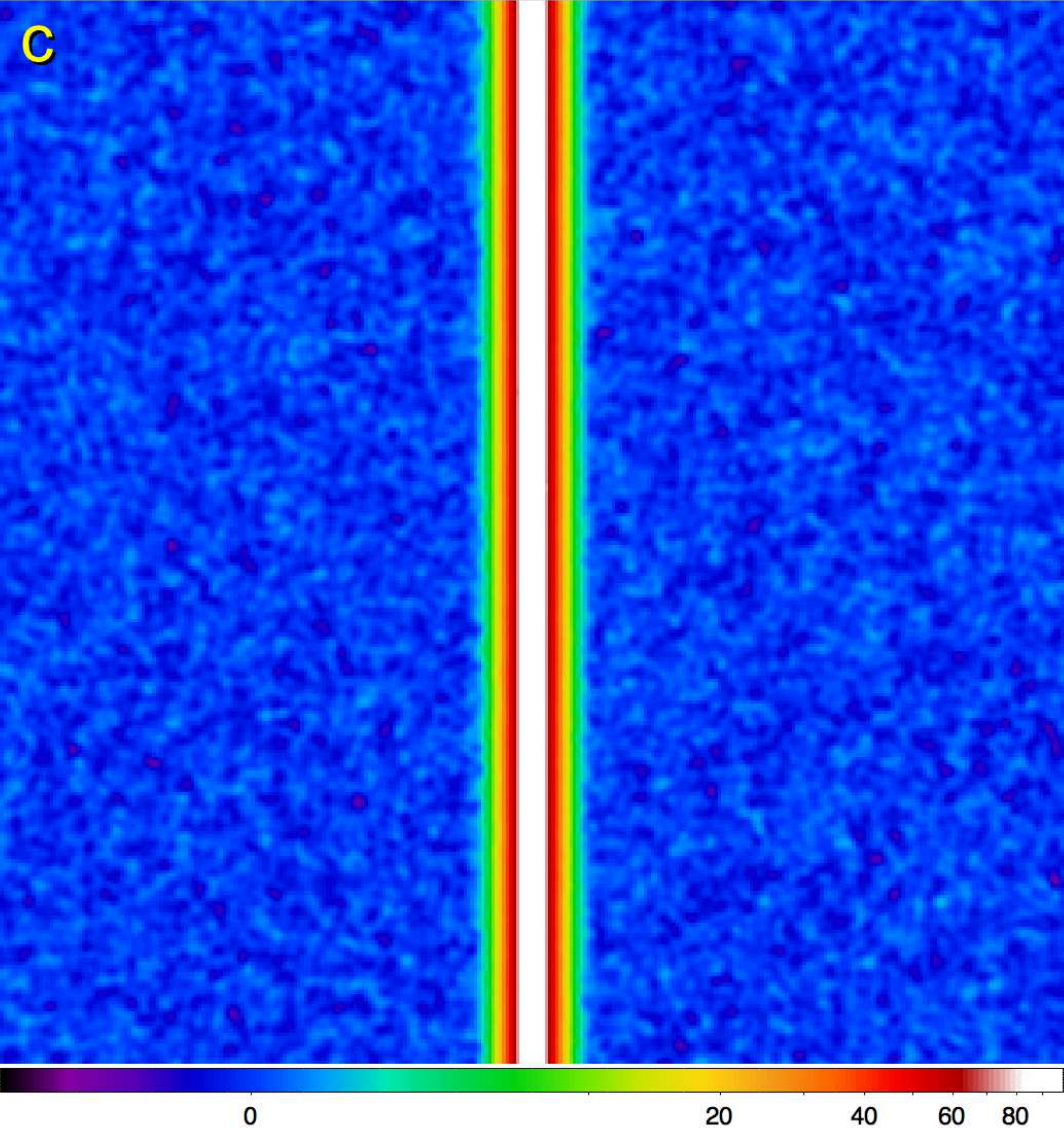}}
            \resizebox{0.246\hsize}{!}{\includegraphics{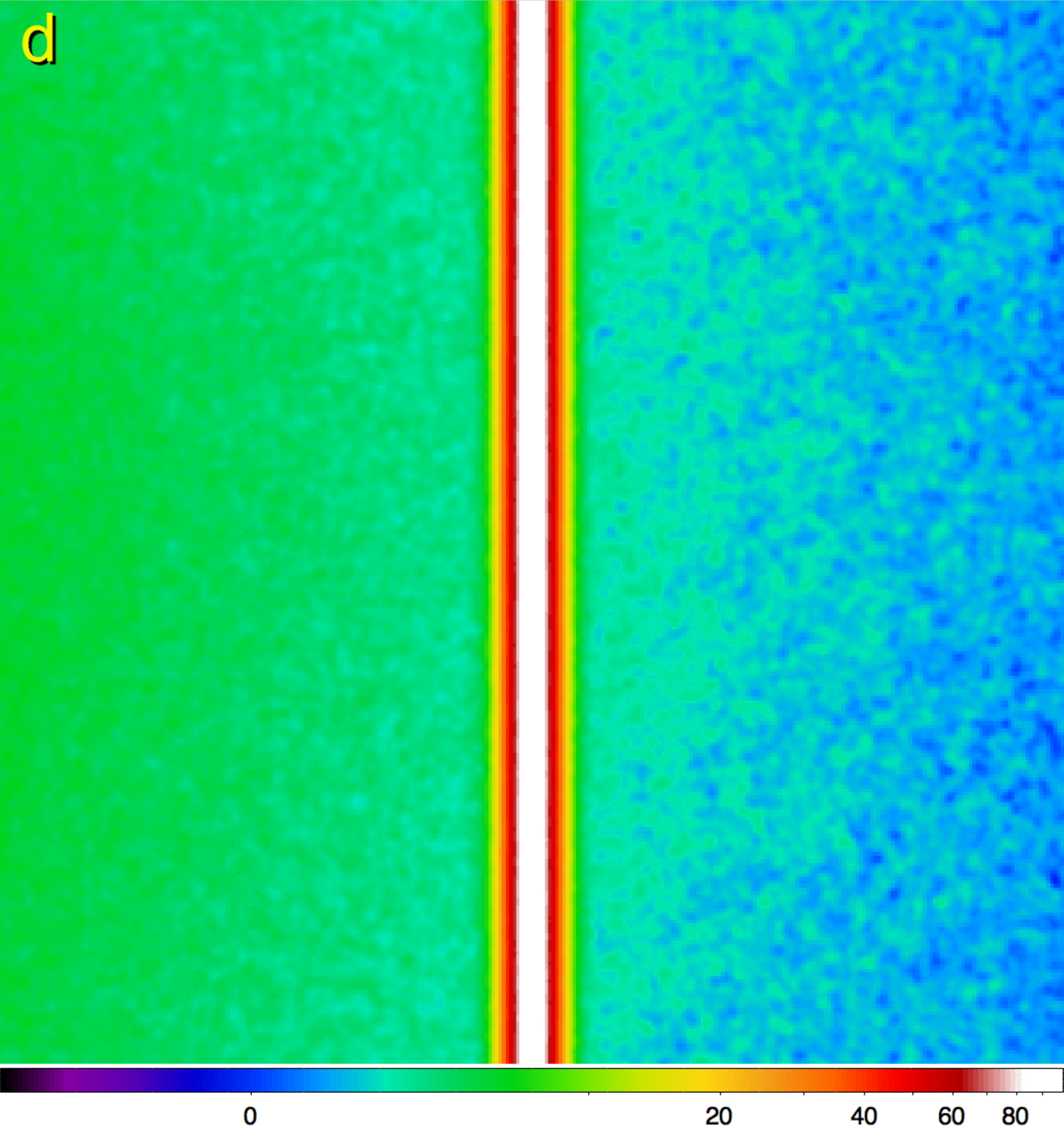}}}
\centerline{\resizebox{0.246\hsize}{!}{\includegraphics{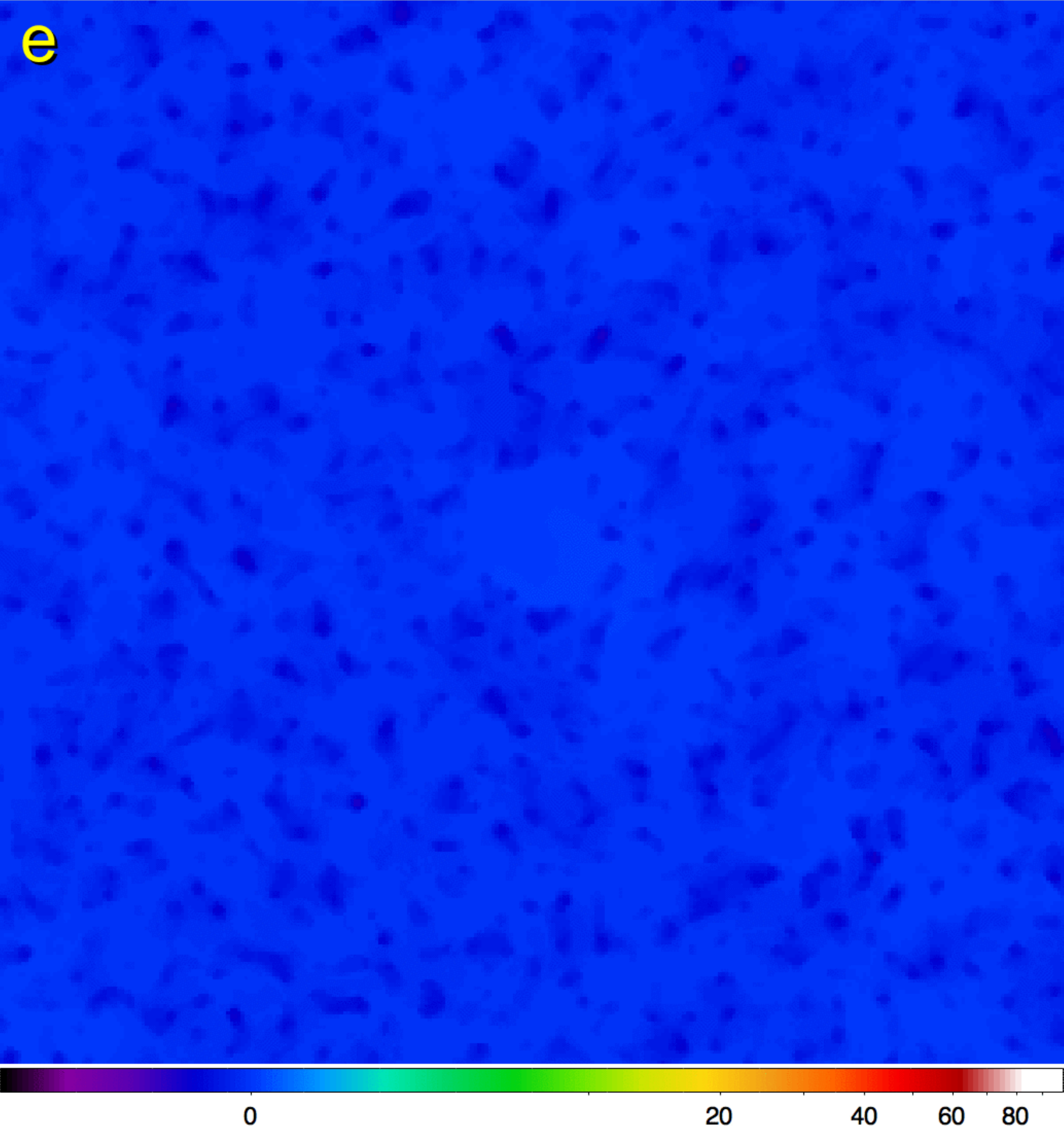}}
            \resizebox{0.246\hsize}{!}{\includegraphics{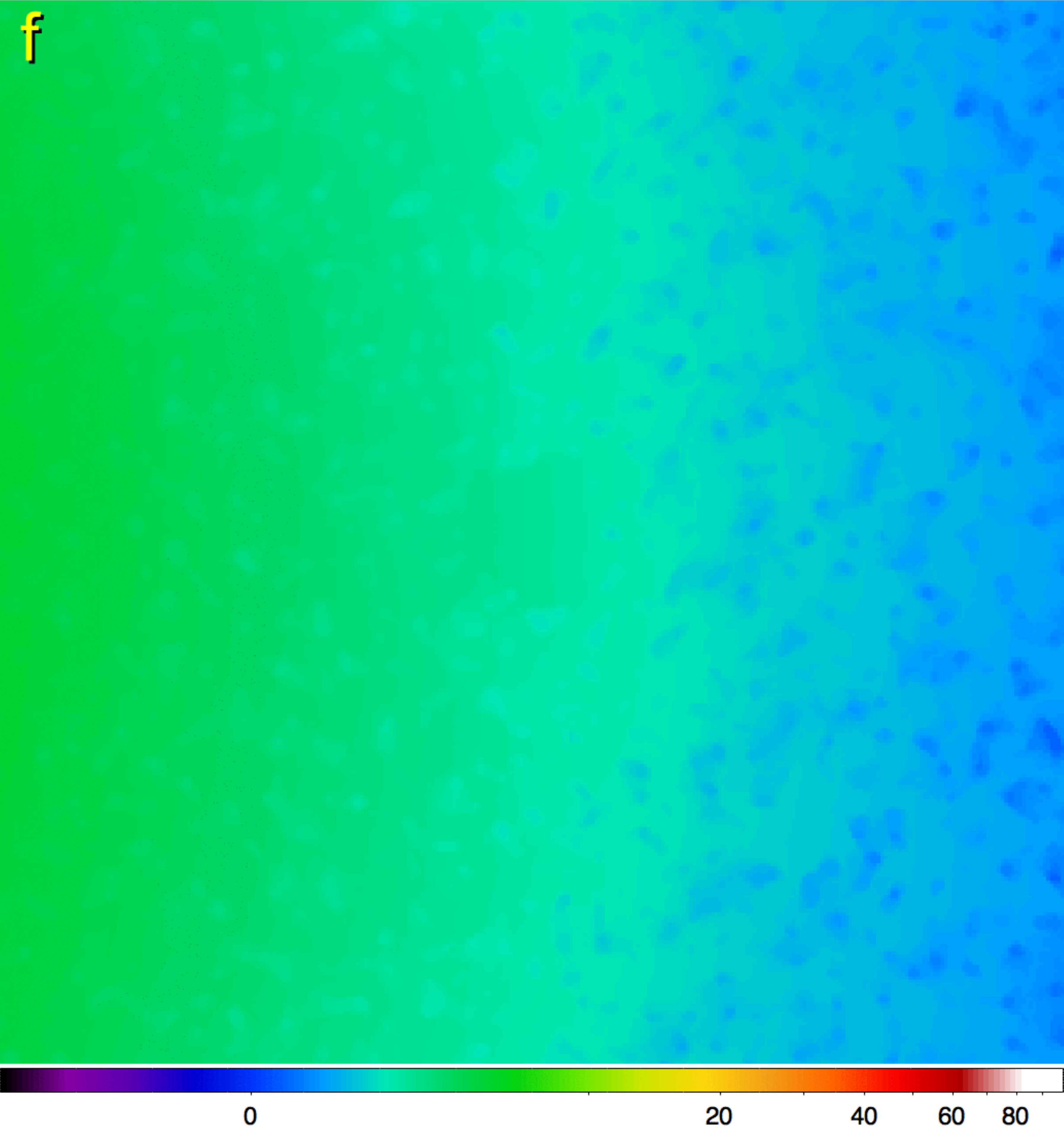}}
            \resizebox{0.246\hsize}{!}{\includegraphics{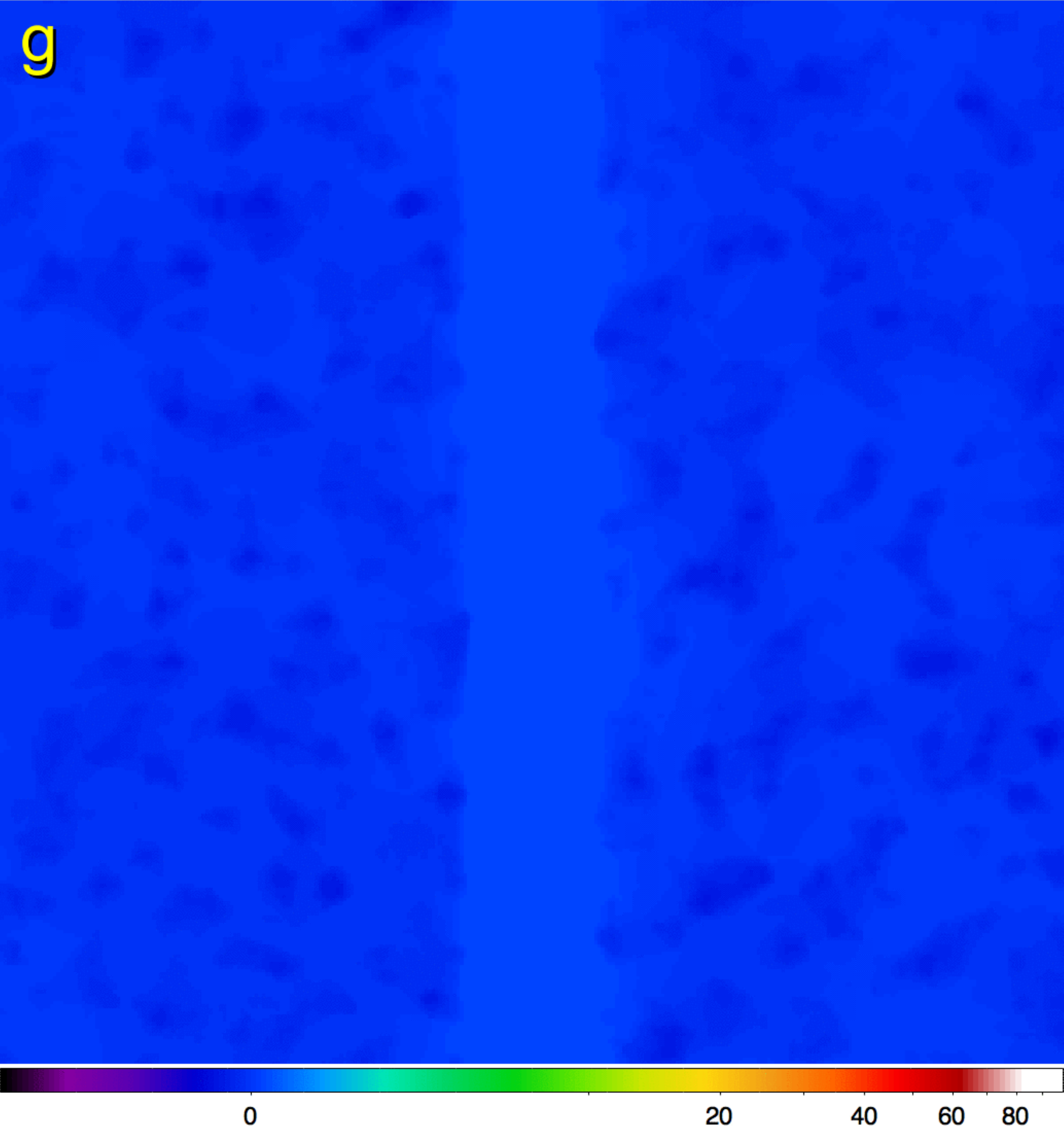}}
            \resizebox{0.246\hsize}{!}{\includegraphics{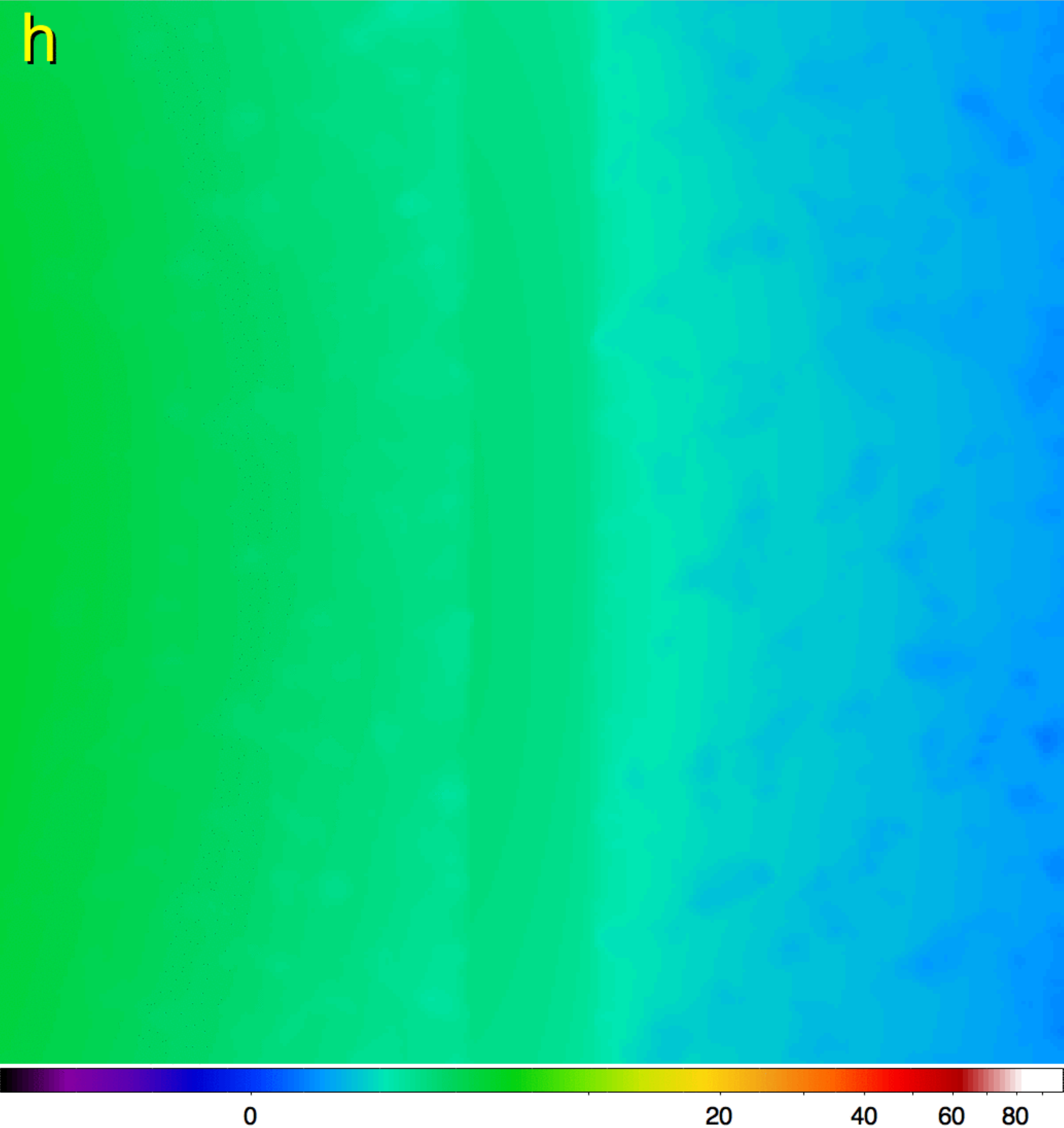}}}
\centerline{\resizebox{0.246\hsize}{!}{\includegraphics{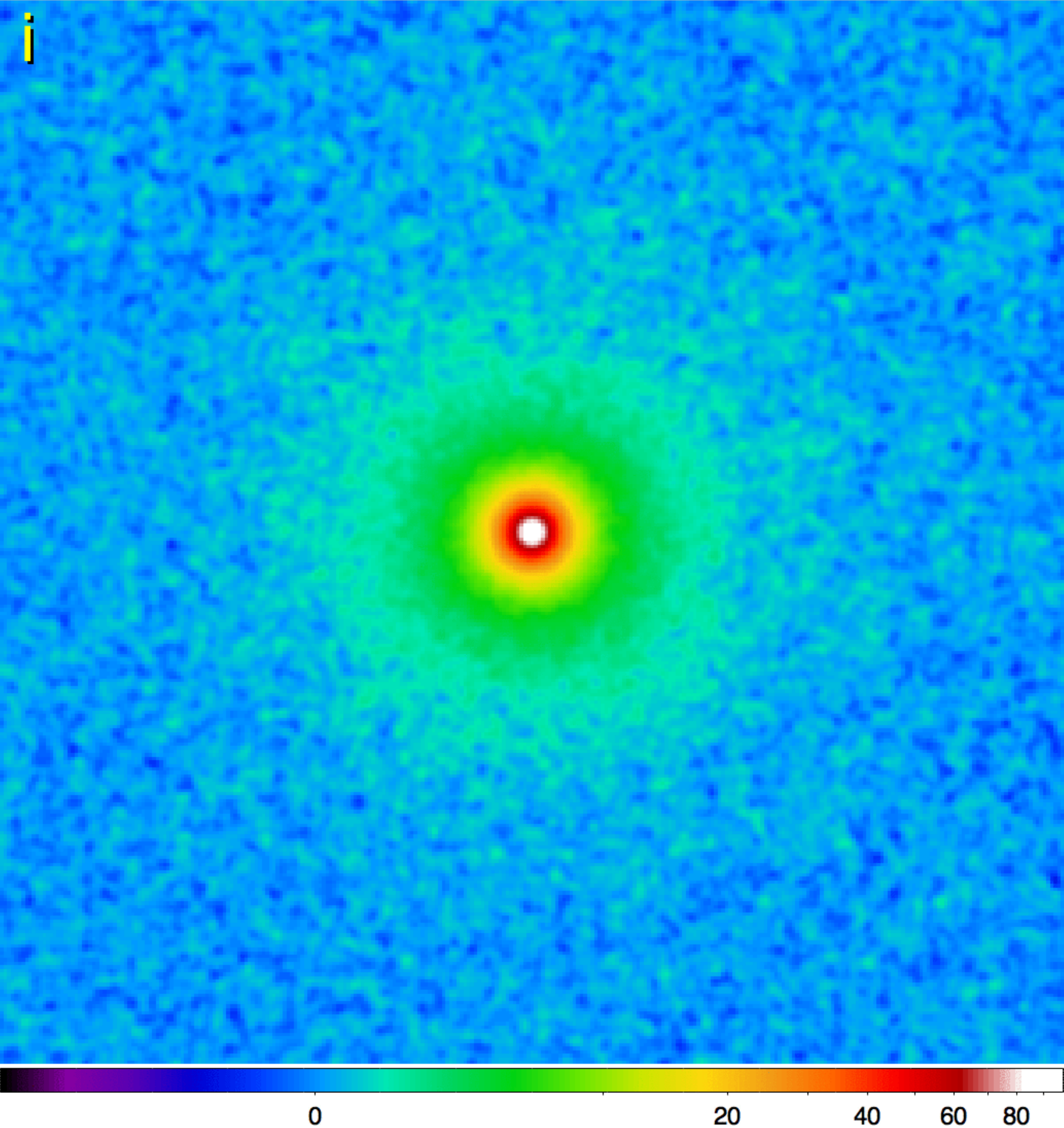}}
            \resizebox{0.246\hsize}{!}{\includegraphics{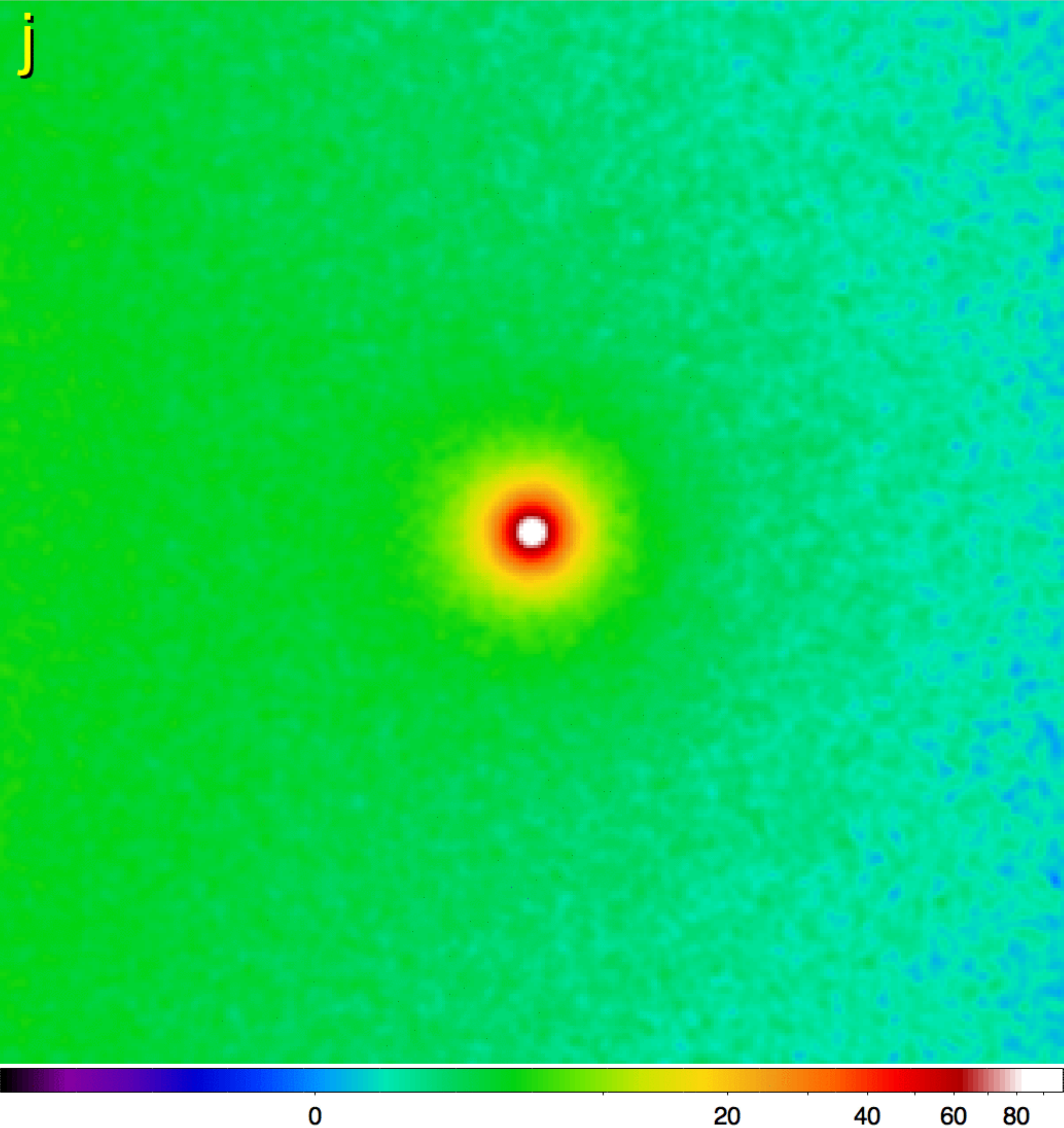}}
            \resizebox{0.246\hsize}{!}{\includegraphics{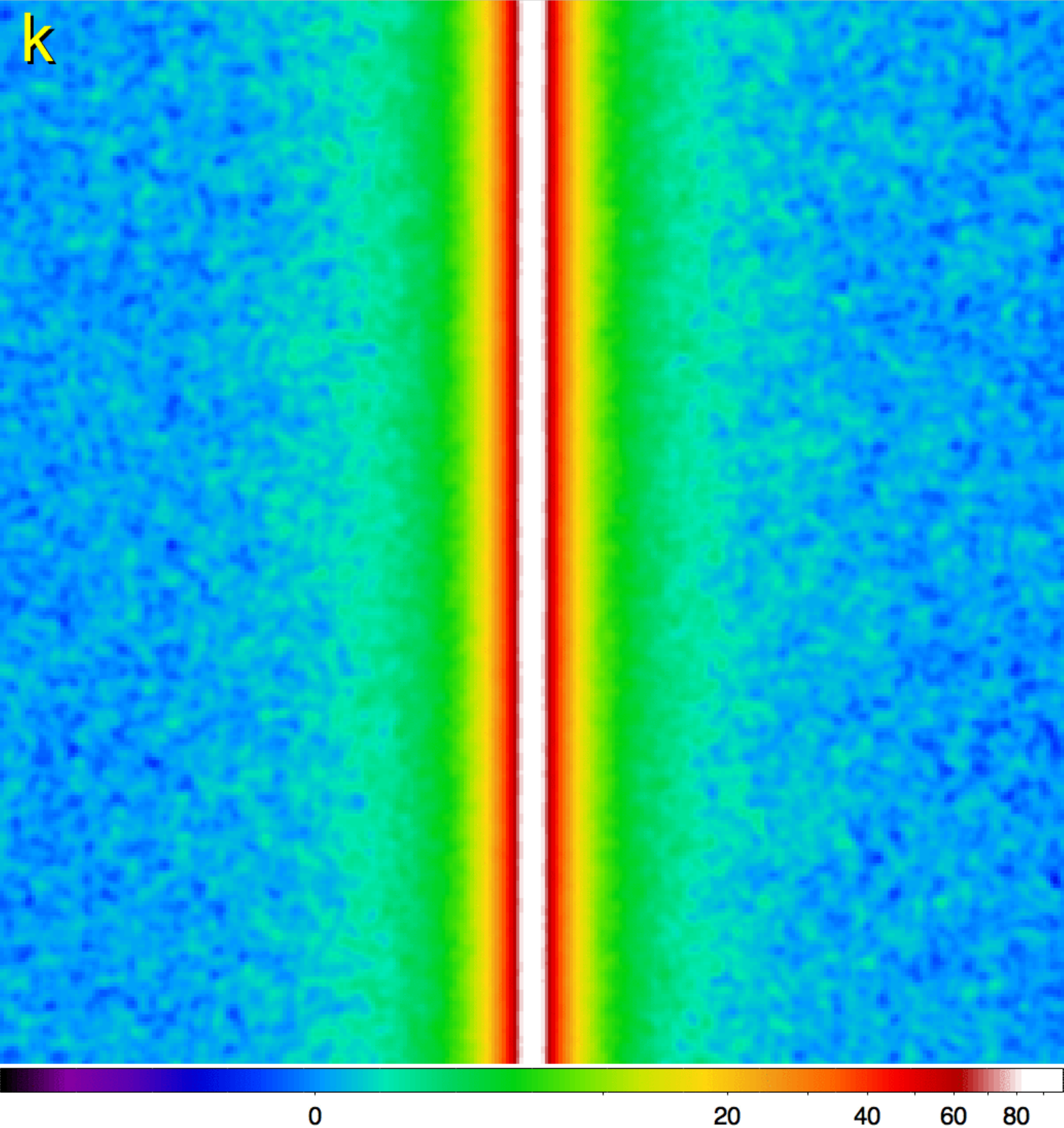}}
            \resizebox{0.246\hsize}{!}{\includegraphics{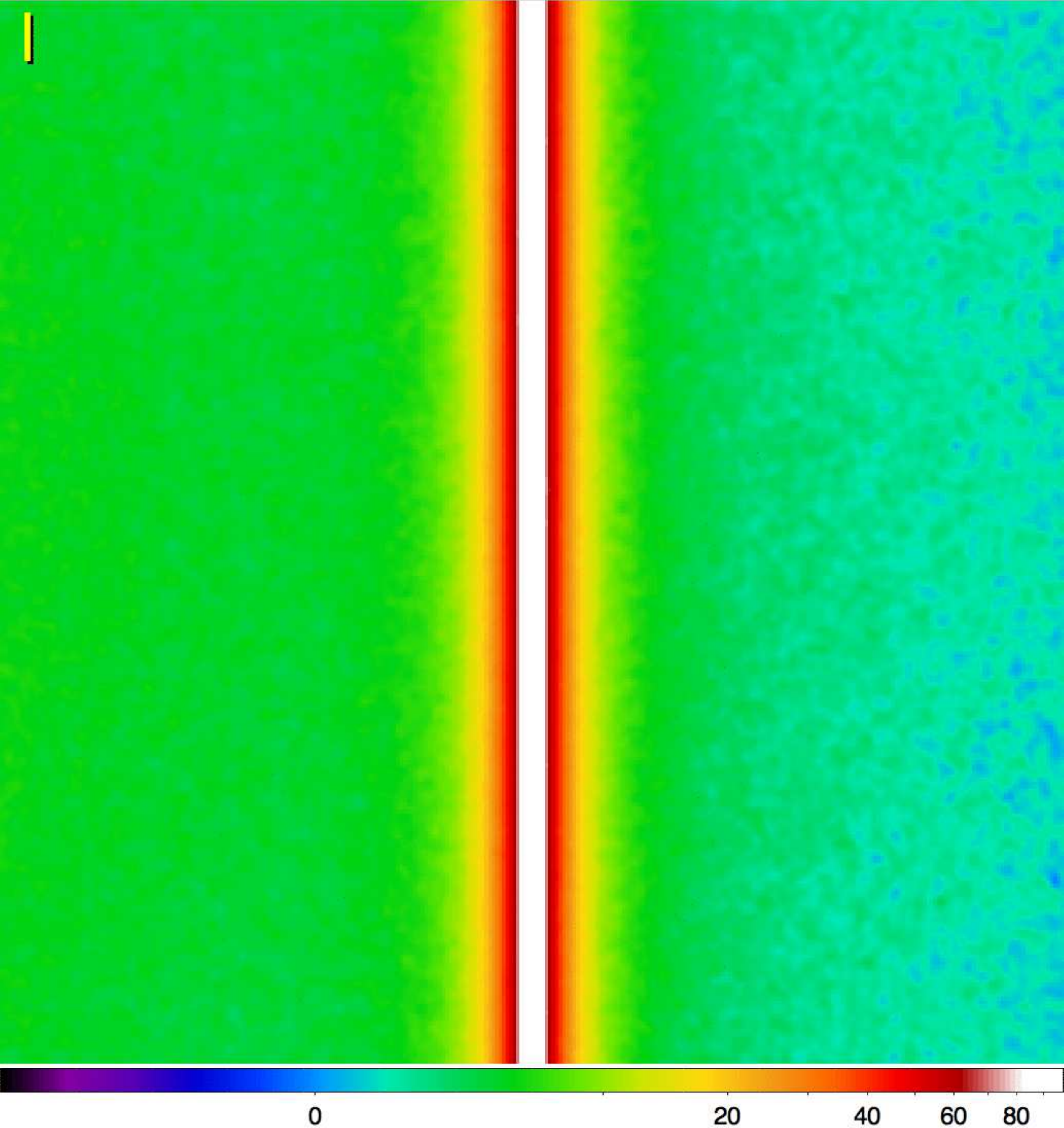}}}
\centerline{\resizebox{0.246\hsize}{!}{\includegraphics{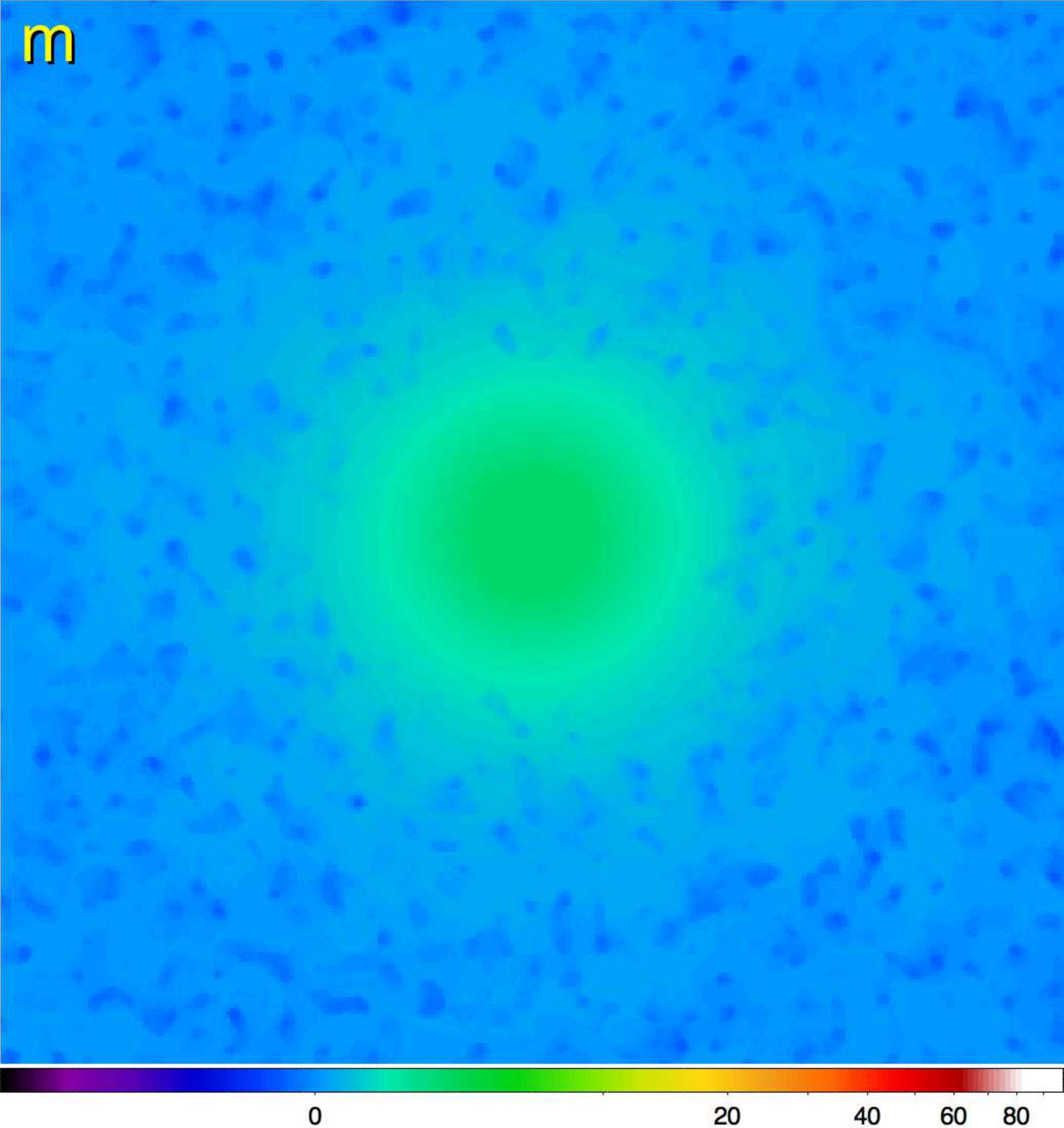}}
            \resizebox{0.246\hsize}{!}{\includegraphics{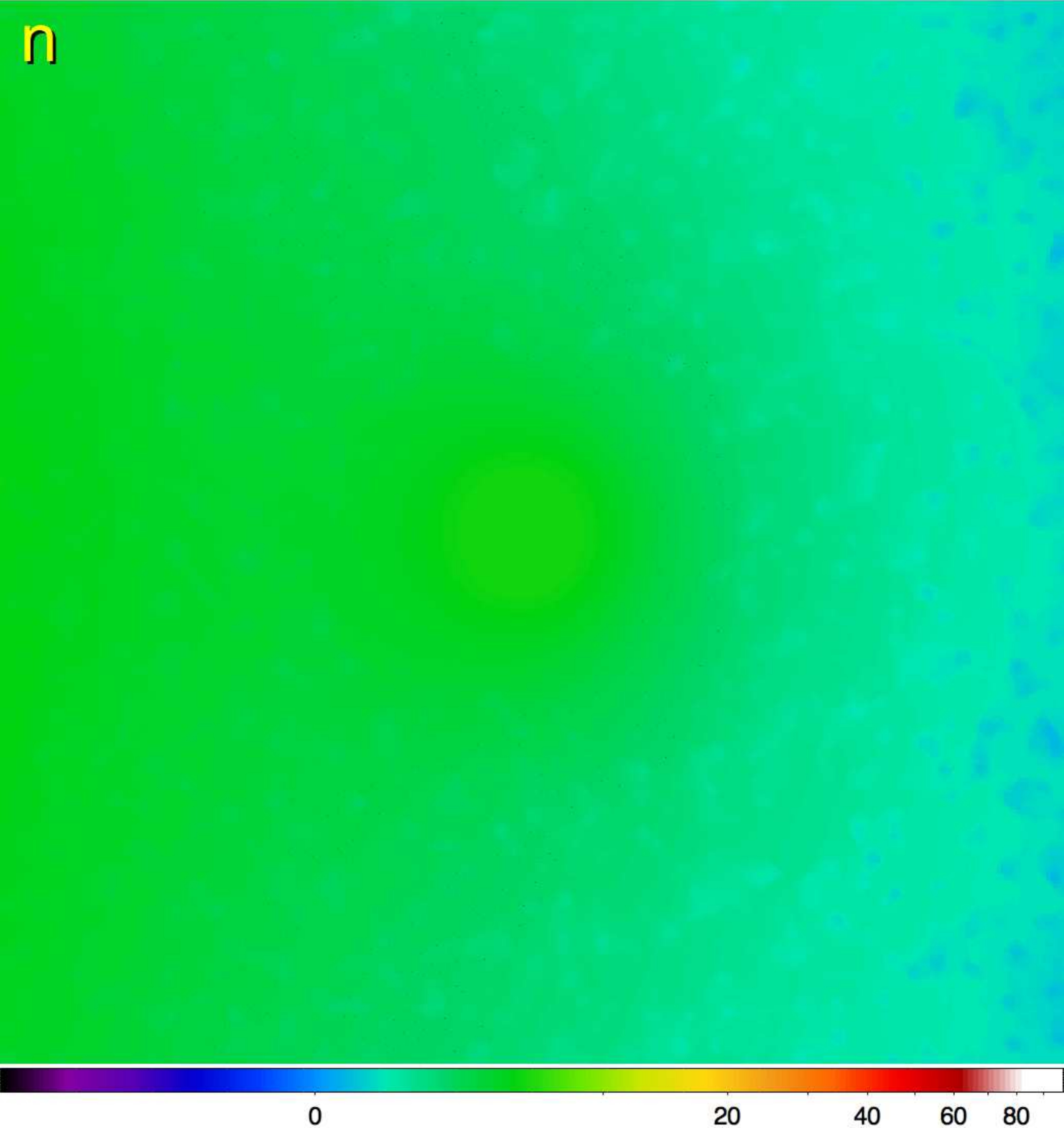}}
            \resizebox{0.246\hsize}{!}{\includegraphics{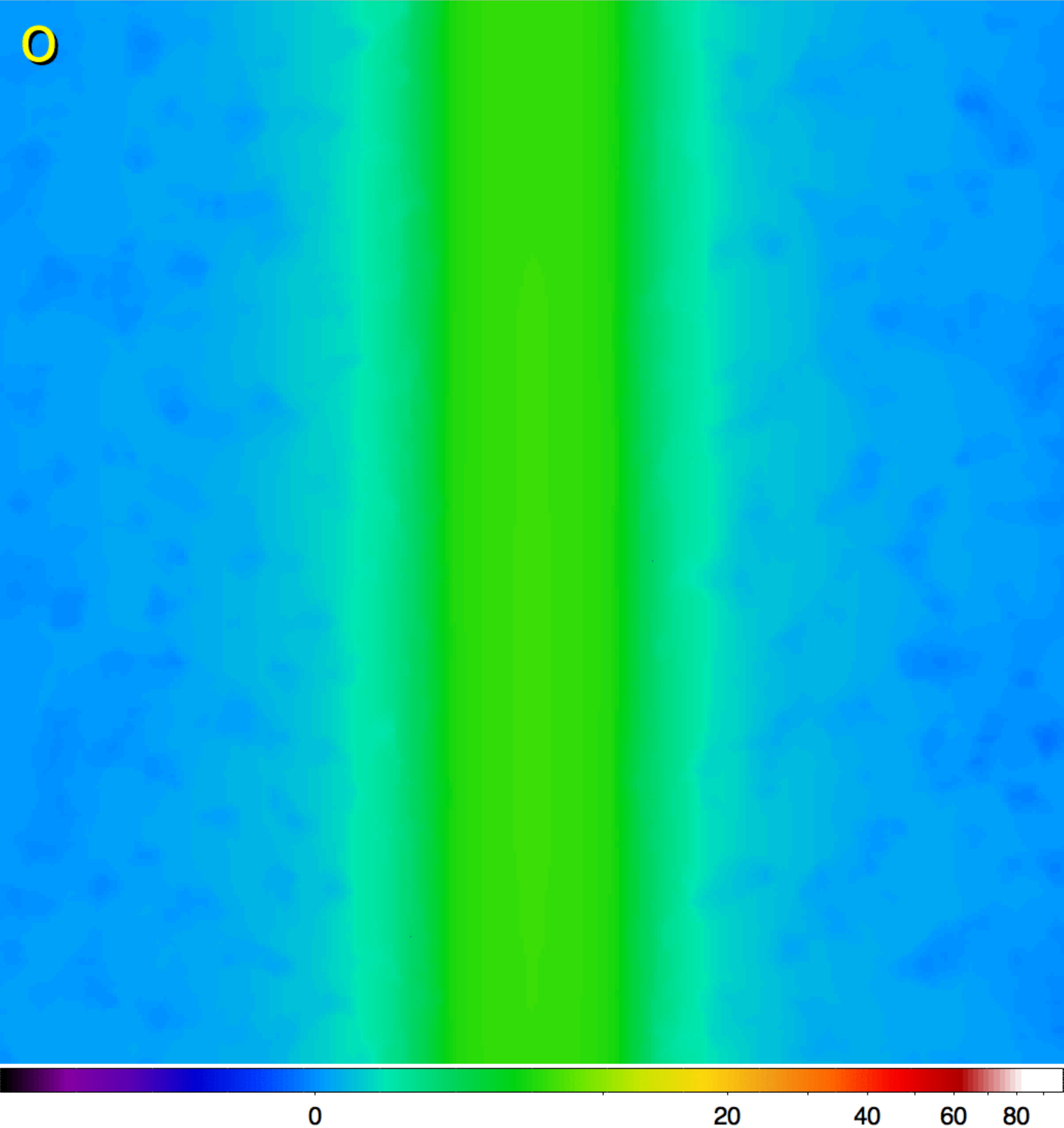}}
            \resizebox{0.246\hsize}{!}{\includegraphics{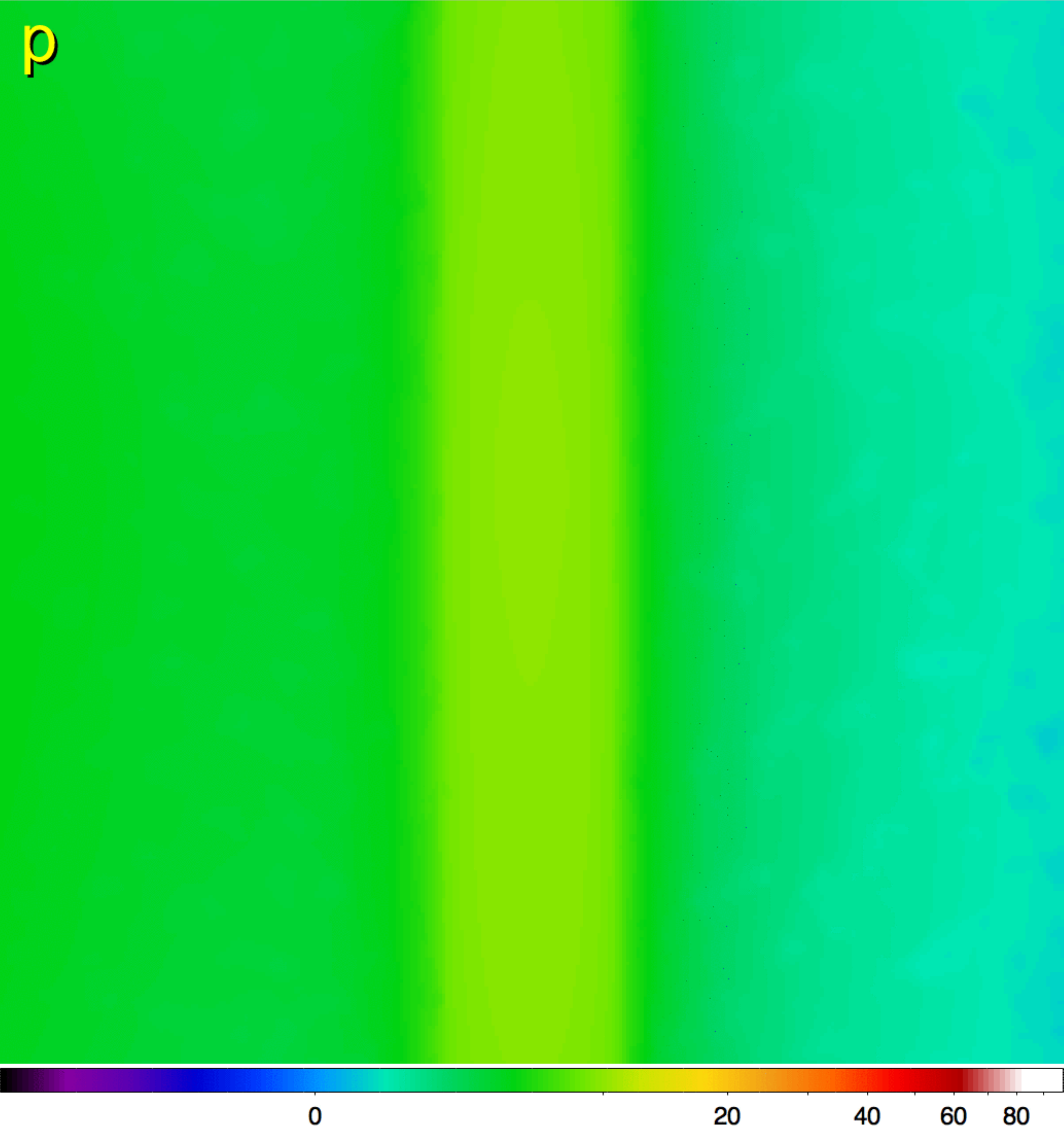}}}
\caption
{ 
Simulated sources and filaments with $H{\,=\,}8${\arcsec} and their median-filtered $W{\,=\,}4$ counterparts (cf.
Sect.~\ref{sourceremoval}). The two upper rows show the simulated Gaussian structures ($\mathcal{S}_\mathrm{G}{+}\mathcal{N}$,
$\mathcal{S}_\mathrm{G}{+}\mathcal{N}{+}\mathcal{B}$, $\mathcal{F}_\mathrm{G}{+}\mathcal{N}$,
$\mathcal{F}_\mathrm{G}{+}\mathcal{N}{+}\mathcal{B}$: panels \emph{a}{\,--\,}\emph{d}) and their median-filtered images
(\emph{e}{\,--\,}\emph{h}). The two bottom rows show the simulated power-law structures ($\mathcal{S}_\mathrm{P}{+}\mathcal{N}$,
$\mathcal{S}_\mathrm{P}{+}\mathcal{N}{+}\mathcal{B}$, $\mathcal{F}_\mathrm{P}{+}\mathcal{N}$,
$\mathcal{F}_\mathrm{P}{+}\mathcal{N}{+}\mathcal{B}$: panels \emph{i}{\,--\,}\emph{l}) and their median-filtered images
(\emph{m}{\,--\,}\emph{p}). The images with an intensity range of $[-0.3, 100]$ and sizes of $200${\arcsec} are profiled in
Fig.~\ref{gausspower}. The images are displayed with logarithmic color scaling.
} 
\label{examples}
\end{figure*}

The structure removal by median filtering described in Sect.~\ref{sourceremoval} is further illustrated in Fig.~\ref{examples} by
images of the simulated sources and filaments with sizes of $H{\,=\,}8${\arcsec} and median-filtered images obtained with
$W{\,=\,}4$. The intensity profiles shown in Fig.~\ref{gausspower} and truncation factors listed in Table~\ref{truncation} were
measured along horizontal lines through the image centers.

\section{Implications for source and filament extraction with \textsl{getsources} and \textsl{getfilaments}}
\label{AppendixC}

\begin{figure}
\centering
\centerline{\resizebox{0.63\hsize}{!}{\includegraphics{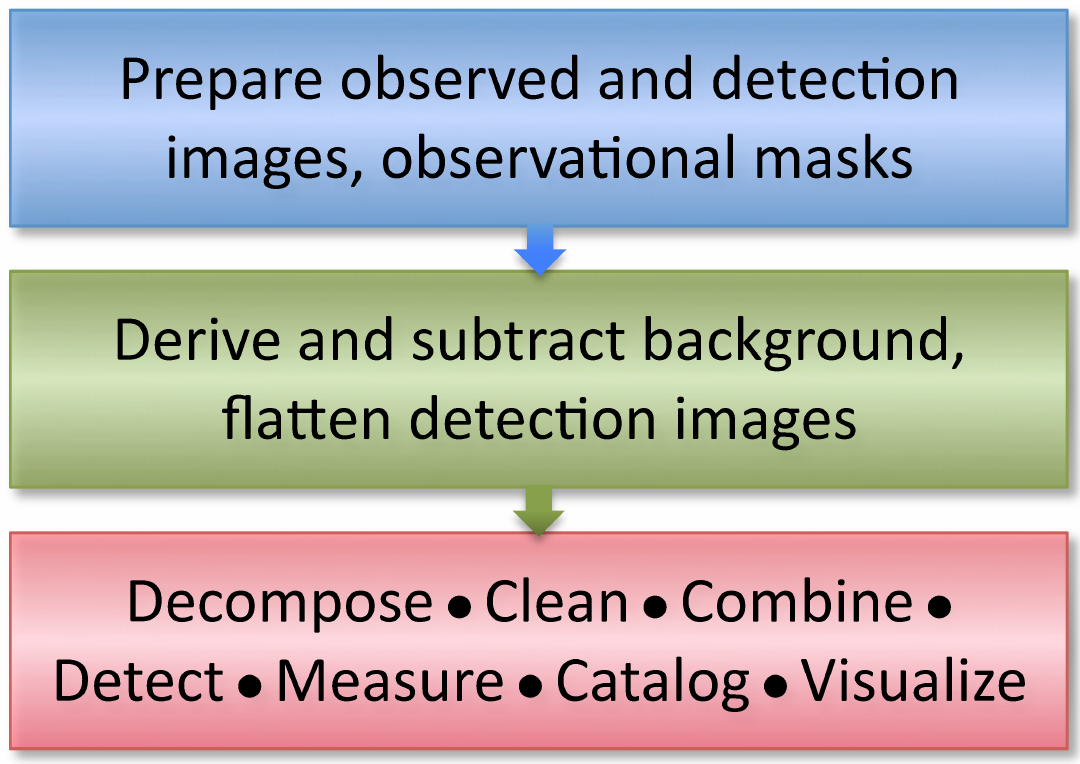}}}
\caption
{ 
Flowchart of the extraction approach with \textsl{getsources} and \textsl{getfilaments} when using the flattened detection images
produced by \textsl{getimages}. The three blocks represent the \textsl{prepareobs} script (\emph{blue}), the \textsl{getimages}
method (\emph{green}), and the \textsl{getsources} (\textsl{getfilaments}) method (\emph{red}).
} 
\label{flowchart}
\end{figure}

This section describes an updated approach to using \textsl{getimages} in combination with the extraction method presented in
Papers I and II. The old strategy involved an image preparation step and two full extractions separated by a flattening step (cf.
Fig.~20 in Paper I and Fig.~1 in Paper II). The new approach, outlined in Fig.~\ref{flowchart}, requires only a single run of
\textsl{getsources} (\textsl{getfilaments}), preceded by the runs of the image preparation script \textsl{prepareobs} (part of
\textsl{getsources}) and of the new \textsl{getimages} method. The latter was coded in a \textsl{Bash} script\footnote{The script
is freely available on \url{http://gouldbelt-herschel.cea.fr/getimages} or \url{http://ascl.net/1705.007}; it can also be obtained
(with support) from the author.} as a series of calls to the FORTRAN utilities of \textsl{getsources} (cf. Appendix G in Paper I),
and it uses a configuration file that is a subset of the \textsl{getsources} configuration file.

With the new approach and a single extraction run, default values of several parameters in the \textsl{getsources} configuration 
file need to be modified. The maximum sizes $X_{\lambda}$ used in the \textsl{getimages} run replace the old default value of 
\texttt{220}. For example, if $X_{\lambda}$ equals twice the \emph{Herschel} beam size $O_{\lambda}$, the values are used as the 
third parameter on the following lines:
\begin{verbatim}
 070  8.4 16.8 0 1 y 0 | .................
 100  9.4 18.8 0 1 y 0 | .................
 160 13.5 27.0 0 1 y 0 | .................
 250 18.2 36.4 0 1 y 0 | .................
 350 24.9 49.8 0 1 y 0 | .................
 500 36.3 72.6 0 1 y 0 | .................
\end{verbatim}
When the flattened detection images are produced in a prior run of \textsl{getimages}, the parameter \texttt{flattening} must 
be set to \texttt{n}:
\begin{verbatim}
 n         <--user.. | flattening dofl ...
\end{verbatim}
The parameters \texttt{sreliable}, \texttt{stentative}, and \texttt{contranoise} must have a plus in front of them, and 
the default values of \texttt{nsigmacutss}, \texttt{nsigmacutfloor} must be set to \texttt{6}:
\begin{verbatim}
+7          ..user.. | sreliable .........
+5          ..user.. | stentative ........
 6          ..expert | nsigmacutss .......
 6          ..expert | nsigmacutfloor ....
+1.3        ..expert | contranoise .......
\end{verbatim}
These are the only differences with respect to the default configuration parameters of \textsl{getsources} used in the original 
strategy described in Paper I.

\end{appendix}


\bibliographystyle{aa}
\bibliography{aamnem99,getimages}

\end{document}